\documentclass[sigconf, table]{acmart}
\setcopyright{none}

\usepackage{xspace}
\usepackage{xcolor}
\usepackage{amsmath}
\usepackage{enumitem}
\usepackage{pifont}
\usepackage[T1]{fontenc}

\definecolor{light-gray}{gray}{0.95}
\newcommand{\code}[1]{\colorbox{light-gray}{\texttt{#1}}}

\usepackage{hyperref}
\hypersetup{linkcolor=black,citecolor=black,anchorcolor=black,filecolor=black,menucolor=black,runcolor=black,urlcolor=black,hidelinks}
\usepackage{breakurl}

\usepackage{booktabs}
\usepackage{multirow}
\usepackage{makecell}
\usepackage{ragged2e}

\usepackage{caption}
\usepackage{subcaption}
\usepackage{graphicx}

\usepackage{listings}

\usepackage{tikz}
\usetikzlibrary{calc}
\usetikzlibrary{shapes.geometric}
\usetikzlibrary{decorations.pathreplacing}
\usetikzlibrary{positioning}
\tikzstyle{every picture}+=[remember picture]
\tikzstyle{na} = [baseline=1ex]

\usepackage{datetime} %
\usepackage{seqsplit} %
\usepackage{textcomp}

\newcommand{\XSpace}[1]{}
\newcommand{\XComment}[1]{}
\newcommand{\Fix}[1]{\textcolor{red}{#1}}
\newcommand{\mg}[1]{\textcolor{purple}{#1}}
\newcommand{\ol}[1]{\textcolor{teal}{#1}}
\newcommand{\yl}[1]{\textcolor{orange}{#1}}
\newcommand{\pn}[1]{\textcolor{blue}{#1}}
\newcommand{\EditAdd}[1]{\textcolor{green}{[#1]}}
\newcommand{\EditRm}[1]{\textcolor{red}{[\sout{#1}]}}
\newcommand{\EditMod}[2]{\textcolor{red}{[\sout{#1}]}\textcolor{green}{[#2]}}
\newcommand{\DefMacro}[2]{\expandafter\newcommand\csname rmk-#1\endcsname{#2}}
\newcommand{\UseMacro}[1]{\csname rmk-#1\endcsname}

\newcommand{\MyPara}[1]{\vspace{1pt}\noindent\textbf{#1}.}
\newcommand{\MyParaOnly}[1]{\noindent\textbf{#1}}
\newcommand{\MyItPara}[1]{\noindent\textit{#1}.}
\newcommand{\reducedstrut}{\vrule width 0pt height .9\ht\strutbox depth
  .9\dp\strutbox\relax}
\newcommand{\InputWithSpace}[1]{\bgroup\def\arraystretch{1.1}\input{#1}\egroup}
\newcommand{\Code}[1]{{\ifmmode{\mathtt{#1}}\else$\mathtt{#1}$\fi}}
\newcommand{\CodeIn}[1]{{\ifmmode{\mathtt{#1}}\else$\mathtt{#1}$\fi}}
\newcommand{\ColorBack}[1]{%
  \begingroup
  \setlength{\fboxsep}{0pt}%
  \colorbox{purple!20}{\reducedstrut#1\/}%
  \endgroup
}

\newcommand{\specialcell}[2][c]{%
  \begin{tabular}[#1]{@{}c@{}}#2\end{tabular}}
\newcolumntype{R}[1]{>{\RaggedLeft\arraybackslash}p{#1}}
\newcolumntype{L}[1]{>{\RaggedRight\arraybackslash}p{#1}}

\newcommand{\RevisionInfo}{\Fix{Time: \today{} at \currenttime{.}}}

\newcommand{\InlineTest}{\textsc{I-Test}\xspace}
\newcommand{\inlinetesting}{inline testing\xspace}
\newcommand{\InlineTesting}{Inline Testing\xspace}
\newcommand{\Inlinetesting}{Inline testing\xspace}
\newcommand{\inlinetest}{inline test\xspace}
\newcommand{\Inlinetest}{Inline test\xspace}
\newcommand{\InlineTTest}{Inline Test\xspace}
\newcommand{\inlinetests}{inline tests\xspace}
\newcommand{\InlineTests}{Inline Tests\xspace}
\newcommand{\Inlinetests}{Inline tests\xspace}
\newcommand{\TrigIt}{\textsc{TrigIt}}
\newcommand{\x}{$x$\xspace}
\newcommand{\Contrib}[1]{$\star$#1}
\newcommand{\printfdebugging}{``\CodeIn{printf} debugging''\xspace}
\newcommand{\RQCostAlone}{RQ1\xspace}
\newcommand{\RQCostWithUnitEnabled}{RQ2\xspace}
\newcommand{\RQCostWithUnitDisabled}{RQ3\xspace}
\newcommand{\pytest}{\CodeIn{pytest}\xspace}
\newcommand{\desiderata}{requirements\xspace}

\newcommand{\cmark}{{\color{blue}{\ding{51}}}}
\newcommand{\pmark}{{\color{blue}{\ding{51}}$^*$}}
\newcommand{\xmark}{{\color{red}{\ding{55}}}}

\newcommand{\ltrue}{\top} %
\newcommand{\lfalse}{\bot} %

\newcommand{\limply}{\rightarrow} %
\newcommand{\liff}{\leftrightarrow} %
\newcommand{\lcofactor}{\downarrow} %

\newcommand{\llimply}{\Rightarrow} %
\newcommand{\lliff}{\Leftrightarrow} %
\newcommand{\lsat}{\vDash} %
\newcommand{\lnsat}{\nvDash} %

\newcommand{\lqall}[1]{\forall #1.~} %
\newcommand{\lqexist}[1]{\exists #1.~} %
\newcommand{\lqallOnly}[1]{\forall #1 } %
\newcommand{\lqexistOnly}[1]{\exists #1 } %

\newcommand{\lAnd}[1]{\bigwedge_{\substack{#1}}} %
\newcommand{\lOr}[1]{\bigvee_{\substack{#1}}} %

\newcommand{\set}[1]{\{#1\}} %
\newcommand{\Set}[1]{\left\{#1\right\}} %
\newcommand{\tuple}[1]{\langle#1\rangle} %
\newcommand{\Tuple}[1]{\left\langle#1\right\rangle} %

\newcommand{\sunion}{\cup} %
\newcommand{\sinter}{\cap} %

\definecolor{gray}{RGB}{211,211,211}
\newcommand{\jbasicstyle}{\small\sffamily} %
\newcommand{\textcode}[1]{{#1}}
\newcommand{\jnumberstyle}{\scriptsize}
\newcommand{\Hilight}{\makebox[0pt][l]{\color{gray}\rule[-3pt]{0.80\linewidth}{9pt}}}

\lstdefinelanguage{pseudo}
{
  morekeywords={},
  keywordstyle=\bfseries,
  lineskip=-0.1em,
  numbers=left, %
  numberstyle=\jnumberstyle,
  numbersep=4pt,
  basicstyle=\jbasicstyle,
  breaklines=true,
  breakautoindent=true,
  tabsize=2,
  columns=fullflexible,
  morecomment=*[l][\textsl]{//},
  mathescape=true,
  xleftmargin=10pt,
}

\lstdefinelanguage{todo-comment}
{
  morekeywords={},
  keywordstyle=\bfseries,
  lineskip=-0.1em,
  numbers=none,
  basicstyle=\jbasicstyle,
  breaklines=true,
  breakautoindent=true,
  tabsize=2,
  columns=fullflexible,
  morecomment=*[l][\textsl]{//},
  mathescape=true,
  xleftmargin=-10pt,
}

\lstdefinelanguage{java-pretty}
{
  language=java,
  numbers=left,
  basicstyle=\scriptsize\ttfamily,
  numberstyle=\scriptsize,
  breaklines=true,
  columns=fullflexible,
  xleftmargin=16pt,
  showstringspaces=false,
  frame=tb,
}

\lstdefinelanguage{python-pretty}{
  language=Python,
  numbers=left,
  basicstyle=\scriptsize\ttfamily,
  numberstyle=\scriptsize,
  breaklines=true,
  breakautoindent=true,
  tabsize=2,
  columns=fullflexible,
  xleftmargin=16pt,
  showstringspaces=false,
  sensitive=false,
  comment=[l]{\#},
  morecomment=[s]{/*}{*/},
  frame=tb,
}

\lstdefinelanguage{java-diff}
{
  language=java,
  numbers=left,
  numbersep=1.2pt,
  basicstyle=\scriptsize\ttfamily,
  numberstyle=\scriptsize,
  breaklines=true,
  frame=lines,
  showstringspaces=false,
  morecomment=[f][\color{red!80!black}]-,         %
  morecomment=[f][\color{green!30!black}]+,       %
}

\definecolor{medgray}{gray}{0.4}
\definecolor{blueshadow}{RGB}{105,115,145}
\definecolor{darkblue}{RGB}{0,0,255}

\newcommand{\Title}{Inline Tests}

\DefMacro{repo}{https://github.com/EngineeringSoftware/inlinetest}

\DefMacro{proj-CorentinJ_Real-Time-Voice-Cloning}{\CodeIn{CorentinJ/Real\text{-}Time\text{-}Voice\text{-}Cloning}\xspace}
\DefMacro{proj-Jack-Cherish_python-spider}{\CodeIn{Jack\text{-}Cherish/python\text{-}spider}\xspace}
\DefMacro{proj-OWASP_CheatSheetSeries}{\CodeIn{OWASP/CheatSheetSeries}\xspace}
\DefMacro{proj-RaRe-Technologies_gensim}{\CodeIn{RaRe\text{-}Technologies/gensim}\xspace}
\DefMacro{proj-Textualize_rich}{\CodeIn{Textualize/rich}\xspace}
\DefMacro{proj-ansible_ansible}{\CodeIn{ansible/ansible}\xspace}
\DefMacro{proj-apprenticeharper_DeDRM_tools}{\CodeIn{apprenticeharper/DeDRM\_tools}\xspace}
\DefMacro{proj-bokeh_bokeh}{\CodeIn{bokeh/bokeh}\xspace}
\DefMacro{proj-chubin_cheat.sh}{\CodeIn{chubin/cheat.sh}\xspace}
\DefMacro{proj-davidsandberg_facenet}{\CodeIn{davidsandberg/facenet}\xspace}
\DefMacro{proj-geekcomputers_Python}{\CodeIn{geekcomputers/Python}\xspace}
\DefMacro{proj-google-research_bert}{\CodeIn{google\text{-}research/bert}\xspace}
\DefMacro{proj-google_jax}{\CodeIn{google/jax}\xspace}
\DefMacro{proj-iperov_DeepFaceLab}{\CodeIn{iperov/DeepFaceLab}\xspace}
\DefMacro{proj-joke2k_faker}{\CodeIn{joke2k/faker}\xspace}
\DefMacro{proj-keras-team_keras}{\CodeIn{keras\text{-}team/keras}\xspace}
\DefMacro{proj-kovidgoyal_kitty}{\CodeIn{kovidgoyal/kitty}\xspace}
\DefMacro{proj-matplotlib_matplotlib}{\CodeIn{matplotlib/matplotlib}\xspace}
\DefMacro{proj-mitmproxy_mitmproxy}{\CodeIn{mitmproxy/mitmproxy}\xspace}
\DefMacro{proj-nicolargo_glances}{\CodeIn{nicolargo/glances}\xspace}
\DefMacro{proj-numpy_numpy}{\CodeIn{numpy/numpy}\xspace}
\DefMacro{proj-pandas-dev_pandas}{\CodeIn{pandas\text{-}dev/pandas}\xspace}
\DefMacro{proj-psf_black}{\CodeIn{psf/black}\xspace}
\DefMacro{proj-pypa_pipenv}{\CodeIn{pypa/pipenv}\xspace}
\DefMacro{proj-python_cpython}{\CodeIn{python/cpython}\xspace}
\DefMacro{proj-rwightman_pytorch-image-models}{\CodeIn{rwightman/pytorch\text{-}image\text{-}models}\xspace}
\DefMacro{proj-scrapy_scrapy}{\CodeIn{scrapy/scrapy}\xspace}
\DefMacro{proj-sebastianruder_NLP-progress}{\CodeIn{sebastianruder/NLP\text{-}progress}\xspace}
\DefMacro{proj-tensorflow_models}{\CodeIn{tensorflow/models}\xspace}
\DefMacro{proj-yt-dlp_yt-dlp}{\CodeIn{yt\text{-}dlp/yt\text{-}dlp}\xspace}
\DefMacro{proj-ytdl-org_youtube-dl}{\CodeIn{ytdl\text{-}org\_youtube\text{-}dl}\xspace}

\DefMacro{proj-NationalSecurityAgency_ghidra}{\CodeIn{NationalSecurityAgency/ghidra}\xspace}
\DefMacro{proj-Tencent_APIJSON}{\CodeIn{Tencent/APIJSON}\xspace}
\DefMacro{proj-alibaba_canal}{\CodeIn{alibaba/canal}\xspace}
\DefMacro{proj-alibaba_druid}{\CodeIn{alibaba/druid}\xspace}
\DefMacro{proj-alibaba_easyexcel}{\CodeIn{alibaba/easyexcel}\xspace}
\DefMacro{proj-alibaba_fastjson}{\CodeIn{alibaba/fastjson}\xspace}
\DefMacro{proj-alibaba_nacos}{\CodeIn{alibaba/nacos}\xspace}
\DefMacro{proj-apache_dubbo}{\CodeIn{apache/dubbo}\xspace}
\DefMacro{proj-apache_flink}{\CodeIn{apache/flink}\xspace}
\DefMacro{proj-apache_hadoop}{\CodeIn{apache/hadoop}\xspace}
\DefMacro{proj-apache_kafka}{\CodeIn{apache/kafka}\xspace}
\DefMacro{proj-apache_shardingsphere}{\CodeIn{apache/shardingsphere}\xspace}
\DefMacro{proj-arduino_Arduino}{\CodeIn{arduino/Arduino}\xspace}
\DefMacro{proj-bumptech_glide}{\CodeIn{bumptech/glide}\xspace}
\DefMacro{proj-dbeaver_dbeaver}{\CodeIn{dbeaver/dbeaver}\xspace}
\DefMacro{proj-didi_DoraemonKit}{\CodeIn{didi/DoraemonKit}\xspace}
\DefMacro{proj-eclipse-vertx_vert.x}{\CodeIn{eclipse\text{-}vertx/vert.x}\xspace}
\DefMacro{proj-eclipse_deeplearning4j}{\CodeIn{eclipse/deeplearning4j}\xspace}
\DefMacro{proj-elastic_elasticsearch}{\CodeIn{elastic/elasticsearch}\xspace}
\DefMacro{proj-google_guava}{\CodeIn{google/guava}\xspace}
\DefMacro{proj-greenrobot_greenDAO}{\CodeIn{greenrobot/greenDAO}\xspace}
\DefMacro{proj-iBotPeaches_Apktool}{\CodeIn{iBotPeaches/Apktool}\xspace}
\DefMacro{proj-jeecgboot_jeecg-boot}{\CodeIn{jeecgboot/jeecg\text{-}boot}\xspace}
\DefMacro{proj-jenkinsci_jenkins}{\CodeIn{jenkinsci/jenkins}\xspace}
\DefMacro{proj-linlinjava_litemall}{\CodeIn{linlinjava/litemall}\xspace}
\DefMacro{proj-oracle_graal}{\CodeIn{oracle/graal}\xspace}
\DefMacro{proj-prestodb_presto}{\CodeIn{prestodb/presto}\xspace}
\DefMacro{proj-redisson_redisson}{\CodeIn{redisson/redisson}\xspace}
\DefMacro{proj-signalapp_Signal-Android}{\CodeIn{signalapp/Signal\text{-}Android}\xspace}
\DefMacro{proj-skylot_jadx}{\CodeIn{skylot/jadx}\xspace}
\DefMacro{proj-spring-projects_spring-boot}{\CodeIn{spring\text{-}projects/spring\text{-}boot}\xspace}
\DefMacro{proj-spring-projects_spring-framework}{\CodeIn{spring\text{-}projects/spring\text{-}framework}\xspace}

\newcommand{\Regex}{Regexes}
\newcommand{\String}{String}
\newcommand{\Bit}{Bit}
\newcommand{\Collection}{Collection}
\newcommand{\Stream}{Streaming}
\newcommand{\HeaderKind}{\textbf{Kind}}
\newcommand{\HeaderCategory}{\textbf{PL}}
\newcommand{\HeaderProjs}{\textbf{\# Projs}}
\newcommand{\HeaderFiles}{\textbf{\# Examples}}
\newcommand{\HeaderTestedStmts}{\textbf{\makecell{\# Target\\ stmts}}}
\newcommand{\HeaderTestedStmtsDesp}{\textbf{\# Target stmts}}
\newcommand{\HeaderInlineTests}{\textbf{\makecell{\# \Inlinetests}}}
\newcommand{\HeaderInlineTestsDesp}{\textbf{\# Inline tests}}
\newcommand{\HeaderPL}{\textbf{\makecell{PL}}}
\newcommand{\HeaderProj}{\textbf{Proj}}
\newcommand{\HeaderNumVanilla}{\textbf{\#UT}}
\newcommand{\HeaderNumITOnly}{\textbf{\#IT}}
\newcommand{\HeaderNumITStandalone}{\textbf{\#IT}}
\newcommand{\HeaderTimeITOnly}{\textbf{$T_{\text{IT}}$ [s]}}
\newcommand{\HeaderTimePerTestITOnly}{\textbf{$t_{\text{IT}}$[s]}}
\newcommand{\HeaderTimeVanilla}{\textbf{$T_{\text{UT}}$ [s]}}
\newcommand{\HeaderTimePerTestVanilla}{\textbf{$t_{\text{UT}}$[s]}}
\newcommand{\HeaderTimeITEnabled}{\textbf{$T_{\text{ITE}}$ [s]}}
\newcommand{\HeaderTimePerTestITEnabled}{\textbf{$t_{\text{ITE}}$[s]}}
\newcommand{\HeaderOverheadITEnabled}{\textbf{$O_{\text{ITE}}$}}
\newcommand{\HeaderTimeITDisabled}{\textbf{$T_{\text{ITD}}$[s]}}
\newcommand{\HeaderTimePerTestITDisabled}{\textbf{$t_{\text{ITD}}$[s]}}
\newcommand{\HeaderOverheadITDisabled}{\textbf{$O_{\text{ITD}}$}}
\newcommand{\HeaderTimeITStandalone}{\textbf{$T_{\text{IT}}$[s]}}
\newcommand{\HeaderTimePerTestITStandalone}{\textbf{$t_{\text{IT}}$[s]}}
\newcommand{\HeaderTask}{\textbf{Task}}
\newcommand{\HeaderUnderstandingTime}{\textbf{$T_{\text{u}}$[min]}}
\newcommand{\HeaderWritingTime}{\textbf{$T_{\text{w}}$[min]}}
\newcommand{\HeaderNumTests}{\textbf{\#IT}}
\newcommand{\HeaderWritingTimePerTest}{\textbf{$T_{\text{w}}$/\#IT [min]}}
\newcommand{\HeaderNumPassedTests}{\textbf{Corr}}
\newcommand{\HeaderBenefits}{\textbf{Adv}}
\newcommand{\HeaderAvg}{avg}
\newcommand{\HeaderMed}{med}
\newcommand{\HeaderSum}{$\Sigma$}
\newcommand{\HeaderNA}{N/A}
\newcommand{\HeaderDup}{\textbf{Dup}\xspace}
\DefMacro{dup-1}{x1\xspace}
\DefMacro{dup-10}{x10\xspace}
\DefMacro{dup-100}{x100\xspace}
\DefMacro{dup-1000}{x1000\xspace}

\newcommand{\CaptionExpItOnly}{Results of only running inline tests for
  Python projects. \HeaderProj = project name,
  \HeaderNumITOnly = total number of inline tests,
  \HeaderTimeITOnly = total time of running inline tests,
  \HeaderTimePerTestITOnly = time of running each inline test.
  \label{tab:exp-it-only}}
\newcommand{\CaptionExpItOnlyPython}{Python\label{tab:exp-it-only-python}}
\newcommand{\CaptionExpItOnlyJava}{Java\label{tab:exp-it-only-java}}
\newcommand{\CaptionExpItStandalone}{Results of standalone experiments. \HeaderDup = duplication count,
  \HeaderNumITStandalone = total no. of \inlinetests,
  \HeaderTimeITStandalone = total \inlinetests running time,
  \HeaderTimePerTestITStandalone = \inlinetest running time.
  \label{tab:exp-it-standalone}}
\newcommand{\CaptionExpItStandalonePython}{Python\label{tab:exp-it-standalone-python}}
\newcommand{\CaptionExpItStandaloneJava}{Java\label{tab:exp-it-standalone-java}}
\newcommand{\CaptionExpBasicStats}{No. of examples and the \inlinetests that we write to guide API design. \HeaderCategory = programming language, \HeaderProjs = no. of projects, \HeaderFiles = no. of examples, \HeaderTestedStmtsDesp = no. of target statements, and \HeaderInlineTestsDesp = no. of \inlinetests.}
\newcommand{\CaptionInlineTestsCategory}{Breakdown of the \inlinetests that we write.}
\newcommand{\CaptionExpUtAndIt}{Results of integrated experiments.
\HeaderProj = project name, \HeaderDup = duplication times,
\HeaderNumVanilla = total no. of unit tests, \HeaderNumITOnly = total
no. of \inlinetests, \HeaderTimePerTestVanilla = time to run each unit
test, \HeaderTimePerTestITOnly = time to run each \inlinetest,
\HeaderTimeITEnabled = total time to run unit tests with \inlinetests
enabled, \HeaderTimePerTestITEnabled = time to run each unit test with
\inlinetests enabled, \HeaderOverheadITEnabled = overhead of running
unit tests with \inlinetests enabled, \HeaderTimeITDisabled = total
time to run unit tests with \inlinetests disabled,
\HeaderTimePerTestITDisabled = time to run each unit test with
\inlinetests disabled, \HeaderOverheadITDisabled = overhead of running
unit tests with \inlinetests disabled.  \label{tab:exp-ut-and-it}}
\newcommand{\CaptionExpUtAndItPython}{Python\label{tab:exp-ut-and-it-python}}
\newcommand{\CaptionExpUtAndItDupsPython}{Python, with duplicating
\inlinetests\label{tab:exp-ut-and-it-dups-python}}
\newcommand{\CaptionExpUtAndItJava}{Java\label{tab:exp-ut-and-it-java}}
\newcommand{\CaptionExpUtAndItDupsJava}{Java, with duplicating
\inlinetests\label{tab:exp-ut-and-it-dups-java}}
  
\newcommand{\TitleUserStudyResults}{User study results.
  \HeaderUnderstandingTime = time to understand each task,
  \HeaderWritingTime = time to write all \inlinetests per task,
  \HeaderNumTests = no. of \inlinetests,
  \HeaderWritingTimePerTest = avg. time to write each \inlinetest,
  \HeaderNumPassedTests = ratio of participants who write passing \inlinetests,
  \HeaderBenefits = ratio of participants who find \inlinetests beneficial. }

\DefMacro{NumUserStudyInitialParticipants}{13}
\DefMacro{NumUserStudyFirstGroup}{five}
\DefMacro{NumUserStudySecondGroup}{six}
\DefMacro{NumUserStudyValidResponses}{nine}
\DefMacro{NumUserStudyInitialGrad}{eight}
\DefMacro{NumUserStudyGrad}{six}
\DefMacro{NumUserStudyInitialUnderGrad}{two}
\DefMacro{NumUserStudyUnderGrad}{two}
\DefMacro{NumUserStudyIndustry}{one}

\DefMacro{NumPythonRegexTestedStmts}{19}
\DefMacro{NumPythonRegexInlineTests}{22}
\DefMacro{NumPythonRegexProjs}{15}
\DefMacro{NumPythonRegexFiles}{17}
\DefMacro{NumPythonStringTestedStmts}{30}
\DefMacro{NumPythonStringInlineTests}{32}
\DefMacro{NumPythonStringProjs}{13}
\DefMacro{NumPythonStringFiles}{14}
\DefMacro{NumPythonBitTestedStmts}{26}
\DefMacro{NumPythonBitInlineTests}{27}
\DefMacro{NumPythonBitProjs}{15}
\DefMacro{NumPythonBitFiles}{15}
\DefMacro{NumPythonCollectionTestedStmts}{5}
\DefMacro{NumPythonCollectionInlineTests}{6}
\DefMacro{NumPythonCollectionProjs}{4}
\DefMacro{NumPythonCollectionFiles}{4}
\DefMacro{NumPythonProjs}{31}
\DefMacro{NumPythonFiles}{50}
\DefMacro{NumPythonInlineTests}{87}
\DefMacro{NumPythonTestedStmts}{80}

\DefMacro{NumJavaRegexTestedStmts}{17}
\DefMacro{NumJavaRegexInlineTests}{17}
\DefMacro{NumJavaRegexProjs}{15}
\DefMacro{NumJavaRegexFiles}{17}
\DefMacro{NumJavaStringTestedStmts}{20}
\DefMacro{NumJavaStringInlineTests}{20}
\DefMacro{NumJavaStringProjs}{15}
\DefMacro{NumJavaStringFiles}{15}
\DefMacro{NumJavaBitTestedStmts}{25}
\DefMacro{NumJavaBitInlineTests}{26}
\DefMacro{NumJavaBitProjs}{16}
\DefMacro{NumJavaBitFiles}{16}
\DefMacro{NumJavaStreamTestedStmts}{2}
\DefMacro{NumJavaStreamInlineTests}{2}
\DefMacro{NumJavaStreamProjs}{2}
\DefMacro{NumJavaStreamFiles}{2}
\DefMacro{NumJavaTestedStmts}{64}
\DefMacro{NumJavaInlineTests}{65}
\DefMacro{NumJavaProjs}{37}
\DefMacro{NumJavaFiles}{50}

\DefMacro{AVG_difficulty_of_learning}{4.2}
\DefMacro{SUM_difficulty_of_learning}{38.0}
\DefMacro{MAX_difficulty_of_learning}{5.0}
\DefMacro{MIN_difficulty_of_learning}{3.0}
\DefMacro{MEDIAN_difficulty_of_learning}{4.0}
\DefMacro{STDEV_difficulty_of_learning}{0.8}
\DefMacro{CNT_difficulty_of_learning}{9.0}
\DefMacro{AVG_difficulty_of_using}{4.1}
\DefMacro{SUM_difficulty_of_using}{37.0}
\DefMacro{MAX_difficulty_of_using}{5.0}
\DefMacro{MIN_difficulty_of_using}{3.0}
\DefMacro{MEDIAN_difficulty_of_using}{4.0}
\DefMacro{STDEV_difficulty_of_using}{0.6}
\DefMacro{CNT_difficulty_of_using}{9.0}
\DefMacro{AVG_years_of_programming}{6.1}
\DefMacro{SUM_years_of_programming}{55.0}
\DefMacro{MAX_years_of_programming}{12.0}
\DefMacro{MIN_years_of_programming}{3.0}
\DefMacro{MEDIAN_years_of_programming}{6.0}
\DefMacro{STDEV_years_of_programming}{2.6}
\DefMacro{CNT_years_of_programming}{9.0}
\DefMacro{AVG_python_expertise}{3.4}
\DefMacro{SUM_python_expertise}{30.5}
\DefMacro{MAX_python_expertise}{4.0}
\DefMacro{MIN_python_expertise}{3.0}
\DefMacro{MEDIAN_python_expertise}{3.0}
\DefMacro{STDEV_python_expertise}{0.5}
\DefMacro{CNT_python_expertise}{9.0}
\DefMacro{AVG_task_1_time_per_test}{2.8}
\DefMacro{SUM_task_1_time_per_test}{24.8}
\DefMacro{MAX_task_1_time_per_test}{9.0}
\DefMacro{MIN_task_1_time_per_test}{0.5}
\DefMacro{MEDIAN_task_1_time_per_test}{2.0}
\DefMacro{STDEV_task_1_time_per_test}{2.6}
\DefMacro{CNT_task_1_time_per_test}{9.0}
\DefMacro{AVG_task_1_understanding_time}{4.0}
\DefMacro{SUM_task_1_understanding_time}{36.0}
\DefMacro{MAX_task_1_understanding_time}{8.0}
\DefMacro{MIN_task_1_understanding_time}{1.0}
\DefMacro{MEDIAN_task_1_understanding_time}{4.0}
\DefMacro{STDEV_task_1_understanding_time}{2.7}
\DefMacro{CNT_task_1_understanding_time}{9.0}
\DefMacro{AVG_task_1_writing_time}{3.7}
\DefMacro{SUM_task_1_writing_time}{33.0}
\DefMacro{MAX_task_1_writing_time}{9.0}
\DefMacro{MIN_task_1_writing_time}{1.0}
\DefMacro{MEDIAN_task_1_writing_time}{3.0}
\DefMacro{STDEV_task_1_writing_time}{2.6}
\DefMacro{CNT_task_1_writing_time}{9.0}
\DefMacro{AVG_task_1_num_tests}{1.7}
\DefMacro{SUM_task_1_num_tests}{15.0}
\DefMacro{MAX_task_1_num_tests}{3.0}
\DefMacro{MIN_task_1_num_tests}{1.0}
\DefMacro{MEDIAN_task_1_num_tests}{1.0}
\DefMacro{STDEV_task_1_num_tests}{0.8}
\DefMacro{CNT_task_1_num_tests}{9.0}
\DefMacro{AVG_task_1_num_passed_tests}{1.7}
\DefMacro{SUM_task_1_num_passed_tests}{15.0}
\DefMacro{MAX_task_1_num_passed_tests}{3.0}
\DefMacro{MIN_task_1_num_passed_tests}{1.0}
\DefMacro{MEDIAN_task_1_num_passed_tests}{1.0}
\DefMacro{STDEV_task_1_num_passed_tests}{0.8}
\DefMacro{CNT_task_1_num_passed_tests}{9.0}
\DefMacro{AVG_task_2_time_per_test}{2.5}
\DefMacro{SUM_task_2_time_per_test}{22.2}
\DefMacro{MAX_task_2_time_per_test}{5.0}
\DefMacro{MIN_task_2_time_per_test}{1.5}
\DefMacro{MEDIAN_task_2_time_per_test}{2.0}
\DefMacro{STDEV_task_2_time_per_test}{1.0}
\DefMacro{CNT_task_2_time_per_test}{9.0}
\DefMacro{AVG_task_2_understanding_time}{1.6}
\DefMacro{SUM_task_2_understanding_time}{14.0}
\DefMacro{MAX_task_2_understanding_time}{3.0}
\DefMacro{MIN_task_2_understanding_time}{1.0}
\DefMacro{MEDIAN_task_2_understanding_time}{1.0}
\DefMacro{STDEV_task_2_understanding_time}{0.7}
\DefMacro{CNT_task_2_understanding_time}{9.0}
\DefMacro{AVG_task_2_writing_time}{3.4}
\DefMacro{SUM_task_2_writing_time}{31.0}
\DefMacro{MAX_task_2_writing_time}{5.0}
\DefMacro{MIN_task_2_writing_time}{2.0}
\DefMacro{MEDIAN_task_2_writing_time}{3.0}
\DefMacro{STDEV_task_2_writing_time}{1.1}
\DefMacro{CNT_task_2_writing_time}{9.0}
\DefMacro{AVG_task_2_num_tests}{1.6}
\DefMacro{SUM_task_2_num_tests}{14.0}
\DefMacro{MAX_task_2_num_tests}{3.0}
\DefMacro{MIN_task_2_num_tests}{1.0}
\DefMacro{MEDIAN_task_2_num_tests}{1.0}
\DefMacro{STDEV_task_2_num_tests}{0.7}
\DefMacro{CNT_task_2_num_tests}{9.0}
\DefMacro{AVG_task_2_num_passed_tests}{1.6}
\DefMacro{SUM_task_2_num_passed_tests}{14.0}
\DefMacro{MAX_task_2_num_passed_tests}{3.0}
\DefMacro{MIN_task_2_num_passed_tests}{1.0}
\DefMacro{MEDIAN_task_2_num_passed_tests}{1.0}
\DefMacro{STDEV_task_2_num_passed_tests}{0.7}
\DefMacro{CNT_task_2_num_passed_tests}{9.0}
\DefMacro{AVG_task_3_time_per_test}{3.0}
\DefMacro{SUM_task_3_time_per_test}{27.0}
\DefMacro{MAX_task_3_time_per_test}{9.0}
\DefMacro{MIN_task_3_time_per_test}{1.0}
\DefMacro{MEDIAN_task_3_time_per_test}{2.0}
\DefMacro{STDEV_task_3_time_per_test}{2.4}
\DefMacro{CNT_task_3_time_per_test}{9.0}
\DefMacro{AVG_task_3_understanding_time}{2.2}
\DefMacro{SUM_task_3_understanding_time}{20.0}
\DefMacro{MAX_task_3_understanding_time}{5.0}
\DefMacro{MIN_task_3_understanding_time}{0.0}
\DefMacro{MEDIAN_task_3_understanding_time}{2.0}
\DefMacro{STDEV_task_3_understanding_time}{1.5}
\DefMacro{CNT_task_3_understanding_time}{9.0}
\DefMacro{AVG_task_3_writing_time}{4.1}
\DefMacro{SUM_task_3_writing_time}{37.0}
\DefMacro{MAX_task_3_writing_time}{9.0}
\DefMacro{MIN_task_3_writing_time}{2.0}
\DefMacro{MEDIAN_task_3_writing_time}{4.0}
\DefMacro{STDEV_task_3_writing_time}{2.1}
\DefMacro{CNT_task_3_writing_time}{9.0}
\DefMacro{AVG_task_3_num_tests}{1.7}
\DefMacro{SUM_task_3_num_tests}{15.0}
\DefMacro{MAX_task_3_num_tests}{2.0}
\DefMacro{MIN_task_3_num_tests}{1.0}
\DefMacro{MEDIAN_task_3_num_tests}{2.0}
\DefMacro{STDEV_task_3_num_tests}{0.5}
\DefMacro{CNT_task_3_num_tests}{9.0}
\DefMacro{AVG_task_3_num_passed_tests}{1.7}
\DefMacro{SUM_task_3_num_passed_tests}{15.0}
\DefMacro{MAX_task_3_num_passed_tests}{2.0}
\DefMacro{MIN_task_3_num_passed_tests}{1.0}
\DefMacro{MEDIAN_task_3_num_passed_tests}{2.0}
\DefMacro{STDEV_task_3_num_passed_tests}{0.5}
\DefMacro{CNT_task_3_num_passed_tests}{9.0}
\DefMacro{AVG_task_4_time_per_test}{1.9}
\DefMacro{SUM_task_4_time_per_test}{17.3}
\DefMacro{MAX_task_4_time_per_test}{5.0}
\DefMacro{MIN_task_4_time_per_test}{0.3}
\DefMacro{MEDIAN_task_4_time_per_test}{1.0}
\DefMacro{STDEV_task_4_time_per_test}{1.7}
\DefMacro{CNT_task_4_time_per_test}{9.0}
\DefMacro{AVG_task_4_understanding_time}{3.3}
\DefMacro{SUM_task_4_understanding_time}{30.0}
\DefMacro{MAX_task_4_understanding_time}{8.0}
\DefMacro{MIN_task_4_understanding_time}{1.0}
\DefMacro{MEDIAN_task_4_understanding_time}{3.0}
\DefMacro{STDEV_task_4_understanding_time}{2.2}
\DefMacro{CNT_task_4_understanding_time}{9.0}
\DefMacro{AVG_task_4_writing_time}{2.8}
\DefMacro{SUM_task_4_writing_time}{25.0}
\DefMacro{MAX_task_4_writing_time}{6.0}
\DefMacro{MIN_task_4_writing_time}{1.0}
\DefMacro{MEDIAN_task_4_writing_time}{2.0}
\DefMacro{STDEV_task_4_writing_time}{1.9}
\DefMacro{CNT_task_4_writing_time}{9.0}
\DefMacro{AVG_task_4_num_tests}{1.8}
\DefMacro{SUM_task_4_num_tests}{16.0}
\DefMacro{MAX_task_4_num_tests}{3.0}
\DefMacro{MIN_task_4_num_tests}{1.0}
\DefMacro{MEDIAN_task_4_num_tests}{2.0}
\DefMacro{STDEV_task_4_num_tests}{0.8}
\DefMacro{CNT_task_4_num_tests}{9.0}
\DefMacro{AVG_task_4_num_passed_tests}{1.8}
\DefMacro{SUM_task_4_num_passed_tests}{16.0}
\DefMacro{MAX_task_4_num_passed_tests}{3.0}
\DefMacro{MIN_task_4_num_passed_tests}{1.0}
\DefMacro{MEDIAN_task_4_num_passed_tests}{2.0}
\DefMacro{STDEV_task_4_num_passed_tests}{0.8}
\DefMacro{CNT_task_4_num_passed_tests}{9.0}
\DefMacro{AVG_time_per_test}{2.5}
\DefMacro{MEDIAN_time_per_test}{2.6}
\DefMacro{AVG_understanding_time}{2.8}
\DefMacro{MEDIAN_understanding_time}{2.8}
\DefMacro{AVG_writing_time}{3.5}
\DefMacro{MEDIAN_writing_time}{3.6}
\DefMacro{AVG_num_tests}{1.7}
\DefMacro{MEDIAN_num_tests}{1.7}
\DefMacro{task_1_benefit}{8}
\DefMacro{task_2_benefit}{8}
\DefMacro{task_3_benefit}{5}
\DefMacro{task_4_benefit}{9}
\DefMacro{task_1_correct_times}{9}
\DefMacro{task_2_correct_times}{9}
\DefMacro{task_3_correct_times}{9}
\DefMacro{task_4_correct_times}{9}
\DefMacro{NumUsers}{9}

\DefMacro{exp-python-num-AVG-dup1-its}{1.74}
\DefMacro{exp-python-num-SUM-dup1-its}{87}
\DefMacro{exp-python-num-MAX-dup1-its}{4}
\DefMacro{exp-python-num-MIN-dup1-its}{1}
\DefMacro{exp-python-num-MEDIAN-dup1-its}{1.00}
\DefMacro{exp-python-num-STDEV-dup1-its}{0.91}
\DefMacro{exp-python-num-CNT-dup1-its}{50}
\DefMacro{exp-python-time-AVG-dup1-its}{0.26}
\DefMacro{exp-python-time-SUM-dup1-its}{12.78}
\DefMacro{exp-python-time-MAX-dup1-its}{0.42}
\DefMacro{exp-python-time-MIN-dup1-its}{0.25}
\DefMacro{exp-python-time-MEDIAN-dup1-its}{0.25}
\DefMacro{exp-python-time-STDEV-dup1-its}{0.02}
\DefMacro{exp-python-time-CNT-dup1-its}{50}
\DefMacro{exp-python-timept-AVG-dup1-its}{0.185}
\DefMacro{exp-python-timept-SUM-dup1-its}{9.245}
\DefMacro{exp-python-timept-MAX-dup1-its}{0.417}
\DefMacro{exp-python-timept-MIN-dup1-its}{0.063}
\DefMacro{exp-python-timept-MEDIAN-dup1-its}{0.248}
\DefMacro{exp-python-timept-STDEV-dup1-its}{0.081}
\DefMacro{exp-python-timept-CNT-dup1-its}{50}
\DefMacro{exp-python-timept-MACROAVG-dup1-its}{0.147}
\DefMacro{exp-python-num-AVG-dup10-its}{17.40}
\DefMacro{exp-python-num-SUM-dup10-its}{870}
\DefMacro{exp-python-num-MAX-dup10-its}{40}
\DefMacro{exp-python-num-MIN-dup10-its}{10}
\DefMacro{exp-python-num-MEDIAN-dup10-its}{10.00}
\DefMacro{exp-python-num-STDEV-dup10-its}{9.12}
\DefMacro{exp-python-num-CNT-dup10-its}{50}
\DefMacro{exp-python-time-AVG-dup10-its}{0.27}
\DefMacro{exp-python-time-SUM-dup10-its}{13.41}
\DefMacro{exp-python-time-MAX-dup10-its}{0.43}
\DefMacro{exp-python-time-MIN-dup10-its}{0.26}
\DefMacro{exp-python-time-MEDIAN-dup10-its}{0.26}
\DefMacro{exp-python-time-STDEV-dup10-its}{0.02}
\DefMacro{exp-python-time-CNT-dup10-its}{50}
\DefMacro{exp-python-timept-AVG-dup10-its}{0.019}
\DefMacro{exp-python-timept-SUM-dup10-its}{0.963}
\DefMacro{exp-python-timept-MAX-dup10-its}{0.043}
\DefMacro{exp-python-timept-MIN-dup10-its}{0.007}
\DefMacro{exp-python-timept-MEDIAN-dup10-its}{0.026}
\DefMacro{exp-python-timept-STDEV-dup10-its}{0.008}
\DefMacro{exp-python-timept-CNT-dup10-its}{50}
\DefMacro{exp-python-timept-MACROAVG-dup10-its}{0.015}
\DefMacro{exp-python-num-AVG-dup100-its}{174.00}
\DefMacro{exp-python-num-SUM-dup100-its}{8,700}
\DefMacro{exp-python-num-MAX-dup100-its}{400}
\DefMacro{exp-python-num-MIN-dup100-its}{100}
\DefMacro{exp-python-num-MEDIAN-dup100-its}{100.00}
\DefMacro{exp-python-num-STDEV-dup100-its}{91.24}
\DefMacro{exp-python-num-CNT-dup100-its}{50}
\DefMacro{exp-python-time-AVG-dup100-its}{0.40}
\DefMacro{exp-python-time-SUM-dup100-its}{19.86}
\DefMacro{exp-python-time-MAX-dup100-its}{0.62}
\DefMacro{exp-python-time-MIN-dup100-its}{0.32}
\DefMacro{exp-python-time-MEDIAN-dup100-its}{0.38}
\DefMacro{exp-python-time-STDEV-dup100-its}{0.08}
\DefMacro{exp-python-time-CNT-dup100-its}{50}
\DefMacro{exp-python-timept-AVG-dup100-its}{0.003}
\DefMacro{exp-python-timept-SUM-dup100-its}{0.134}
\DefMacro{exp-python-timept-MAX-dup100-its}{0.006}
\DefMacro{exp-python-timept-MIN-dup100-its}{0.001}
\DefMacro{exp-python-timept-MEDIAN-dup100-its}{0.003}
\DefMacro{exp-python-timept-STDEV-dup100-its}{0.001}
\DefMacro{exp-python-timept-CNT-dup100-its}{50}
\DefMacro{exp-python-timept-MACROAVG-dup100-its}{0.002}
\DefMacro{exp-python-num-AVG-dup1000-its}{1,740.00}
\DefMacro{exp-python-num-SUM-dup1000-its}{87,000}
\DefMacro{exp-python-num-MAX-dup1000-its}{4,000}
\DefMacro{exp-python-num-MIN-dup1000-its}{1,000}
\DefMacro{exp-python-num-MEDIAN-dup1000-its}{1,000.00}
\DefMacro{exp-python-num-STDEV-dup1000-its}{912.36}
\DefMacro{exp-python-num-CNT-dup1000-its}{50}
\DefMacro{exp-python-time-AVG-dup1000-its}{2.50}
\DefMacro{exp-python-time-SUM-dup1000-its}{124.92}
\DefMacro{exp-python-time-MAX-dup1000-its}{6.25}
\DefMacro{exp-python-time-MIN-dup1000-its}{1.21}
\DefMacro{exp-python-time-MEDIAN-dup1000-its}{2.19}
\DefMacro{exp-python-time-STDEV-dup1000-its}{1.26}
\DefMacro{exp-python-time-CNT-dup1000-its}{50}
\DefMacro{exp-python-timept-AVG-dup1000-its}{0.001}
\DefMacro{exp-python-timept-SUM-dup1000-its}{0.074}
\DefMacro{exp-python-timept-MAX-dup1000-its}{0.003}
\DefMacro{exp-python-timept-MIN-dup1000-its}{0.001}
\DefMacro{exp-python-timept-MEDIAN-dup1000-its}{0.001}
\DefMacro{exp-python-timept-STDEV-dup1000-its}{0.000}
\DefMacro{exp-python-timept-CNT-dup1000-its}{50}
\DefMacro{exp-python-timept-MACROAVG-dup1000-its}{0.001}

\DefMacro{exp-java-num-AVG-dup1-its}{1.30}
\DefMacro{exp-java-num-SUM-dup1-its}{65}
\DefMacro{exp-java-num-MAX-dup1-its}{5}
\DefMacro{exp-java-num-MIN-dup1-its}{1}
\DefMacro{exp-java-num-MEDIAN-dup1-its}{1.00}
\DefMacro{exp-java-num-STDEV-dup1-its}{0.67}
\DefMacro{exp-java-num-CNT-dup1-its}{50}
\DefMacro{exp-java-time-AVG-dup1-its}{0.46}
\DefMacro{exp-java-time-SUM-dup1-its}{23.08}
\DefMacro{exp-java-time-MAX-dup1-its}{0.51}
\DefMacro{exp-java-time-MIN-dup1-its}{0.44}
\DefMacro{exp-java-time-MEDIAN-dup1-its}{0.46}
\DefMacro{exp-java-time-STDEV-dup1-its}{0.01}
\DefMacro{exp-java-time-CNT-dup1-its}{50}
\DefMacro{exp-java-timept-AVG-dup1-its}{0.403}
\DefMacro{exp-java-timept-SUM-dup1-its}{20.175}
\DefMacro{exp-java-timept-MAX-dup1-its}{0.514}
\DefMacro{exp-java-timept-MIN-dup1-its}{0.092}
\DefMacro{exp-java-timept-MEDIAN-dup1-its}{0.454}
\DefMacro{exp-java-timept-STDEV-dup1-its}{0.106}
\DefMacro{exp-java-timept-CNT-dup1-its}{50}
\DefMacro{exp-java-timept-MACROAVG-dup1-its}{0.355}
\DefMacro{exp-java-num-AVG-dup10-its}{13.00}
\DefMacro{exp-java-num-SUM-dup10-its}{650}
\DefMacro{exp-java-num-MAX-dup10-its}{50}
\DefMacro{exp-java-num-MIN-dup10-its}{10}
\DefMacro{exp-java-num-MEDIAN-dup10-its}{10.00}
\DefMacro{exp-java-num-STDEV-dup10-its}{6.71}
\DefMacro{exp-java-num-CNT-dup10-its}{50}
\DefMacro{exp-java-time-AVG-dup10-its}{0.50}
\DefMacro{exp-java-time-SUM-dup10-its}{24.92}
\DefMacro{exp-java-time-MAX-dup10-its}{0.64}
\DefMacro{exp-java-time-MIN-dup10-its}{0.47}
\DefMacro{exp-java-time-MEDIAN-dup10-its}{0.49}
\DefMacro{exp-java-time-STDEV-dup10-its}{0.03}
\DefMacro{exp-java-time-CNT-dup10-its}{50}
\DefMacro{exp-java-timept-AVG-dup10-its}{0.043}
\DefMacro{exp-java-timept-SUM-dup10-its}{2.172}
\DefMacro{exp-java-timept-MAX-dup10-its}{0.064}
\DefMacro{exp-java-timept-MIN-dup10-its}{0.011}
\DefMacro{exp-java-timept-MEDIAN-dup10-its}{0.048}
\DefMacro{exp-java-timept-STDEV-dup10-its}{0.011}
\DefMacro{exp-java-timept-CNT-dup10-its}{50}
\DefMacro{exp-java-timept-MACROAVG-dup10-its}{0.038}
\DefMacro{exp-java-num-AVG-dup100-its}{130.00}
\DefMacro{exp-java-num-SUM-dup100-its}{6,500}
\DefMacro{exp-java-num-MAX-dup100-its}{500}
\DefMacro{exp-java-num-MIN-dup100-its}{100}
\DefMacro{exp-java-num-MEDIAN-dup100-its}{100.00}
\DefMacro{exp-java-num-STDEV-dup100-its}{67.08}
\DefMacro{exp-java-num-CNT-dup100-its}{50}
\DefMacro{exp-java-time-AVG-dup100-its}{0.68}
\DefMacro{exp-java-time-SUM-dup100-its}{34.21}
\DefMacro{exp-java-time-MAX-dup100-its}{1.03}
\DefMacro{exp-java-time-MIN-dup100-its}{0.58}
\DefMacro{exp-java-time-MEDIAN-dup100-its}{0.67}
\DefMacro{exp-java-time-STDEV-dup100-its}{0.08}
\DefMacro{exp-java-time-CNT-dup100-its}{50}
\DefMacro{exp-java-timept-AVG-dup100-its}{0.006}
\DefMacro{exp-java-timept-SUM-dup100-its}{0.296}
\DefMacro{exp-java-timept-MAX-dup100-its}{0.010}
\DefMacro{exp-java-timept-MIN-dup100-its}{0.002}
\DefMacro{exp-java-timept-MEDIAN-dup100-its}{0.006}
\DefMacro{exp-java-timept-STDEV-dup100-its}{0.002}
\DefMacro{exp-java-timept-CNT-dup100-its}{50}
\DefMacro{exp-java-timept-MACROAVG-dup100-its}{0.005}
\DefMacro{exp-java-num-AVG-dup1000-its}{1,300.00}
\DefMacro{exp-java-num-SUM-dup1000-its}{65,000}
\DefMacro{exp-java-num-MAX-dup1000-its}{5,000}
\DefMacro{exp-java-num-MIN-dup1000-its}{1,000}
\DefMacro{exp-java-num-MEDIAN-dup1000-its}{1,000.00}
\DefMacro{exp-java-num-STDEV-dup1000-its}{670.82}
\DefMacro{exp-java-num-CNT-dup1000-its}{50}
\DefMacro{exp-java-time-AVG-dup1000-its}{1.36}
\DefMacro{exp-java-time-SUM-dup1000-its}{67.87}
\DefMacro{exp-java-time-MAX-dup1000-its}{3.03}
\DefMacro{exp-java-time-MIN-dup1000-its}{0.97}
\DefMacro{exp-java-time-MEDIAN-dup1000-its}{1.25}
\DefMacro{exp-java-time-STDEV-dup1000-its}{0.40}
\DefMacro{exp-java-time-CNT-dup1000-its}{50}
\DefMacro{exp-java-timept-AVG-dup1000-its}{0.001}
\DefMacro{exp-java-timept-SUM-dup1000-its}{0.057}
\DefMacro{exp-java-timept-MAX-dup1000-its}{0.003}
\DefMacro{exp-java-timept-MIN-dup1000-its}{0.001}
\DefMacro{exp-java-timept-MEDIAN-dup1000-its}{0.001}
\DefMacro{exp-java-timept-STDEV-dup1000-its}{0.000}
\DefMacro{exp-java-timept-CNT-dup1000-its}{50}
\DefMacro{exp-java-timept-MACROAVG-dup1000-its}{0.001}

\DefMacro{exp-python-proj-initial}{31}
\DefMacro{exp-python-skipped}{2}
\DefMacro{exp-python-proj-can-run}{29}
\DefMacro{exp-python-skipped-import-error}{2}
\DefMacro{exp-python-skipped-unit-tests}{15}
\DefMacro{exp-python-proj-can-run-unit-tests}{14}
\DefMacro{exp-python-skipped-unit-tests-does-not-use-pytest}{5}
\DefMacro{exp-python-skipped-unit-tests-no-tests}{5}
\DefMacro{exp-python-skipped-unit-tests-too-many-test-failures}{5}

\DefMacro{exp-java-proj-initial}{37}
\DefMacro{exp-java-skipped}{10}
\DefMacro{exp-java-proj-can-run}{27}
\DefMacro{exp-java-skipped-compilation-failure}{10}
\DefMacro{exp-java-skipped-unit-tests}{20}
\DefMacro{exp-java-proj-can-run-unit-tests}{7}
\DefMacro{exp-java-skipped-unit-tests-too-many-test-failures}{17}
\DefMacro{exp-java-skipped-unit-tests-no-tests}{3}

\DefMacro{exp-python-RaRe-Technologies_gensim-time-vanilla-dup1}{225.92}
\DefMacro{exp-python-RaRe-Technologies_gensim-time-unit-dup1}{226.35}
\DefMacro{exp-python-RaRe-Technologies_gensim-time-inline-dup1}{0.55}
\DefMacro{exp-python-RaRe-Technologies_gensim-time-unit-and-inline-dup1}{226.90}
\DefMacro{exp-python-RaRe-Technologies_gensim-num-vanilla-dup1}{968}
\DefMacro{exp-python-RaRe-Technologies_gensim-num-unit-dup1}{968}
\DefMacro{exp-python-RaRe-Technologies_gensim-num-inline-dup1}{2}
\DefMacro{exp-python-RaRe-Technologies_gensim-num-unit-and-inline-dup1}{970}
\DefMacro{exp-python-RaRe-Technologies_gensim-timept-vanilla-dup1}{0.233}
\DefMacro{exp-python-RaRe-Technologies_gensim-timept-unit-dup1}{0.234}
\DefMacro{exp-python-RaRe-Technologies_gensim-timept-inline-dup1}{0.274}
\DefMacro{exp-python-RaRe-Technologies_gensim-timept-unit-and-inline-dup1}{0.234}
\DefMacro{exp-python-RaRe-Technologies_gensim-overhead-unit-dup1}{0.002}
\DefMacro{exp-python-RaRe-Technologies_gensim-overhead-unit-and-inline-dup1}{0.004}
\DefMacro{exp-python-Textualize_rich-time-vanilla-dup1}{3.71}
\DefMacro{exp-python-Textualize_rich-time-unit-dup1}{3.72}
\DefMacro{exp-python-Textualize_rich-time-inline-dup1}{0.23}
\DefMacro{exp-python-Textualize_rich-time-unit-and-inline-dup1}{3.94}
\DefMacro{exp-python-Textualize_rich-num-vanilla-dup1}{622}
\DefMacro{exp-python-Textualize_rich-num-unit-dup1}{622}
\DefMacro{exp-python-Textualize_rich-num-inline-dup1}{2}
\DefMacro{exp-python-Textualize_rich-num-unit-and-inline-dup1}{624}
\DefMacro{exp-python-Textualize_rich-timept-vanilla-dup1}{0.006}
\DefMacro{exp-python-Textualize_rich-timept-unit-dup1}{0.006}
\DefMacro{exp-python-Textualize_rich-timept-inline-dup1}{0.113}
\DefMacro{exp-python-Textualize_rich-timept-unit-and-inline-dup1}{0.006}
\DefMacro{exp-python-Textualize_rich-overhead-unit-dup1}{0.002}
\DefMacro{exp-python-Textualize_rich-overhead-unit-and-inline-dup1}{0.063}
\DefMacro{exp-python-bokeh_bokeh-time-vanilla-dup1}{49.63}
\DefMacro{exp-python-bokeh_bokeh-time-unit-dup1}{50.13}
\DefMacro{exp-python-bokeh_bokeh-time-inline-dup1}{0.78}
\DefMacro{exp-python-bokeh_bokeh-time-unit-and-inline-dup1}{50.91}
\DefMacro{exp-python-bokeh_bokeh-num-vanilla-dup1}{8,616}
\DefMacro{exp-python-bokeh_bokeh-num-unit-dup1}{8,616}
\DefMacro{exp-python-bokeh_bokeh-num-inline-dup1}{8}
\DefMacro{exp-python-bokeh_bokeh-num-unit-and-inline-dup1}{8,624}
\DefMacro{exp-python-bokeh_bokeh-timept-vanilla-dup1}{0.006}
\DefMacro{exp-python-bokeh_bokeh-timept-unit-dup1}{0.006}
\DefMacro{exp-python-bokeh_bokeh-timept-inline-dup1}{0.098}
\DefMacro{exp-python-bokeh_bokeh-timept-unit-and-inline-dup1}{0.006}
\DefMacro{exp-python-bokeh_bokeh-overhead-unit-dup1}{0.010}
\DefMacro{exp-python-bokeh_bokeh-overhead-unit-and-inline-dup1}{0.026}
\DefMacro{exp-python-chubin_cheat.sh-time-vanilla-dup1}{0.34}
\DefMacro{exp-python-chubin_cheat.sh-time-unit-dup1}{0.33}
\DefMacro{exp-python-chubin_cheat.sh-time-inline-dup1}{0.41}
\DefMacro{exp-python-chubin_cheat.sh-time-unit-and-inline-dup1}{0.74}
\DefMacro{exp-python-chubin_cheat.sh-num-vanilla-dup1}{1}
\DefMacro{exp-python-chubin_cheat.sh-num-unit-dup1}{1}
\DefMacro{exp-python-chubin_cheat.sh-num-inline-dup1}{3}
\DefMacro{exp-python-chubin_cheat.sh-num-unit-and-inline-dup1}{4}
\DefMacro{exp-python-chubin_cheat.sh-timept-vanilla-dup1}{0.337}
\DefMacro{exp-python-chubin_cheat.sh-timept-unit-dup1}{0.334}
\DefMacro{exp-python-chubin_cheat.sh-timept-inline-dup1}{0.136}
\DefMacro{exp-python-chubin_cheat.sh-timept-unit-and-inline-dup1}{0.186}
\DefMacro{exp-python-chubin_cheat.sh-overhead-unit-dup1}{-0.010}
\DefMacro{exp-python-chubin_cheat.sh-overhead-unit-and-inline-dup1}{1.204}
\DefMacro{exp-python-davidsandberg_facenet-time-vanilla-dup1}{0.97}
\DefMacro{exp-python-davidsandberg_facenet-time-unit-dup1}{0.98}
\DefMacro{exp-python-davidsandberg_facenet-time-inline-dup1}{0.86}
\DefMacro{exp-python-davidsandberg_facenet-time-unit-and-inline-dup1}{1.83}
\DefMacro{exp-python-davidsandberg_facenet-num-vanilla-dup1}{3}
\DefMacro{exp-python-davidsandberg_facenet-num-unit-dup1}{3}
\DefMacro{exp-python-davidsandberg_facenet-num-inline-dup1}{1}
\DefMacro{exp-python-davidsandberg_facenet-num-unit-and-inline-dup1}{4}
\DefMacro{exp-python-davidsandberg_facenet-timept-vanilla-dup1}{0.323}
\DefMacro{exp-python-davidsandberg_facenet-timept-unit-dup1}{0.325}
\DefMacro{exp-python-davidsandberg_facenet-timept-inline-dup1}{0.855}
\DefMacro{exp-python-davidsandberg_facenet-timept-unit-and-inline-dup1}{0.458}
\DefMacro{exp-python-davidsandberg_facenet-overhead-unit-dup1}{0.006}
\DefMacro{exp-python-davidsandberg_facenet-overhead-unit-and-inline-dup1}{0.888}
\DefMacro{exp-python-geekcomputers_Python-time-vanilla-dup1}{0.17}
\DefMacro{exp-python-geekcomputers_Python-time-unit-dup1}{0.18}
\DefMacro{exp-python-geekcomputers_Python-time-inline-dup1}{0.20}
\DefMacro{exp-python-geekcomputers_Python-time-unit-and-inline-dup1}{0.38}
\DefMacro{exp-python-geekcomputers_Python-num-vanilla-dup1}{1}
\DefMacro{exp-python-geekcomputers_Python-num-unit-dup1}{1}
\DefMacro{exp-python-geekcomputers_Python-num-inline-dup1}{4}
\DefMacro{exp-python-geekcomputers_Python-num-unit-and-inline-dup1}{5}
\DefMacro{exp-python-geekcomputers_Python-timept-vanilla-dup1}{0.169}
\DefMacro{exp-python-geekcomputers_Python-timept-unit-dup1}{0.179}
\DefMacro{exp-python-geekcomputers_Python-timept-inline-dup1}{0.049}
\DefMacro{exp-python-geekcomputers_Python-timept-unit-and-inline-dup1}{0.075}
\DefMacro{exp-python-geekcomputers_Python-overhead-unit-dup1}{0.058}
\DefMacro{exp-python-geekcomputers_Python-overhead-unit-and-inline-dup1}{1.217}
\DefMacro{exp-python-google-research_bert-time-vanilla-dup1}{2.05}
\DefMacro{exp-python-google-research_bert-time-unit-dup1}{2.07}
\DefMacro{exp-python-google-research_bert-time-inline-dup1}{0.62}
\DefMacro{exp-python-google-research_bert-time-unit-and-inline-dup1}{2.69}
\DefMacro{exp-python-google-research_bert-num-vanilla-dup1}{15}
\DefMacro{exp-python-google-research_bert-num-unit-dup1}{15}
\DefMacro{exp-python-google-research_bert-num-inline-dup1}{1}
\DefMacro{exp-python-google-research_bert-num-unit-and-inline-dup1}{16}
\DefMacro{exp-python-google-research_bert-timept-vanilla-dup1}{0.137}
\DefMacro{exp-python-google-research_bert-timept-unit-dup1}{0.138}
\DefMacro{exp-python-google-research_bert-timept-inline-dup1}{0.621}
\DefMacro{exp-python-google-research_bert-timept-unit-and-inline-dup1}{0.168}
\DefMacro{exp-python-google-research_bert-overhead-unit-dup1}{0.011}
\DefMacro{exp-python-google-research_bert-overhead-unit-and-inline-dup1}{0.314}
\DefMacro{exp-python-joke2k_faker-time-vanilla-dup1}{16.73}
\DefMacro{exp-python-joke2k_faker-time-unit-dup1}{16.64}
\DefMacro{exp-python-joke2k_faker-time-inline-dup1}{0.27}
\DefMacro{exp-python-joke2k_faker-time-unit-and-inline-dup1}{16.91}
\DefMacro{exp-python-joke2k_faker-num-vanilla-dup1}{1,596}
\DefMacro{exp-python-joke2k_faker-num-unit-dup1}{1,596}
\DefMacro{exp-python-joke2k_faker-num-inline-dup1}{4}
\DefMacro{exp-python-joke2k_faker-num-unit-and-inline-dup1}{1,600}
\DefMacro{exp-python-joke2k_faker-timept-vanilla-dup1}{0.010}
\DefMacro{exp-python-joke2k_faker-timept-unit-dup1}{0.010}
\DefMacro{exp-python-joke2k_faker-timept-inline-dup1}{0.068}
\DefMacro{exp-python-joke2k_faker-timept-unit-and-inline-dup1}{0.011}
\DefMacro{exp-python-joke2k_faker-overhead-unit-dup1}{-0.006}
\DefMacro{exp-python-joke2k_faker-overhead-unit-and-inline-dup1}{0.011}
\DefMacro{exp-python-mitmproxy_mitmproxy-time-vanilla-dup1}{7.50}
\DefMacro{exp-python-mitmproxy_mitmproxy-time-unit-dup1}{7.45}
\DefMacro{exp-python-mitmproxy_mitmproxy-time-inline-dup1}{0.40}
\DefMacro{exp-python-mitmproxy_mitmproxy-time-unit-and-inline-dup1}{7.85}
\DefMacro{exp-python-mitmproxy_mitmproxy-num-vanilla-dup1}{1,287}
\DefMacro{exp-python-mitmproxy_mitmproxy-num-unit-dup1}{1,287}
\DefMacro{exp-python-mitmproxy_mitmproxy-num-inline-dup1}{1}
\DefMacro{exp-python-mitmproxy_mitmproxy-num-unit-and-inline-dup1}{1,288}
\DefMacro{exp-python-mitmproxy_mitmproxy-timept-vanilla-dup1}{0.006}
\DefMacro{exp-python-mitmproxy_mitmproxy-timept-unit-dup1}{0.006}
\DefMacro{exp-python-mitmproxy_mitmproxy-timept-inline-dup1}{0.400}
\DefMacro{exp-python-mitmproxy_mitmproxy-timept-unit-and-inline-dup1}{0.006}
\DefMacro{exp-python-mitmproxy_mitmproxy-overhead-unit-dup1}{-0.007}
\DefMacro{exp-python-mitmproxy_mitmproxy-overhead-unit-and-inline-dup1}{0.046}
\DefMacro{exp-python-numpy_numpy-time-vanilla-dup1}{147.82}
\DefMacro{exp-python-numpy_numpy-time-unit-dup1}{145.36}
\DefMacro{exp-python-numpy_numpy-time-inline-dup1}{0.52}
\DefMacro{exp-python-numpy_numpy-time-unit-and-inline-dup1}{145.88}
\DefMacro{exp-python-numpy_numpy-num-vanilla-dup1}{19,644}
\DefMacro{exp-python-numpy_numpy-num-unit-dup1}{19,644}
\DefMacro{exp-python-numpy_numpy-num-inline-dup1}{2}
\DefMacro{exp-python-numpy_numpy-num-unit-and-inline-dup1}{19,646}
\DefMacro{exp-python-numpy_numpy-timept-vanilla-dup1}{0.008}
\DefMacro{exp-python-numpy_numpy-timept-unit-dup1}{0.007}
\DefMacro{exp-python-numpy_numpy-timept-inline-dup1}{0.258}
\DefMacro{exp-python-numpy_numpy-timept-unit-and-inline-dup1}{0.007}
\DefMacro{exp-python-numpy_numpy-overhead-unit-dup1}{-0.017}
\DefMacro{exp-python-numpy_numpy-overhead-unit-and-inline-dup1}{-0.013}
\DefMacro{exp-python-pandas-dev_pandas-time-vanilla-dup1}{278.43}
\DefMacro{exp-python-pandas-dev_pandas-time-unit-dup1}{278.88}
\DefMacro{exp-python-pandas-dev_pandas-time-inline-dup1}{0.93}
\DefMacro{exp-python-pandas-dev_pandas-time-unit-and-inline-dup1}{279.81}
\DefMacro{exp-python-pandas-dev_pandas-num-vanilla-dup1}{147,307}
\DefMacro{exp-python-pandas-dev_pandas-num-unit-dup1}{147,307}
\DefMacro{exp-python-pandas-dev_pandas-num-inline-dup1}{2}
\DefMacro{exp-python-pandas-dev_pandas-num-unit-and-inline-dup1}{147,309}
\DefMacro{exp-python-pandas-dev_pandas-timept-vanilla-dup1}{0.002}
\DefMacro{exp-python-pandas-dev_pandas-timept-unit-dup1}{0.002}
\DefMacro{exp-python-pandas-dev_pandas-timept-inline-dup1}{0.467}
\DefMacro{exp-python-pandas-dev_pandas-timept-unit-and-inline-dup1}{0.002}
\DefMacro{exp-python-pandas-dev_pandas-overhead-unit-dup1}{0.002}
\DefMacro{exp-python-pandas-dev_pandas-overhead-unit-and-inline-dup1}{0.005}
\DefMacro{exp-python-psf_black-time-vanilla-dup1}{6.96}
\DefMacro{exp-python-psf_black-time-unit-dup1}{7.02}
\DefMacro{exp-python-psf_black-time-inline-dup1}{0.27}
\DefMacro{exp-python-psf_black-time-unit-and-inline-dup1}{7.29}
\DefMacro{exp-python-psf_black-num-vanilla-dup1}{236}
\DefMacro{exp-python-psf_black-num-unit-dup1}{236}
\DefMacro{exp-python-psf_black-num-inline-dup1}{1}
\DefMacro{exp-python-psf_black-num-unit-and-inline-dup1}{237}
\DefMacro{exp-python-psf_black-timept-vanilla-dup1}{0.029}
\DefMacro{exp-python-psf_black-timept-unit-dup1}{0.030}
\DefMacro{exp-python-psf_black-timept-inline-dup1}{0.271}
\DefMacro{exp-python-psf_black-timept-unit-and-inline-dup1}{0.031}
\DefMacro{exp-python-psf_black-overhead-unit-dup1}{0.009}
\DefMacro{exp-python-psf_black-overhead-unit-and-inline-dup1}{0.048}
\DefMacro{exp-python-pypa_pipenv-time-vanilla-dup1}{3.63}
\DefMacro{exp-python-pypa_pipenv-time-unit-dup1}{3.64}
\DefMacro{exp-python-pypa_pipenv-time-inline-dup1}{0.54}
\DefMacro{exp-python-pypa_pipenv-time-unit-and-inline-dup1}{4.17}
\DefMacro{exp-python-pypa_pipenv-num-vanilla-dup1}{106}
\DefMacro{exp-python-pypa_pipenv-num-unit-dup1}{106}
\DefMacro{exp-python-pypa_pipenv-num-inline-dup1}{1}
\DefMacro{exp-python-pypa_pipenv-num-unit-and-inline-dup1}{107}
\DefMacro{exp-python-pypa_pipenv-timept-vanilla-dup1}{0.034}
\DefMacro{exp-python-pypa_pipenv-timept-unit-dup1}{0.034}
\DefMacro{exp-python-pypa_pipenv-timept-inline-dup1}{0.537}
\DefMacro{exp-python-pypa_pipenv-timept-unit-and-inline-dup1}{0.039}
\DefMacro{exp-python-pypa_pipenv-overhead-unit-dup1}{0.003}
\DefMacro{exp-python-pypa_pipenv-overhead-unit-and-inline-dup1}{0.151}
\DefMacro{exp-python-scrapy_scrapy-time-vanilla-dup1}{130.07}
\DefMacro{exp-python-scrapy_scrapy-time-unit-dup1}{130.42}
\DefMacro{exp-python-scrapy_scrapy-time-inline-dup1}{0.51}
\DefMacro{exp-python-scrapy_scrapy-time-unit-and-inline-dup1}{130.93}
\DefMacro{exp-python-scrapy_scrapy-num-vanilla-dup1}{2,246}
\DefMacro{exp-python-scrapy_scrapy-num-unit-dup1}{2,246}
\DefMacro{exp-python-scrapy_scrapy-num-inline-dup1}{2}
\DefMacro{exp-python-scrapy_scrapy-num-unit-and-inline-dup1}{2,248}
\DefMacro{exp-python-scrapy_scrapy-timept-vanilla-dup1}{0.058}
\DefMacro{exp-python-scrapy_scrapy-timept-unit-dup1}{0.058}
\DefMacro{exp-python-scrapy_scrapy-timept-inline-dup1}{0.256}
\DefMacro{exp-python-scrapy_scrapy-timept-unit-and-inline-dup1}{0.058}
\DefMacro{exp-python-scrapy_scrapy-overhead-unit-dup1}{0.003}
\DefMacro{exp-python-scrapy_scrapy-overhead-unit-and-inline-dup1}{0.007}
\DefMacro{exp-python-time-vanilla-AVG-dup1}{62.42}
\DefMacro{exp-python-time-vanilla-SUM-dup1}{873.93}
\DefMacro{exp-python-time-vanilla-MAX-dup1}{278.43}
\DefMacro{exp-python-time-vanilla-MIN-dup1}{0.17}
\DefMacro{exp-python-time-vanilla-MEDIAN-dup1}{7.23}
\DefMacro{exp-python-time-vanilla-STDEV-dup1}{90.89}
\DefMacro{exp-python-time-vanilla-CNT-dup1}{14}
\DefMacro{exp-python-time-unit-AVG-dup1}{62.37}
\DefMacro{exp-python-time-unit-SUM-dup1}{873.16}
\DefMacro{exp-python-time-unit-MAX-dup1}{278.88}
\DefMacro{exp-python-time-unit-MIN-dup1}{0.18}
\DefMacro{exp-python-time-unit-MEDIAN-dup1}{7.23}
\DefMacro{exp-python-time-unit-STDEV-dup1}{90.88}
\DefMacro{exp-python-time-unit-CNT-dup1}{14}
\DefMacro{exp-python-time-inline-AVG-dup1}{0.51}
\DefMacro{exp-python-time-inline-SUM-dup1}{7.08}
\DefMacro{exp-python-time-inline-MAX-dup1}{0.93}
\DefMacro{exp-python-time-inline-MIN-dup1}{0.20}
\DefMacro{exp-python-time-inline-MEDIAN-dup1}{0.51}
\DefMacro{exp-python-time-inline-STDEV-dup1}{0.22}
\DefMacro{exp-python-time-inline-CNT-dup1}{14}
\DefMacro{exp-python-time-unit-and-inline-AVG-dup1}{62.87}
\DefMacro{exp-python-time-unit-and-inline-SUM-dup1}{880.24}
\DefMacro{exp-python-time-unit-and-inline-MAX-dup1}{279.81}
\DefMacro{exp-python-time-unit-and-inline-MIN-dup1}{0.38}
\DefMacro{exp-python-time-unit-and-inline-MEDIAN-dup1}{7.57}
\DefMacro{exp-python-time-unit-and-inline-STDEV-dup1}{90.98}
\DefMacro{exp-python-time-unit-and-inline-CNT-dup1}{14}
\DefMacro{exp-python-num-vanilla-AVG-dup1}{13,046.29}
\DefMacro{exp-python-num-vanilla-SUM-dup1}{182,648}
\DefMacro{exp-python-num-vanilla-MAX-dup1}{147,307}
\DefMacro{exp-python-num-vanilla-MIN-dup1}{1}
\DefMacro{exp-python-num-vanilla-MEDIAN-dup1}{795.00}
\DefMacro{exp-python-num-vanilla-STDEV-dup1}{37,594.87}
\DefMacro{exp-python-num-vanilla-CNT-dup1}{14}
\DefMacro{exp-python-num-unit-AVG-dup1}{13,046.29}
\DefMacro{exp-python-num-unit-SUM-dup1}{182,648}
\DefMacro{exp-python-num-unit-MAX-dup1}{147,307}
\DefMacro{exp-python-num-unit-MIN-dup1}{1}
\DefMacro{exp-python-num-unit-MEDIAN-dup1}{795.00}
\DefMacro{exp-python-num-unit-STDEV-dup1}{37,594.87}
\DefMacro{exp-python-num-unit-CNT-dup1}{14}
\DefMacro{exp-python-num-inline-AVG-dup1}{2.43}
\DefMacro{exp-python-num-inline-SUM-dup1}{34}
\DefMacro{exp-python-num-inline-MAX-dup1}{8}
\DefMacro{exp-python-num-inline-MIN-dup1}{1}
\DefMacro{exp-python-num-inline-MEDIAN-dup1}{2.00}
\DefMacro{exp-python-num-inline-STDEV-dup1}{1.84}
\DefMacro{exp-python-num-inline-CNT-dup1}{14}
\DefMacro{exp-python-num-unit-and-inline-AVG-dup1}{13,048.71}
\DefMacro{exp-python-num-unit-and-inline-SUM-dup1}{182,682}
\DefMacro{exp-python-num-unit-and-inline-MAX-dup1}{147,309}
\DefMacro{exp-python-num-unit-and-inline-MIN-dup1}{4}
\DefMacro{exp-python-num-unit-and-inline-MEDIAN-dup1}{797.00}
\DefMacro{exp-python-num-unit-and-inline-STDEV-dup1}{37,594.82}
\DefMacro{exp-python-num-unit-and-inline-CNT-dup1}{14}
\DefMacro{exp-python-timept-vanilla-AVG-dup1}{0.097}
\DefMacro{exp-python-timept-vanilla-SUM-dup1}{1.359}
\DefMacro{exp-python-timept-vanilla-MAX-dup1}{0.337}
\DefMacro{exp-python-timept-vanilla-MIN-dup1}{0.002}
\DefMacro{exp-python-timept-vanilla-MEDIAN-dup1}{0.032}
\DefMacro{exp-python-timept-vanilla-STDEV-dup1}{0.118}
\DefMacro{exp-python-timept-vanilla-CNT-dup1}{14}
\DefMacro{exp-python-timept-unit-AVG-dup1}{0.098}
\DefMacro{exp-python-timept-unit-SUM-dup1}{1.369}
\DefMacro{exp-python-timept-unit-MAX-dup1}{0.334}
\DefMacro{exp-python-timept-unit-MIN-dup1}{0.002}
\DefMacro{exp-python-timept-unit-MEDIAN-dup1}{0.032}
\DefMacro{exp-python-timept-unit-STDEV-dup1}{0.118}
\DefMacro{exp-python-timept-unit-CNT-dup1}{14}
\DefMacro{exp-python-timept-inline-AVG-dup1}{0.315}
\DefMacro{exp-python-timept-inline-SUM-dup1}{4.403}
\DefMacro{exp-python-timept-inline-MAX-dup1}{0.855}
\DefMacro{exp-python-timept-inline-MIN-dup1}{0.049}
\DefMacro{exp-python-timept-inline-MEDIAN-dup1}{0.264}
\DefMacro{exp-python-timept-inline-STDEV-dup1}{0.228}
\DefMacro{exp-python-timept-inline-CNT-dup1}{14}
\DefMacro{exp-python-timept-unit-and-inline-AVG-dup1}{0.092}
\DefMacro{exp-python-timept-unit-and-inline-SUM-dup1}{1.287}
\DefMacro{exp-python-timept-unit-and-inline-MAX-dup1}{0.458}
\DefMacro{exp-python-timept-unit-and-inline-MIN-dup1}{0.002}
\DefMacro{exp-python-timept-unit-and-inline-MEDIAN-dup1}{0.035}
\DefMacro{exp-python-timept-unit-and-inline-STDEV-dup1}{0.125}
\DefMacro{exp-python-timept-unit-and-inline-CNT-dup1}{14}
\DefMacro{exp-python-overhead-unit-AVG-dup1}{0.005}
\DefMacro{exp-python-overhead-unit-SUM-dup1}{0.066}
\DefMacro{exp-python-overhead-unit-MAX-dup1}{0.058}
\DefMacro{exp-python-overhead-unit-MIN-dup1}{-0.017}
\DefMacro{exp-python-overhead-unit-MEDIAN-dup1}{0.002}
\DefMacro{exp-python-overhead-unit-STDEV-dup1}{0.017}
\DefMacro{exp-python-overhead-unit-CNT-dup1}{14}
\DefMacro{exp-python-overhead-unit-and-inline-AVG-dup1}{0.284}
\DefMacro{exp-python-overhead-unit-and-inline-SUM-dup1}{3.971}
\DefMacro{exp-python-overhead-unit-and-inline-MAX-dup1}{1.217}
\DefMacro{exp-python-overhead-unit-and-inline-MIN-dup1}{-0.013}
\DefMacro{exp-python-overhead-unit-and-inline-MEDIAN-dup1}{0.047}
\DefMacro{exp-python-overhead-unit-and-inline-STDEV-dup1}{0.441}
\DefMacro{exp-python-overhead-unit-and-inline-CNT-dup1}{14}
\DefMacro{exp-python-timept-vanilla-MACROAVG-dup1}{0.005}
\DefMacro{exp-python-timept-unit-MACROAVG-dup1}{0.005}
\DefMacro{exp-python-timept-inline-MACROAVG-dup1}{0.208}
\DefMacro{exp-python-timept-unit-and-inline-MACROAVG-dup1}{0.005}
\DefMacro{exp-python-overhead-unit-MACROAVG-dup1}{-0.001}
\DefMacro{exp-python-overhead-unit-and-inline-MACROAVG-dup1}{0.007}
\DefMacro{exp-python-RaRe-Technologies_gensim-time-vanilla-dup10}{226.40}
\DefMacro{exp-python-RaRe-Technologies_gensim-time-unit-dup10}{225.90}
\DefMacro{exp-python-RaRe-Technologies_gensim-time-inline-dup10}{0.57}
\DefMacro{exp-python-RaRe-Technologies_gensim-time-unit-and-inline-dup10}{226.47}
\DefMacro{exp-python-RaRe-Technologies_gensim-num-vanilla-dup10}{968}
\DefMacro{exp-python-RaRe-Technologies_gensim-num-unit-dup10}{968}
\DefMacro{exp-python-RaRe-Technologies_gensim-num-inline-dup10}{20}
\DefMacro{exp-python-RaRe-Technologies_gensim-num-unit-and-inline-dup10}{988}
\DefMacro{exp-python-RaRe-Technologies_gensim-timept-vanilla-dup10}{0.234}
\DefMacro{exp-python-RaRe-Technologies_gensim-timept-unit-dup10}{0.233}
\DefMacro{exp-python-RaRe-Technologies_gensim-timept-inline-dup10}{0.028}
\DefMacro{exp-python-RaRe-Technologies_gensim-timept-unit-and-inline-dup10}{0.229}
\DefMacro{exp-python-RaRe-Technologies_gensim-overhead-unit-dup10}{-0.002}
\DefMacro{exp-python-RaRe-Technologies_gensim-overhead-unit-and-inline-dup10}{0.000}
\DefMacro{exp-python-Textualize_rich-time-vanilla-dup10}{3.71}
\DefMacro{exp-python-Textualize_rich-time-unit-dup10}{3.71}
\DefMacro{exp-python-Textualize_rich-time-inline-dup10}{0.24}
\DefMacro{exp-python-Textualize_rich-time-unit-and-inline-dup10}{3.95}
\DefMacro{exp-python-Textualize_rich-num-vanilla-dup10}{622}
\DefMacro{exp-python-Textualize_rich-num-unit-dup10}{622}
\DefMacro{exp-python-Textualize_rich-num-inline-dup10}{20}
\DefMacro{exp-python-Textualize_rich-num-unit-and-inline-dup10}{642}
\DefMacro{exp-python-Textualize_rich-timept-vanilla-dup10}{0.006}
\DefMacro{exp-python-Textualize_rich-timept-unit-dup10}{0.006}
\DefMacro{exp-python-Textualize_rich-timept-inline-dup10}{0.012}
\DefMacro{exp-python-Textualize_rich-timept-unit-and-inline-dup10}{0.006}
\DefMacro{exp-python-Textualize_rich-overhead-unit-dup10}{0.002}
\DefMacro{exp-python-Textualize_rich-overhead-unit-and-inline-dup10}{0.067}
\DefMacro{exp-python-bokeh_bokeh-time-vanilla-dup10}{49.72}
\DefMacro{exp-python-bokeh_bokeh-time-unit-dup10}{91.37}
\DefMacro{exp-python-bokeh_bokeh-time-inline-dup10}{0.85}
\DefMacro{exp-python-bokeh_bokeh-time-unit-and-inline-dup10}{92.22}
\DefMacro{exp-python-bokeh_bokeh-num-vanilla-dup10}{8,616}
\DefMacro{exp-python-bokeh_bokeh-num-unit-dup10}{8,616}
\DefMacro{exp-python-bokeh_bokeh-num-inline-dup10}{80}
\DefMacro{exp-python-bokeh_bokeh-num-unit-and-inline-dup10}{8,696}
\DefMacro{exp-python-bokeh_bokeh-timept-vanilla-dup10}{0.006}
\DefMacro{exp-python-bokeh_bokeh-timept-unit-dup10}{0.011}
\DefMacro{exp-python-bokeh_bokeh-timept-inline-dup10}{0.011}
\DefMacro{exp-python-bokeh_bokeh-timept-unit-and-inline-dup10}{0.011}
\DefMacro{exp-python-bokeh_bokeh-overhead-unit-dup10}{0.838}
\DefMacro{exp-python-bokeh_bokeh-overhead-unit-and-inline-dup10}{0.855}
\DefMacro{exp-python-chubin_cheat.sh-time-vanilla-dup10}{0.34}
\DefMacro{exp-python-chubin_cheat.sh-time-unit-dup10}{0.33}
\DefMacro{exp-python-chubin_cheat.sh-time-inline-dup10}{0.43}
\DefMacro{exp-python-chubin_cheat.sh-time-unit-and-inline-dup10}{0.76}
\DefMacro{exp-python-chubin_cheat.sh-num-vanilla-dup10}{1}
\DefMacro{exp-python-chubin_cheat.sh-num-unit-dup10}{1}
\DefMacro{exp-python-chubin_cheat.sh-num-inline-dup10}{30}
\DefMacro{exp-python-chubin_cheat.sh-num-unit-and-inline-dup10}{31}
\DefMacro{exp-python-chubin_cheat.sh-timept-vanilla-dup10}{0.338}
\DefMacro{exp-python-chubin_cheat.sh-timept-unit-dup10}{0.332}
\DefMacro{exp-python-chubin_cheat.sh-timept-inline-dup10}{0.014}
\DefMacro{exp-python-chubin_cheat.sh-timept-unit-and-inline-dup10}{0.024}
\DefMacro{exp-python-chubin_cheat.sh-overhead-unit-dup10}{-0.018}
\DefMacro{exp-python-chubin_cheat.sh-overhead-unit-and-inline-dup10}{1.242}
\DefMacro{exp-python-davidsandberg_facenet-time-vanilla-dup10}{0.97}
\DefMacro{exp-python-davidsandberg_facenet-time-unit-dup10}{0.97}
\DefMacro{exp-python-davidsandberg_facenet-time-inline-dup10}{0.86}
\DefMacro{exp-python-davidsandberg_facenet-time-unit-and-inline-dup10}{1.84}
\DefMacro{exp-python-davidsandberg_facenet-num-vanilla-dup10}{3}
\DefMacro{exp-python-davidsandberg_facenet-num-unit-dup10}{3}
\DefMacro{exp-python-davidsandberg_facenet-num-inline-dup10}{10}
\DefMacro{exp-python-davidsandberg_facenet-num-unit-and-inline-dup10}{13}
\DefMacro{exp-python-davidsandberg_facenet-timept-vanilla-dup10}{0.323}
\DefMacro{exp-python-davidsandberg_facenet-timept-unit-dup10}{0.324}
\DefMacro{exp-python-davidsandberg_facenet-timept-inline-dup10}{0.086}
\DefMacro{exp-python-davidsandberg_facenet-timept-unit-and-inline-dup10}{0.141}
\DefMacro{exp-python-davidsandberg_facenet-overhead-unit-dup10}{0.005}
\DefMacro{exp-python-davidsandberg_facenet-overhead-unit-and-inline-dup10}{0.897}
\DefMacro{exp-python-geekcomputers_Python-time-vanilla-dup10}{0.17}
\DefMacro{exp-python-geekcomputers_Python-time-unit-dup10}{0.18}
\DefMacro{exp-python-geekcomputers_Python-time-inline-dup10}{0.22}
\DefMacro{exp-python-geekcomputers_Python-time-unit-and-inline-dup10}{0.40}
\DefMacro{exp-python-geekcomputers_Python-num-vanilla-dup10}{1}
\DefMacro{exp-python-geekcomputers_Python-num-unit-dup10}{1}
\DefMacro{exp-python-geekcomputers_Python-num-inline-dup10}{40}
\DefMacro{exp-python-geekcomputers_Python-num-unit-and-inline-dup10}{41}
\DefMacro{exp-python-geekcomputers_Python-timept-vanilla-dup10}{0.172}
\DefMacro{exp-python-geekcomputers_Python-timept-unit-dup10}{0.179}
\DefMacro{exp-python-geekcomputers_Python-timept-inline-dup10}{0.006}
\DefMacro{exp-python-geekcomputers_Python-timept-unit-and-inline-dup10}{0.010}
\DefMacro{exp-python-geekcomputers_Python-overhead-unit-dup10}{0.037}
\DefMacro{exp-python-geekcomputers_Python-overhead-unit-and-inline-dup10}{1.319}
\DefMacro{exp-python-google-research_bert-time-vanilla-dup10}{2.06}
\DefMacro{exp-python-google-research_bert-time-unit-dup10}{2.07}
\DefMacro{exp-python-google-research_bert-time-inline-dup10}{0.63}
\DefMacro{exp-python-google-research_bert-time-unit-and-inline-dup10}{2.70}
\DefMacro{exp-python-google-research_bert-num-vanilla-dup10}{15}
\DefMacro{exp-python-google-research_bert-num-unit-dup10}{15}
\DefMacro{exp-python-google-research_bert-num-inline-dup10}{10}
\DefMacro{exp-python-google-research_bert-num-unit-and-inline-dup10}{25}
\DefMacro{exp-python-google-research_bert-timept-vanilla-dup10}{0.138}
\DefMacro{exp-python-google-research_bert-timept-unit-dup10}{0.138}
\DefMacro{exp-python-google-research_bert-timept-inline-dup10}{0.063}
\DefMacro{exp-python-google-research_bert-timept-unit-and-inline-dup10}{0.108}
\DefMacro{exp-python-google-research_bert-overhead-unit-dup10}{0.004}
\DefMacro{exp-python-google-research_bert-overhead-unit-and-inline-dup10}{0.309}
\DefMacro{exp-python-joke2k_faker-time-vanilla-dup10}{16.70}
\DefMacro{exp-python-joke2k_faker-time-unit-dup10}{16.63}
\DefMacro{exp-python-joke2k_faker-time-inline-dup10}{0.30}
\DefMacro{exp-python-joke2k_faker-time-unit-and-inline-dup10}{16.93}
\DefMacro{exp-python-joke2k_faker-num-vanilla-dup10}{1,596}
\DefMacro{exp-python-joke2k_faker-num-unit-dup10}{1,596}
\DefMacro{exp-python-joke2k_faker-num-inline-dup10}{40}
\DefMacro{exp-python-joke2k_faker-num-unit-and-inline-dup10}{1,636}
\DefMacro{exp-python-joke2k_faker-timept-vanilla-dup10}{0.010}
\DefMacro{exp-python-joke2k_faker-timept-unit-dup10}{0.010}
\DefMacro{exp-python-joke2k_faker-timept-inline-dup10}{0.007}
\DefMacro{exp-python-joke2k_faker-timept-unit-and-inline-dup10}{0.010}
\DefMacro{exp-python-joke2k_faker-overhead-unit-dup10}{-0.004}
\DefMacro{exp-python-joke2k_faker-overhead-unit-and-inline-dup10}{0.014}
\DefMacro{exp-python-mitmproxy_mitmproxy-time-vanilla-dup10}{7.47}
\DefMacro{exp-python-mitmproxy_mitmproxy-time-unit-dup10}{7.48}
\DefMacro{exp-python-mitmproxy_mitmproxy-time-inline-dup10}{0.40}
\DefMacro{exp-python-mitmproxy_mitmproxy-time-unit-and-inline-dup10}{7.88}
\DefMacro{exp-python-mitmproxy_mitmproxy-num-vanilla-dup10}{1,287}
\DefMacro{exp-python-mitmproxy_mitmproxy-num-unit-dup10}{1,287}
\DefMacro{exp-python-mitmproxy_mitmproxy-num-inline-dup10}{10}
\DefMacro{exp-python-mitmproxy_mitmproxy-num-unit-and-inline-dup10}{1,297}
\DefMacro{exp-python-mitmproxy_mitmproxy-timept-vanilla-dup10}{0.006}
\DefMacro{exp-python-mitmproxy_mitmproxy-timept-unit-dup10}{0.006}
\DefMacro{exp-python-mitmproxy_mitmproxy-timept-inline-dup10}{0.040}
\DefMacro{exp-python-mitmproxy_mitmproxy-timept-unit-and-inline-dup10}{0.006}
\DefMacro{exp-python-mitmproxy_mitmproxy-overhead-unit-dup10}{0.001}
\DefMacro{exp-python-mitmproxy_mitmproxy-overhead-unit-and-inline-dup10}{0.055}
\DefMacro{exp-python-numpy_numpy-time-vanilla-dup10}{146.29}
\DefMacro{exp-python-numpy_numpy-time-unit-dup10}{147.92}
\DefMacro{exp-python-numpy_numpy-time-inline-dup10}{0.54}
\DefMacro{exp-python-numpy_numpy-time-unit-and-inline-dup10}{148.45}
\DefMacro{exp-python-numpy_numpy-num-vanilla-dup10}{19,644}
\DefMacro{exp-python-numpy_numpy-num-unit-dup10}{19,644}
\DefMacro{exp-python-numpy_numpy-num-inline-dup10}{20}
\DefMacro{exp-python-numpy_numpy-num-unit-and-inline-dup10}{19,664}
\DefMacro{exp-python-numpy_numpy-timept-vanilla-dup10}{0.007}
\DefMacro{exp-python-numpy_numpy-timept-unit-dup10}{0.008}
\DefMacro{exp-python-numpy_numpy-timept-inline-dup10}{0.027}
\DefMacro{exp-python-numpy_numpy-timept-unit-and-inline-dup10}{0.008}
\DefMacro{exp-python-numpy_numpy-overhead-unit-dup10}{0.011}
\DefMacro{exp-python-numpy_numpy-overhead-unit-and-inline-dup10}{0.015}
\DefMacro{exp-python-pandas-dev_pandas-time-vanilla-dup10}{277.40}
\DefMacro{exp-python-pandas-dev_pandas-time-unit-dup10}{277.36}
\DefMacro{exp-python-pandas-dev_pandas-time-inline-dup10}{0.95}
\DefMacro{exp-python-pandas-dev_pandas-time-unit-and-inline-dup10}{278.31}
\DefMacro{exp-python-pandas-dev_pandas-num-vanilla-dup10}{147,306}
\DefMacro{exp-python-pandas-dev_pandas-num-unit-dup10}{147,305}
\DefMacro{exp-python-pandas-dev_pandas-num-inline-dup10}{20}
\DefMacro{exp-python-pandas-dev_pandas-num-unit-and-inline-dup10}{147,325}
\DefMacro{exp-python-pandas-dev_pandas-timept-vanilla-dup10}{0.002}
\DefMacro{exp-python-pandas-dev_pandas-timept-unit-dup10}{0.002}
\DefMacro{exp-python-pandas-dev_pandas-timept-inline-dup10}{0.048}
\DefMacro{exp-python-pandas-dev_pandas-timept-unit-and-inline-dup10}{0.002}
\DefMacro{exp-python-pandas-dev_pandas-overhead-unit-dup10}{-0.000}
\DefMacro{exp-python-pandas-dev_pandas-overhead-unit-and-inline-dup10}{0.003}
\DefMacro{exp-python-psf_black-time-vanilla-dup10}{6.95}
\DefMacro{exp-python-psf_black-time-unit-dup10}{7.02}
\DefMacro{exp-python-psf_black-time-inline-dup10}{0.28}
\DefMacro{exp-python-psf_black-time-unit-and-inline-dup10}{7.29}
\DefMacro{exp-python-psf_black-num-vanilla-dup10}{236}
\DefMacro{exp-python-psf_black-num-unit-dup10}{236}
\DefMacro{exp-python-psf_black-num-inline-dup10}{10}
\DefMacro{exp-python-psf_black-num-unit-and-inline-dup10}{246}
\DefMacro{exp-python-psf_black-timept-vanilla-dup10}{0.029}
\DefMacro{exp-python-psf_black-timept-unit-dup10}{0.030}
\DefMacro{exp-python-psf_black-timept-inline-dup10}{0.028}
\DefMacro{exp-python-psf_black-timept-unit-and-inline-dup10}{0.030}
\DefMacro{exp-python-psf_black-overhead-unit-dup10}{0.010}
\DefMacro{exp-python-psf_black-overhead-unit-and-inline-dup10}{0.050}
\DefMacro{exp-python-pypa_pipenv-time-vanilla-dup10}{3.65}
\DefMacro{exp-python-pypa_pipenv-time-unit-dup10}{3.62}
\DefMacro{exp-python-pypa_pipenv-time-inline-dup10}{0.54}
\DefMacro{exp-python-pypa_pipenv-time-unit-and-inline-dup10}{4.17}
\DefMacro{exp-python-pypa_pipenv-num-vanilla-dup10}{106}
\DefMacro{exp-python-pypa_pipenv-num-unit-dup10}{106}
\DefMacro{exp-python-pypa_pipenv-num-inline-dup10}{10}
\DefMacro{exp-python-pypa_pipenv-num-unit-and-inline-dup10}{116}
\DefMacro{exp-python-pypa_pipenv-timept-vanilla-dup10}{0.034}
\DefMacro{exp-python-pypa_pipenv-timept-unit-dup10}{0.034}
\DefMacro{exp-python-pypa_pipenv-timept-inline-dup10}{0.054}
\DefMacro{exp-python-pypa_pipenv-timept-unit-and-inline-dup10}{0.036}
\DefMacro{exp-python-pypa_pipenv-overhead-unit-dup10}{-0.007}
\DefMacro{exp-python-pypa_pipenv-overhead-unit-and-inline-dup10}{0.142}
\DefMacro{exp-python-scrapy_scrapy-time-vanilla-dup10}{129.90}
\DefMacro{exp-python-scrapy_scrapy-time-unit-dup10}{130.11}
\DefMacro{exp-python-scrapy_scrapy-time-inline-dup10}{0.54}
\DefMacro{exp-python-scrapy_scrapy-time-unit-and-inline-dup10}{130.65}
\DefMacro{exp-python-scrapy_scrapy-num-vanilla-dup10}{2,246}
\DefMacro{exp-python-scrapy_scrapy-num-unit-dup10}{2,246}
\DefMacro{exp-python-scrapy_scrapy-num-inline-dup10}{20}
\DefMacro{exp-python-scrapy_scrapy-num-unit-and-inline-dup10}{2,266}
\DefMacro{exp-python-scrapy_scrapy-timept-vanilla-dup10}{0.058}
\DefMacro{exp-python-scrapy_scrapy-timept-unit-dup10}{0.058}
\DefMacro{exp-python-scrapy_scrapy-timept-inline-dup10}{0.027}
\DefMacro{exp-python-scrapy_scrapy-timept-unit-and-inline-dup10}{0.058}
\DefMacro{exp-python-scrapy_scrapy-overhead-unit-dup10}{0.002}
\DefMacro{exp-python-scrapy_scrapy-overhead-unit-and-inline-dup10}{0.006}
\DefMacro{exp-python-time-vanilla-AVG-dup10}{62.27}
\DefMacro{exp-python-time-vanilla-SUM-dup10}{871.73}
\DefMacro{exp-python-time-vanilla-MAX-dup10}{277.40}
\DefMacro{exp-python-time-vanilla-MIN-dup10}{0.17}
\DefMacro{exp-python-time-vanilla-MEDIAN-dup10}{7.21}
\DefMacro{exp-python-time-vanilla-STDEV-dup10}{90.67}
\DefMacro{exp-python-time-vanilla-CNT-dup10}{14}
\DefMacro{exp-python-time-unit-AVG-dup10}{65.33}
\DefMacro{exp-python-time-unit-SUM-dup10}{914.68}
\DefMacro{exp-python-time-unit-MAX-dup10}{277.36}
\DefMacro{exp-python-time-unit-MIN-dup10}{0.18}
\DefMacro{exp-python-time-unit-MEDIAN-dup10}{7.25}
\DefMacro{exp-python-time-unit-STDEV-dup10}{90.94}
\DefMacro{exp-python-time-unit-CNT-dup10}{14}
\DefMacro{exp-python-time-inline-AVG-dup10}{0.53}
\DefMacro{exp-python-time-inline-SUM-dup10}{7.35}
\DefMacro{exp-python-time-inline-MAX-dup10}{0.95}
\DefMacro{exp-python-time-inline-MIN-dup10}{0.22}
\DefMacro{exp-python-time-inline-MEDIAN-dup10}{0.54}
\DefMacro{exp-python-time-inline-STDEV-dup10}{0.23}
\DefMacro{exp-python-time-inline-CNT-dup10}{14}
\DefMacro{exp-python-time-unit-and-inline-AVG-dup10}{65.86}
\DefMacro{exp-python-time-unit-and-inline-SUM-dup10}{922.03}
\DefMacro{exp-python-time-unit-and-inline-MAX-dup10}{278.31}
\DefMacro{exp-python-time-unit-and-inline-MIN-dup10}{0.40}
\DefMacro{exp-python-time-unit-and-inline-MEDIAN-dup10}{7.59}
\DefMacro{exp-python-time-unit-and-inline-STDEV-dup10}{91.06}
\DefMacro{exp-python-time-unit-and-inline-CNT-dup10}{14}
\DefMacro{exp-python-num-vanilla-AVG-dup10}{13,046.21}
\DefMacro{exp-python-num-vanilla-SUM-dup10}{182,647}
\DefMacro{exp-python-num-vanilla-MAX-dup10}{147,306}
\DefMacro{exp-python-num-vanilla-MIN-dup10}{1}
\DefMacro{exp-python-num-vanilla-MEDIAN-dup10}{795.00}
\DefMacro{exp-python-num-vanilla-STDEV-dup10}{37,594.61}
\DefMacro{exp-python-num-vanilla-CNT-dup10}{14}
\DefMacro{exp-python-num-unit-AVG-dup10}{13,046.14}
\DefMacro{exp-python-num-unit-SUM-dup10}{182,646}
\DefMacro{exp-python-num-unit-MAX-dup10}{147,305}
\DefMacro{exp-python-num-unit-MIN-dup10}{1}
\DefMacro{exp-python-num-unit-MEDIAN-dup10}{795.00}
\DefMacro{exp-python-num-unit-STDEV-dup10}{37,594.36}
\DefMacro{exp-python-num-unit-CNT-dup10}{14}
\DefMacro{exp-python-num-inline-AVG-dup10}{24.29}
\DefMacro{exp-python-num-inline-SUM-dup10}{340}
\DefMacro{exp-python-num-inline-MAX-dup10}{80}
\DefMacro{exp-python-num-inline-MIN-dup10}{10}
\DefMacro{exp-python-num-inline-MEDIAN-dup10}{20.00}
\DefMacro{exp-python-num-inline-STDEV-dup10}{18.41}
\DefMacro{exp-python-num-inline-CNT-dup10}{14}
\DefMacro{exp-python-num-unit-and-inline-AVG-dup10}{13,070.43}
\DefMacro{exp-python-num-unit-and-inline-SUM-dup10}{182,986}
\DefMacro{exp-python-num-unit-and-inline-MAX-dup10}{147,325}
\DefMacro{exp-python-num-unit-and-inline-MIN-dup10}{13}
\DefMacro{exp-python-num-unit-and-inline-MEDIAN-dup10}{815.00}
\DefMacro{exp-python-num-unit-and-inline-STDEV-dup10}{37,593.89}
\DefMacro{exp-python-num-unit-and-inline-CNT-dup10}{14}
\DefMacro{exp-python-timept-vanilla-AVG-dup10}{0.097}
\DefMacro{exp-python-timept-vanilla-SUM-dup10}{1.364}
\DefMacro{exp-python-timept-vanilla-MAX-dup10}{0.338}
\DefMacro{exp-python-timept-vanilla-MIN-dup10}{0.002}
\DefMacro{exp-python-timept-vanilla-MEDIAN-dup10}{0.032}
\DefMacro{exp-python-timept-vanilla-STDEV-dup10}{0.118}
\DefMacro{exp-python-timept-vanilla-CNT-dup10}{14}
\DefMacro{exp-python-timept-unit-AVG-dup10}{0.098}
\DefMacro{exp-python-timept-unit-SUM-dup10}{1.371}
\DefMacro{exp-python-timept-unit-MAX-dup10}{0.332}
\DefMacro{exp-python-timept-unit-MIN-dup10}{0.002}
\DefMacro{exp-python-timept-unit-MEDIAN-dup10}{0.032}
\DefMacro{exp-python-timept-unit-STDEV-dup10}{0.117}
\DefMacro{exp-python-timept-unit-CNT-dup10}{14}
\DefMacro{exp-python-timept-inline-AVG-dup10}{0.032}
\DefMacro{exp-python-timept-inline-SUM-dup10}{0.452}
\DefMacro{exp-python-timept-inline-MAX-dup10}{0.086}
\DefMacro{exp-python-timept-inline-MIN-dup10}{0.006}
\DefMacro{exp-python-timept-inline-MEDIAN-dup10}{0.027}
\DefMacro{exp-python-timept-inline-STDEV-dup10}{0.023}
\DefMacro{exp-python-timept-inline-CNT-dup10}{14}
\DefMacro{exp-python-timept-unit-and-inline-AVG-dup10}{0.048}
\DefMacro{exp-python-timept-unit-and-inline-SUM-dup10}{0.679}
\DefMacro{exp-python-timept-unit-and-inline-MAX-dup10}{0.229}
\DefMacro{exp-python-timept-unit-and-inline-MIN-dup10}{0.002}
\DefMacro{exp-python-timept-unit-and-inline-MEDIAN-dup10}{0.018}
\DefMacro{exp-python-timept-unit-and-inline-STDEV-dup10}{0.064}
\DefMacro{exp-python-timept-unit-and-inline-CNT-dup10}{14}
\DefMacro{exp-python-overhead-unit-AVG-dup10}{0.063}
\DefMacro{exp-python-overhead-unit-SUM-dup10}{0.877}
\DefMacro{exp-python-overhead-unit-MAX-dup10}{0.838}
\DefMacro{exp-python-overhead-unit-MIN-dup10}{-0.018}
\DefMacro{exp-python-overhead-unit-MEDIAN-dup10}{0.002}
\DefMacro{exp-python-overhead-unit-STDEV-dup10}{0.215}
\DefMacro{exp-python-overhead-unit-CNT-dup10}{14}
\DefMacro{exp-python-overhead-unit-and-inline-AVG-dup10}{0.355}
\DefMacro{exp-python-overhead-unit-and-inline-SUM-dup10}{4.973}
\DefMacro{exp-python-overhead-unit-and-inline-MAX-dup10}{1.319}
\DefMacro{exp-python-overhead-unit-and-inline-MIN-dup10}{0.000}
\DefMacro{exp-python-overhead-unit-and-inline-MEDIAN-dup10}{0.061}
\DefMacro{exp-python-overhead-unit-and-inline-STDEV-dup10}{0.476}
\DefMacro{exp-python-overhead-unit-and-inline-CNT-dup10}{14}
\DefMacro{exp-python-timept-vanilla-MACROAVG-dup10}{0.005}
\DefMacro{exp-python-timept-unit-MACROAVG-dup10}{0.005}
\DefMacro{exp-python-timept-inline-MACROAVG-dup10}{0.022}
\DefMacro{exp-python-timept-unit-and-inline-MACROAVG-dup10}{0.005}
\DefMacro{exp-python-overhead-unit-MACROAVG-dup10}{0.049}
\DefMacro{exp-python-overhead-unit-and-inline-MACROAVG-dup10}{0.058}
\DefMacro{exp-python-RaRe-Technologies_gensim-time-vanilla-dup100}{227.16}
\DefMacro{exp-python-RaRe-Technologies_gensim-time-unit-dup100}{226.82}
\DefMacro{exp-python-RaRe-Technologies_gensim-time-inline-dup100}{0.72}
\DefMacro{exp-python-RaRe-Technologies_gensim-time-unit-and-inline-dup100}{227.54}
\DefMacro{exp-python-RaRe-Technologies_gensim-num-vanilla-dup100}{968}
\DefMacro{exp-python-RaRe-Technologies_gensim-num-unit-dup100}{968}
\DefMacro{exp-python-RaRe-Technologies_gensim-num-inline-dup100}{200}
\DefMacro{exp-python-RaRe-Technologies_gensim-num-unit-and-inline-dup100}{1,168}
\DefMacro{exp-python-RaRe-Technologies_gensim-timept-vanilla-dup100}{0.235}
\DefMacro{exp-python-RaRe-Technologies_gensim-timept-unit-dup100}{0.234}
\DefMacro{exp-python-RaRe-Technologies_gensim-timept-inline-dup100}{0.004}
\DefMacro{exp-python-RaRe-Technologies_gensim-timept-unit-and-inline-dup100}{0.195}
\DefMacro{exp-python-RaRe-Technologies_gensim-overhead-unit-dup100}{-0.002}
\DefMacro{exp-python-RaRe-Technologies_gensim-overhead-unit-and-inline-dup100}{0.002}
\DefMacro{exp-python-Textualize_rich-time-vanilla-dup100}{3.71}
\DefMacro{exp-python-Textualize_rich-time-unit-dup100}{3.72}
\DefMacro{exp-python-Textualize_rich-time-inline-dup100}{0.41}
\DefMacro{exp-python-Textualize_rich-time-unit-and-inline-dup100}{4.13}
\DefMacro{exp-python-Textualize_rich-num-vanilla-dup100}{622}
\DefMacro{exp-python-Textualize_rich-num-unit-dup100}{622}
\DefMacro{exp-python-Textualize_rich-num-inline-dup100}{200}
\DefMacro{exp-python-Textualize_rich-num-unit-and-inline-dup100}{822}
\DefMacro{exp-python-Textualize_rich-timept-vanilla-dup100}{0.006}
\DefMacro{exp-python-Textualize_rich-timept-unit-dup100}{0.006}
\DefMacro{exp-python-Textualize_rich-timept-inline-dup100}{0.002}
\DefMacro{exp-python-Textualize_rich-timept-unit-and-inline-dup100}{0.005}
\DefMacro{exp-python-Textualize_rich-overhead-unit-dup100}{0.003}
\DefMacro{exp-python-Textualize_rich-overhead-unit-and-inline-dup100}{0.112}
\DefMacro{exp-python-bokeh_bokeh-time-vanilla-dup100}{51.35}
\DefMacro{exp-python-bokeh_bokeh-time-unit-dup100}{49.15}
\DefMacro{exp-python-bokeh_bokeh-time-inline-dup100}{1.70}
\DefMacro{exp-python-bokeh_bokeh-time-unit-and-inline-dup100}{50.85}
\DefMacro{exp-python-bokeh_bokeh-num-vanilla-dup100}{8,616}
\DefMacro{exp-python-bokeh_bokeh-num-unit-dup100}{8,616}
\DefMacro{exp-python-bokeh_bokeh-num-inline-dup100}{800}
\DefMacro{exp-python-bokeh_bokeh-num-unit-and-inline-dup100}{9,416}
\DefMacro{exp-python-bokeh_bokeh-timept-vanilla-dup100}{0.006}
\DefMacro{exp-python-bokeh_bokeh-timept-unit-dup100}{0.006}
\DefMacro{exp-python-bokeh_bokeh-timept-inline-dup100}{0.002}
\DefMacro{exp-python-bokeh_bokeh-timept-unit-and-inline-dup100}{0.005}
\DefMacro{exp-python-bokeh_bokeh-overhead-unit-dup100}{-0.043}
\DefMacro{exp-python-bokeh_bokeh-overhead-unit-and-inline-dup100}{-0.010}
\DefMacro{exp-python-chubin_cheat.sh-time-vanilla-dup100}{0.34}
\DefMacro{exp-python-chubin_cheat.sh-time-unit-dup100}{0.34}
\DefMacro{exp-python-chubin_cheat.sh-time-inline-dup100}{0.64}
\DefMacro{exp-python-chubin_cheat.sh-time-unit-and-inline-dup100}{0.97}
\DefMacro{exp-python-chubin_cheat.sh-num-vanilla-dup100}{1}
\DefMacro{exp-python-chubin_cheat.sh-num-unit-dup100}{1}
\DefMacro{exp-python-chubin_cheat.sh-num-inline-dup100}{300}
\DefMacro{exp-python-chubin_cheat.sh-num-unit-and-inline-dup100}{301}
\DefMacro{exp-python-chubin_cheat.sh-timept-vanilla-dup100}{0.337}
\DefMacro{exp-python-chubin_cheat.sh-timept-unit-dup100}{0.335}
\DefMacro{exp-python-chubin_cheat.sh-timept-inline-dup100}{0.002}
\DefMacro{exp-python-chubin_cheat.sh-timept-unit-and-inline-dup100}{0.003}
\DefMacro{exp-python-chubin_cheat.sh-overhead-unit-dup100}{-0.007}
\DefMacro{exp-python-chubin_cheat.sh-overhead-unit-and-inline-dup100}{1.878}
\DefMacro{exp-python-davidsandberg_facenet-time-vanilla-dup100}{0.97}
\DefMacro{exp-python-davidsandberg_facenet-time-unit-dup100}{0.98}
\DefMacro{exp-python-davidsandberg_facenet-time-inline-dup100}{0.97}
\DefMacro{exp-python-davidsandberg_facenet-time-unit-and-inline-dup100}{1.95}
\DefMacro{exp-python-davidsandberg_facenet-num-vanilla-dup100}{3}
\DefMacro{exp-python-davidsandberg_facenet-num-unit-dup100}{3}
\DefMacro{exp-python-davidsandberg_facenet-num-inline-dup100}{100}
\DefMacro{exp-python-davidsandberg_facenet-num-unit-and-inline-dup100}{103}
\DefMacro{exp-python-davidsandberg_facenet-timept-vanilla-dup100}{0.323}
\DefMacro{exp-python-davidsandberg_facenet-timept-unit-dup100}{0.326}
\DefMacro{exp-python-davidsandberg_facenet-timept-inline-dup100}{0.010}
\DefMacro{exp-python-davidsandberg_facenet-timept-unit-and-inline-dup100}{0.019}
\DefMacro{exp-python-davidsandberg_facenet-overhead-unit-dup100}{0.011}
\DefMacro{exp-python-davidsandberg_facenet-overhead-unit-and-inline-dup100}{1.008}
\DefMacro{exp-python-geekcomputers_Python-time-vanilla-dup100}{0.17}
\DefMacro{exp-python-geekcomputers_Python-time-unit-dup100}{0.18}
\DefMacro{exp-python-geekcomputers_Python-time-inline-dup100}{0.51}
\DefMacro{exp-python-geekcomputers_Python-time-unit-and-inline-dup100}{0.69}
\DefMacro{exp-python-geekcomputers_Python-num-vanilla-dup100}{1}
\DefMacro{exp-python-geekcomputers_Python-num-unit-dup100}{1}
\DefMacro{exp-python-geekcomputers_Python-num-inline-dup100}{400}
\DefMacro{exp-python-geekcomputers_Python-num-unit-and-inline-dup100}{401}
\DefMacro{exp-python-geekcomputers_Python-timept-vanilla-dup100}{0.172}
\DefMacro{exp-python-geekcomputers_Python-timept-unit-dup100}{0.180}
\DefMacro{exp-python-geekcomputers_Python-timept-inline-dup100}{0.001}
\DefMacro{exp-python-geekcomputers_Python-timept-unit-and-inline-dup100}{0.002}
\DefMacro{exp-python-geekcomputers_Python-overhead-unit-dup100}{0.044}
\DefMacro{exp-python-geekcomputers_Python-overhead-unit-and-inline-dup100}{3.029}
\DefMacro{exp-python-google-research_bert-time-vanilla-dup100}{2.06}
\DefMacro{exp-python-google-research_bert-time-unit-dup100}{2.07}
\DefMacro{exp-python-google-research_bert-time-inline-dup100}{0.72}
\DefMacro{exp-python-google-research_bert-time-unit-and-inline-dup100}{2.79}
\DefMacro{exp-python-google-research_bert-num-vanilla-dup100}{15}
\DefMacro{exp-python-google-research_bert-num-unit-dup100}{15}
\DefMacro{exp-python-google-research_bert-num-inline-dup100}{100}
\DefMacro{exp-python-google-research_bert-num-unit-and-inline-dup100}{115}
\DefMacro{exp-python-google-research_bert-timept-vanilla-dup100}{0.137}
\DefMacro{exp-python-google-research_bert-timept-unit-dup100}{0.138}
\DefMacro{exp-python-google-research_bert-timept-inline-dup100}{0.007}
\DefMacro{exp-python-google-research_bert-timept-unit-and-inline-dup100}{0.024}
\DefMacro{exp-python-google-research_bert-overhead-unit-dup100}{0.006}
\DefMacro{exp-python-google-research_bert-overhead-unit-and-inline-dup100}{0.355}
\DefMacro{exp-python-joke2k_faker-time-vanilla-dup100}{16.71}
\DefMacro{exp-python-joke2k_faker-time-unit-dup100}{16.83}
\DefMacro{exp-python-joke2k_faker-time-inline-dup100}{0.60}
\DefMacro{exp-python-joke2k_faker-time-unit-and-inline-dup100}{17.43}
\DefMacro{exp-python-joke2k_faker-num-vanilla-dup100}{1,596}
\DefMacro{exp-python-joke2k_faker-num-unit-dup100}{1,596}
\DefMacro{exp-python-joke2k_faker-num-inline-dup100}{400}
\DefMacro{exp-python-joke2k_faker-num-unit-and-inline-dup100}{1,996}
\DefMacro{exp-python-joke2k_faker-timept-vanilla-dup100}{0.010}
\DefMacro{exp-python-joke2k_faker-timept-unit-dup100}{0.011}
\DefMacro{exp-python-joke2k_faker-timept-inline-dup100}{0.002}
\DefMacro{exp-python-joke2k_faker-timept-unit-and-inline-dup100}{0.009}
\DefMacro{exp-python-joke2k_faker-overhead-unit-dup100}{0.007}
\DefMacro{exp-python-joke2k_faker-overhead-unit-and-inline-dup100}{0.043}
\DefMacro{exp-python-mitmproxy_mitmproxy-time-vanilla-dup100}{7.45}
\DefMacro{exp-python-mitmproxy_mitmproxy-time-unit-dup100}{7.47}
\DefMacro{exp-python-mitmproxy_mitmproxy-time-inline-dup100}{0.47}
\DefMacro{exp-python-mitmproxy_mitmproxy-time-unit-and-inline-dup100}{7.94}
\DefMacro{exp-python-mitmproxy_mitmproxy-num-vanilla-dup100}{1,287}
\DefMacro{exp-python-mitmproxy_mitmproxy-num-unit-dup100}{1,287}
\DefMacro{exp-python-mitmproxy_mitmproxy-num-inline-dup100}{100}
\DefMacro{exp-python-mitmproxy_mitmproxy-num-unit-and-inline-dup100}{1,387}
\DefMacro{exp-python-mitmproxy_mitmproxy-timept-vanilla-dup100}{0.006}
\DefMacro{exp-python-mitmproxy_mitmproxy-timept-unit-dup100}{0.006}
\DefMacro{exp-python-mitmproxy_mitmproxy-timept-inline-dup100}{0.005}
\DefMacro{exp-python-mitmproxy_mitmproxy-timept-unit-and-inline-dup100}{0.006}
\DefMacro{exp-python-mitmproxy_mitmproxy-overhead-unit-dup100}{0.002}
\DefMacro{exp-python-mitmproxy_mitmproxy-overhead-unit-and-inline-dup100}{0.065}
\DefMacro{exp-python-numpy_numpy-time-vanilla-dup100}{146.65}
\DefMacro{exp-python-numpy_numpy-time-unit-dup100}{145.87}
\DefMacro{exp-python-numpy_numpy-time-inline-dup100}{0.71}
\DefMacro{exp-python-numpy_numpy-time-unit-and-inline-dup100}{146.58}
\DefMacro{exp-python-numpy_numpy-num-vanilla-dup100}{19,644}
\DefMacro{exp-python-numpy_numpy-num-unit-dup100}{19,644}
\DefMacro{exp-python-numpy_numpy-num-inline-dup100}{200}
\DefMacro{exp-python-numpy_numpy-num-unit-and-inline-dup100}{19,844}
\DefMacro{exp-python-numpy_numpy-timept-vanilla-dup100}{0.007}
\DefMacro{exp-python-numpy_numpy-timept-unit-dup100}{0.007}
\DefMacro{exp-python-numpy_numpy-timept-inline-dup100}{0.004}
\DefMacro{exp-python-numpy_numpy-timept-unit-and-inline-dup100}{0.007}
\DefMacro{exp-python-numpy_numpy-overhead-unit-dup100}{-0.005}
\DefMacro{exp-python-numpy_numpy-overhead-unit-and-inline-dup100}{-0.000}
\DefMacro{exp-python-pandas-dev_pandas-time-vanilla-dup100}{278.66}
\DefMacro{exp-python-pandas-dev_pandas-time-unit-dup100}{278.87}
\DefMacro{exp-python-pandas-dev_pandas-time-inline-dup100}{1.12}
\DefMacro{exp-python-pandas-dev_pandas-time-unit-and-inline-dup100}{279.99}
\DefMacro{exp-python-pandas-dev_pandas-num-vanilla-dup100}{147,307}
\DefMacro{exp-python-pandas-dev_pandas-num-unit-dup100}{147,306}
\DefMacro{exp-python-pandas-dev_pandas-num-inline-dup100}{200}
\DefMacro{exp-python-pandas-dev_pandas-num-unit-and-inline-dup100}{147,506}
\DefMacro{exp-python-pandas-dev_pandas-timept-vanilla-dup100}{0.002}
\DefMacro{exp-python-pandas-dev_pandas-timept-unit-dup100}{0.002}
\DefMacro{exp-python-pandas-dev_pandas-timept-inline-dup100}{0.006}
\DefMacro{exp-python-pandas-dev_pandas-timept-unit-and-inline-dup100}{0.002}
\DefMacro{exp-python-pandas-dev_pandas-overhead-unit-dup100}{0.001}
\DefMacro{exp-python-pandas-dev_pandas-overhead-unit-and-inline-dup100}{0.005}
\DefMacro{exp-python-psf_black-time-vanilla-dup100}{6.98}
\DefMacro{exp-python-psf_black-time-unit-dup100}{7.03}
\DefMacro{exp-python-psf_black-time-inline-dup100}{0.35}
\DefMacro{exp-python-psf_black-time-unit-and-inline-dup100}{7.38}
\DefMacro{exp-python-psf_black-num-vanilla-dup100}{236}
\DefMacro{exp-python-psf_black-num-unit-dup100}{236}
\DefMacro{exp-python-psf_black-num-inline-dup100}{100}
\DefMacro{exp-python-psf_black-num-unit-and-inline-dup100}{336}
\DefMacro{exp-python-psf_black-timept-vanilla-dup100}{0.030}
\DefMacro{exp-python-psf_black-timept-unit-dup100}{0.030}
\DefMacro{exp-python-psf_black-timept-inline-dup100}{0.004}
\DefMacro{exp-python-psf_black-timept-unit-and-inline-dup100}{0.022}
\DefMacro{exp-python-psf_black-overhead-unit-dup100}{0.007}
\DefMacro{exp-python-psf_black-overhead-unit-and-inline-dup100}{0.058}
\DefMacro{exp-python-pypa_pipenv-time-vanilla-dup100}{3.63}
\DefMacro{exp-python-pypa_pipenv-time-unit-dup100}{3.66}
\DefMacro{exp-python-pypa_pipenv-time-inline-dup100}{0.62}
\DefMacro{exp-python-pypa_pipenv-time-unit-and-inline-dup100}{4.29}
\DefMacro{exp-python-pypa_pipenv-num-vanilla-dup100}{106}
\DefMacro{exp-python-pypa_pipenv-num-unit-dup100}{106}
\DefMacro{exp-python-pypa_pipenv-num-inline-dup100}{100}
\DefMacro{exp-python-pypa_pipenv-num-unit-and-inline-dup100}{206}
\DefMacro{exp-python-pypa_pipenv-timept-vanilla-dup100}{0.034}
\DefMacro{exp-python-pypa_pipenv-timept-unit-dup100}{0.035}
\DefMacro{exp-python-pypa_pipenv-timept-inline-dup100}{0.006}
\DefMacro{exp-python-pypa_pipenv-timept-unit-and-inline-dup100}{0.021}
\DefMacro{exp-python-pypa_pipenv-overhead-unit-dup100}{0.010}
\DefMacro{exp-python-pypa_pipenv-overhead-unit-and-inline-dup100}{0.181}
\DefMacro{exp-python-scrapy_scrapy-time-vanilla-dup100}{130.30}
\DefMacro{exp-python-scrapy_scrapy-time-unit-dup100}{130.67}
\DefMacro{exp-python-scrapy_scrapy-time-inline-dup100}{0.97}
\DefMacro{exp-python-scrapy_scrapy-time-unit-and-inline-dup100}{131.63}
\DefMacro{exp-python-scrapy_scrapy-num-vanilla-dup100}{2,246}
\DefMacro{exp-python-scrapy_scrapy-num-unit-dup100}{2,246}
\DefMacro{exp-python-scrapy_scrapy-num-inline-dup100}{200}
\DefMacro{exp-python-scrapy_scrapy-num-unit-and-inline-dup100}{2,446}
\DefMacro{exp-python-scrapy_scrapy-timept-vanilla-dup100}{0.058}
\DefMacro{exp-python-scrapy_scrapy-timept-unit-dup100}{0.058}
\DefMacro{exp-python-scrapy_scrapy-timept-inline-dup100}{0.005}
\DefMacro{exp-python-scrapy_scrapy-timept-unit-and-inline-dup100}{0.054}
\DefMacro{exp-python-scrapy_scrapy-overhead-unit-dup100}{0.003}
\DefMacro{exp-python-scrapy_scrapy-overhead-unit-and-inline-dup100}{0.010}
\DefMacro{exp-python-time-vanilla-AVG-dup100}{62.58}
\DefMacro{exp-python-time-vanilla-SUM-dup100}{876.13}
\DefMacro{exp-python-time-vanilla-MAX-dup100}{278.66}
\DefMacro{exp-python-time-vanilla-MIN-dup100}{0.17}
\DefMacro{exp-python-time-vanilla-MEDIAN-dup100}{7.22}
\DefMacro{exp-python-time-vanilla-STDEV-dup100}{91.01}
\DefMacro{exp-python-time-vanilla-CNT-dup100}{14}
\DefMacro{exp-python-time-unit-AVG-dup100}{62.40}
\DefMacro{exp-python-time-unit-SUM-dup100}{873.65}
\DefMacro{exp-python-time-unit-MAX-dup100}{278.87}
\DefMacro{exp-python-time-unit-MIN-dup100}{0.18}
\DefMacro{exp-python-time-unit-MEDIAN-dup100}{7.25}
\DefMacro{exp-python-time-unit-STDEV-dup100}{90.98}
\DefMacro{exp-python-time-unit-CNT-dup100}{14}
\DefMacro{exp-python-time-inline-AVG-dup100}{0.75}
\DefMacro{exp-python-time-inline-SUM-dup100}{10.51}
\DefMacro{exp-python-time-inline-MAX-dup100}{1.70}
\DefMacro{exp-python-time-inline-MIN-dup100}{0.35}
\DefMacro{exp-python-time-inline-MEDIAN-dup100}{0.67}
\DefMacro{exp-python-time-inline-STDEV-dup100}{0.34}
\DefMacro{exp-python-time-inline-CNT-dup100}{14}
\DefMacro{exp-python-time-unit-and-inline-AVG-dup100}{63.15}
\DefMacro{exp-python-time-unit-and-inline-SUM-dup100}{884.16}
\DefMacro{exp-python-time-unit-and-inline-MAX-dup100}{279.99}
\DefMacro{exp-python-time-unit-and-inline-MIN-dup100}{0.69}
\DefMacro{exp-python-time-unit-and-inline-MEDIAN-dup100}{7.66}
\DefMacro{exp-python-time-unit-and-inline-STDEV-dup100}{91.10}
\DefMacro{exp-python-time-unit-and-inline-CNT-dup100}{14}
\DefMacro{exp-python-num-vanilla-AVG-dup100}{13,046.29}
\DefMacro{exp-python-num-vanilla-SUM-dup100}{182,648}
\DefMacro{exp-python-num-vanilla-MAX-dup100}{147,307}
\DefMacro{exp-python-num-vanilla-MIN-dup100}{1}
\DefMacro{exp-python-num-vanilla-MEDIAN-dup100}{795.00}
\DefMacro{exp-python-num-vanilla-STDEV-dup100}{37,594.87}
\DefMacro{exp-python-num-vanilla-CNT-dup100}{14}
\DefMacro{exp-python-num-unit-AVG-dup100}{13,046.21}
\DefMacro{exp-python-num-unit-SUM-dup100}{182,647}
\DefMacro{exp-python-num-unit-MAX-dup100}{147,306}
\DefMacro{exp-python-num-unit-MIN-dup100}{1}
\DefMacro{exp-python-num-unit-MEDIAN-dup100}{795.00}
\DefMacro{exp-python-num-unit-STDEV-dup100}{37,594.61}
\DefMacro{exp-python-num-unit-CNT-dup100}{14}
\DefMacro{exp-python-num-inline-AVG-dup100}{242.86}
\DefMacro{exp-python-num-inline-SUM-dup100}{3,400}
\DefMacro{exp-python-num-inline-MAX-dup100}{800}
\DefMacro{exp-python-num-inline-MIN-dup100}{100}
\DefMacro{exp-python-num-inline-MEDIAN-dup100}{200.00}
\DefMacro{exp-python-num-inline-STDEV-dup100}{184.06}
\DefMacro{exp-python-num-inline-CNT-dup100}{14}
\DefMacro{exp-python-num-unit-and-inline-AVG-dup100}{13,289.07}
\DefMacro{exp-python-num-unit-and-inline-SUM-dup100}{186,047}
\DefMacro{exp-python-num-unit-and-inline-MAX-dup100}{147,506}
\DefMacro{exp-python-num-unit-and-inline-MIN-dup100}{103}
\DefMacro{exp-python-num-unit-and-inline-MEDIAN-dup100}{995.00}
\DefMacro{exp-python-num-unit-and-inline-STDEV-dup100}{37,590.31}
\DefMacro{exp-python-num-unit-and-inline-CNT-dup100}{14}
\DefMacro{exp-python-timept-vanilla-AVG-dup100}{0.097}
\DefMacro{exp-python-timept-vanilla-SUM-dup100}{1.363}
\DefMacro{exp-python-timept-vanilla-MAX-dup100}{0.337}
\DefMacro{exp-python-timept-vanilla-MIN-dup100}{0.002}
\DefMacro{exp-python-timept-vanilla-MEDIAN-dup100}{0.032}
\DefMacro{exp-python-timept-vanilla-STDEV-dup100}{0.118}
\DefMacro{exp-python-timept-vanilla-CNT-dup100}{14}
\DefMacro{exp-python-timept-unit-AVG-dup100}{0.098}
\DefMacro{exp-python-timept-unit-SUM-dup100}{1.373}
\DefMacro{exp-python-timept-unit-MAX-dup100}{0.335}
\DefMacro{exp-python-timept-unit-MIN-dup100}{0.002}
\DefMacro{exp-python-timept-unit-MEDIAN-dup100}{0.032}
\DefMacro{exp-python-timept-unit-STDEV-dup100}{0.118}
\DefMacro{exp-python-timept-unit-CNT-dup100}{14}
\DefMacro{exp-python-timept-inline-AVG-dup100}{0.004}
\DefMacro{exp-python-timept-inline-SUM-dup100}{0.058}
\DefMacro{exp-python-timept-inline-MAX-dup100}{0.010}
\DefMacro{exp-python-timept-inline-MIN-dup100}{0.001}
\DefMacro{exp-python-timept-inline-MEDIAN-dup100}{0.004}
\DefMacro{exp-python-timept-inline-STDEV-dup100}{0.002}
\DefMacro{exp-python-timept-inline-CNT-dup100}{14}
\DefMacro{exp-python-timept-unit-and-inline-AVG-dup100}{0.027}
\DefMacro{exp-python-timept-unit-and-inline-SUM-dup100}{0.374}
\DefMacro{exp-python-timept-unit-and-inline-MAX-dup100}{0.195}
\DefMacro{exp-python-timept-unit-and-inline-MIN-dup100}{0.002}
\DefMacro{exp-python-timept-unit-and-inline-MEDIAN-dup100}{0.008}
\DefMacro{exp-python-timept-unit-and-inline-STDEV-dup100}{0.049}
\DefMacro{exp-python-timept-unit-and-inline-CNT-dup100}{14}
\DefMacro{exp-python-overhead-unit-AVG-dup100}{0.003}
\DefMacro{exp-python-overhead-unit-SUM-dup100}{0.037}
\DefMacro{exp-python-overhead-unit-MAX-dup100}{0.044}
\DefMacro{exp-python-overhead-unit-MIN-dup100}{-0.043}
\DefMacro{exp-python-overhead-unit-MEDIAN-dup100}{0.003}
\DefMacro{exp-python-overhead-unit-STDEV-dup100}{0.017}
\DefMacro{exp-python-overhead-unit-CNT-dup100}{14}
\DefMacro{exp-python-overhead-unit-and-inline-AVG-dup100}{0.481}
\DefMacro{exp-python-overhead-unit-and-inline-SUM-dup100}{6.737}
\DefMacro{exp-python-overhead-unit-and-inline-MAX-dup100}{3.029}
\DefMacro{exp-python-overhead-unit-and-inline-MIN-dup100}{-0.010}
\DefMacro{exp-python-overhead-unit-and-inline-MEDIAN-dup100}{0.062}
\DefMacro{exp-python-overhead-unit-and-inline-STDEV-dup100}{0.873}
\DefMacro{exp-python-overhead-unit-and-inline-CNT-dup100}{14}
\DefMacro{exp-python-timept-vanilla-MACROAVG-dup100}{0.005}
\DefMacro{exp-python-timept-unit-MACROAVG-dup100}{0.005}
\DefMacro{exp-python-timept-inline-MACROAVG-dup100}{0.003}
\DefMacro{exp-python-timept-unit-and-inline-MACROAVG-dup100}{0.005}
\DefMacro{exp-python-overhead-unit-MACROAVG-dup100}{-0.003}
\DefMacro{exp-python-overhead-unit-and-inline-MACROAVG-dup100}{0.009}
\DefMacro{exp-python-RaRe-Technologies_gensim-time-vanilla-dup1000}{226.72}
\DefMacro{exp-python-RaRe-Technologies_gensim-time-unit-dup1000}{238.83}
\DefMacro{exp-python-RaRe-Technologies_gensim-time-inline-dup1000}{3.11}
\DefMacro{exp-python-RaRe-Technologies_gensim-time-unit-and-inline-dup1000}{241.94}
\DefMacro{exp-python-RaRe-Technologies_gensim-num-vanilla-dup1000}{968}
\DefMacro{exp-python-RaRe-Technologies_gensim-num-unit-dup1000}{968}
\DefMacro{exp-python-RaRe-Technologies_gensim-num-inline-dup1000}{2,000}
\DefMacro{exp-python-RaRe-Technologies_gensim-num-unit-and-inline-dup1000}{2,968}
\DefMacro{exp-python-RaRe-Technologies_gensim-timept-vanilla-dup1000}{0.234}
\DefMacro{exp-python-RaRe-Technologies_gensim-timept-unit-dup1000}{0.247}
\DefMacro{exp-python-RaRe-Technologies_gensim-timept-inline-dup1000}{0.002}
\DefMacro{exp-python-RaRe-Technologies_gensim-timept-unit-and-inline-dup1000}{0.082}
\DefMacro{exp-python-RaRe-Technologies_gensim-overhead-unit-dup1000}{0.053}
\DefMacro{exp-python-RaRe-Technologies_gensim-overhead-unit-and-inline-dup1000}{0.067}
\DefMacro{exp-python-Textualize_rich-time-vanilla-dup1000}{3.71}
\DefMacro{exp-python-Textualize_rich-time-unit-dup1000}{3.70}
\DefMacro{exp-python-Textualize_rich-time-inline-dup1000}{2.96}
\DefMacro{exp-python-Textualize_rich-time-unit-and-inline-dup1000}{6.67}
\DefMacro{exp-python-Textualize_rich-num-vanilla-dup1000}{622}
\DefMacro{exp-python-Textualize_rich-num-unit-dup1000}{622}
\DefMacro{exp-python-Textualize_rich-num-inline-dup1000}{2,000}
\DefMacro{exp-python-Textualize_rich-num-unit-and-inline-dup1000}{2,622}
\DefMacro{exp-python-Textualize_rich-timept-vanilla-dup1000}{0.006}
\DefMacro{exp-python-Textualize_rich-timept-unit-dup1000}{0.006}
\DefMacro{exp-python-Textualize_rich-timept-inline-dup1000}{0.001}
\DefMacro{exp-python-Textualize_rich-timept-unit-and-inline-dup1000}{0.003}
\DefMacro{exp-python-Textualize_rich-overhead-unit-dup1000}{-0.001}
\DefMacro{exp-python-Textualize_rich-overhead-unit-and-inline-dup1000}{0.798}
\DefMacro{exp-python-bokeh_bokeh-time-vanilla-dup1000}{49.51}
\DefMacro{exp-python-bokeh_bokeh-time-unit-dup1000}{49.15}
\DefMacro{exp-python-bokeh_bokeh-time-inline-dup1000}{15.53}
\DefMacro{exp-python-bokeh_bokeh-time-unit-and-inline-dup1000}{64.68}
\DefMacro{exp-python-bokeh_bokeh-num-vanilla-dup1000}{8,616}
\DefMacro{exp-python-bokeh_bokeh-num-unit-dup1000}{8,616}
\DefMacro{exp-python-bokeh_bokeh-num-inline-dup1000}{8,000}
\DefMacro{exp-python-bokeh_bokeh-num-unit-and-inline-dup1000}{16,616}
\DefMacro{exp-python-bokeh_bokeh-timept-vanilla-dup1000}{0.006}
\DefMacro{exp-python-bokeh_bokeh-timept-unit-dup1000}{0.006}
\DefMacro{exp-python-bokeh_bokeh-timept-inline-dup1000}{0.002}
\DefMacro{exp-python-bokeh_bokeh-timept-unit-and-inline-dup1000}{0.004}
\DefMacro{exp-python-bokeh_bokeh-overhead-unit-dup1000}{-0.007}
\DefMacro{exp-python-bokeh_bokeh-overhead-unit-and-inline-dup1000}{0.306}
\DefMacro{exp-python-chubin_cheat.sh-time-vanilla-dup1000}{0.34}
\DefMacro{exp-python-chubin_cheat.sh-time-unit-dup1000}{0.34}
\DefMacro{exp-python-chubin_cheat.sh-time-inline-dup1000}{3.79}
\DefMacro{exp-python-chubin_cheat.sh-time-unit-and-inline-dup1000}{4.13}
\DefMacro{exp-python-chubin_cheat.sh-num-vanilla-dup1000}{1}
\DefMacro{exp-python-chubin_cheat.sh-num-unit-dup1000}{1}
\DefMacro{exp-python-chubin_cheat.sh-num-inline-dup1000}{3,000}
\DefMacro{exp-python-chubin_cheat.sh-num-unit-and-inline-dup1000}{3,001}
\DefMacro{exp-python-chubin_cheat.sh-timept-vanilla-dup1000}{0.338}
\DefMacro{exp-python-chubin_cheat.sh-timept-unit-dup1000}{0.335}
\DefMacro{exp-python-chubin_cheat.sh-timept-inline-dup1000}{0.001}
\DefMacro{exp-python-chubin_cheat.sh-timept-unit-and-inline-dup1000}{0.001}
\DefMacro{exp-python-chubin_cheat.sh-overhead-unit-dup1000}{-0.009}
\DefMacro{exp-python-chubin_cheat.sh-overhead-unit-and-inline-dup1000}{11.211}
\DefMacro{exp-python-davidsandberg_facenet-time-vanilla-dup1000}{0.97}
\DefMacro{exp-python-davidsandberg_facenet-time-unit-dup1000}{0.98}
\DefMacro{exp-python-davidsandberg_facenet-time-inline-dup1000}{2.43}
\DefMacro{exp-python-davidsandberg_facenet-time-unit-and-inline-dup1000}{3.41}
\DefMacro{exp-python-davidsandberg_facenet-num-vanilla-dup1000}{3}
\DefMacro{exp-python-davidsandberg_facenet-num-unit-dup1000}{3}
\DefMacro{exp-python-davidsandberg_facenet-num-inline-dup1000}{1,000}
\DefMacro{exp-python-davidsandberg_facenet-num-unit-and-inline-dup1000}{1,003}
\DefMacro{exp-python-davidsandberg_facenet-timept-vanilla-dup1000}{0.325}
\DefMacro{exp-python-davidsandberg_facenet-timept-unit-dup1000}{0.325}
\DefMacro{exp-python-davidsandberg_facenet-timept-inline-dup1000}{0.002}
\DefMacro{exp-python-davidsandberg_facenet-timept-unit-and-inline-dup1000}{0.003}
\DefMacro{exp-python-davidsandberg_facenet-overhead-unit-dup1000}{0.002}
\DefMacro{exp-python-davidsandberg_facenet-overhead-unit-and-inline-dup1000}{2.497}
\DefMacro{exp-python-geekcomputers_Python-time-vanilla-dup1000}{0.17}
\DefMacro{exp-python-geekcomputers_Python-time-unit-dup1000}{0.20}
\DefMacro{exp-python-geekcomputers_Python-time-inline-dup1000}{5.60}
\DefMacro{exp-python-geekcomputers_Python-time-unit-and-inline-dup1000}{5.80}
\DefMacro{exp-python-geekcomputers_Python-num-vanilla-dup1000}{1}
\DefMacro{exp-python-geekcomputers_Python-num-unit-dup1000}{1}
\DefMacro{exp-python-geekcomputers_Python-num-inline-dup1000}{4,000}
\DefMacro{exp-python-geekcomputers_Python-num-unit-and-inline-dup1000}{4,001}
\DefMacro{exp-python-geekcomputers_Python-timept-vanilla-dup1000}{0.171}
\DefMacro{exp-python-geekcomputers_Python-timept-unit-dup1000}{0.197}
\DefMacro{exp-python-geekcomputers_Python-timept-inline-dup1000}{0.001}
\DefMacro{exp-python-geekcomputers_Python-timept-unit-and-inline-dup1000}{0.001}
\DefMacro{exp-python-geekcomputers_Python-overhead-unit-dup1000}{0.150}
\DefMacro{exp-python-geekcomputers_Python-overhead-unit-and-inline-dup1000}{32.877}
\DefMacro{exp-python-google-research_bert-time-vanilla-dup1000}{2.05}
\DefMacro{exp-python-google-research_bert-time-unit-dup1000}{2.07}
\DefMacro{exp-python-google-research_bert-time-inline-dup1000}{2.21}
\DefMacro{exp-python-google-research_bert-time-unit-and-inline-dup1000}{4.29}
\DefMacro{exp-python-google-research_bert-num-vanilla-dup1000}{15}
\DefMacro{exp-python-google-research_bert-num-unit-dup1000}{15}
\DefMacro{exp-python-google-research_bert-num-inline-dup1000}{1,000}
\DefMacro{exp-python-google-research_bert-num-unit-and-inline-dup1000}{1,015}
\DefMacro{exp-python-google-research_bert-timept-vanilla-dup1000}{0.137}
\DefMacro{exp-python-google-research_bert-timept-unit-dup1000}{0.138}
\DefMacro{exp-python-google-research_bert-timept-inline-dup1000}{0.002}
\DefMacro{exp-python-google-research_bert-timept-unit-and-inline-dup1000}{0.004}
\DefMacro{exp-python-google-research_bert-overhead-unit-dup1000}{0.012}
\DefMacro{exp-python-google-research_bert-overhead-unit-and-inline-dup1000}{1.093}
\DefMacro{exp-python-joke2k_faker-time-vanilla-dup1000}{16.67}
\DefMacro{exp-python-joke2k_faker-time-unit-dup1000}{17.53}
\DefMacro{exp-python-joke2k_faker-time-inline-dup1000}{5.42}
\DefMacro{exp-python-joke2k_faker-time-unit-and-inline-dup1000}{22.96}
\DefMacro{exp-python-joke2k_faker-num-vanilla-dup1000}{1,596}
\DefMacro{exp-python-joke2k_faker-num-unit-dup1000}{1,596}
\DefMacro{exp-python-joke2k_faker-num-inline-dup1000}{4,000}
\DefMacro{exp-python-joke2k_faker-num-unit-and-inline-dup1000}{5,596}
\DefMacro{exp-python-joke2k_faker-timept-vanilla-dup1000}{0.010}
\DefMacro{exp-python-joke2k_faker-timept-unit-dup1000}{0.011}
\DefMacro{exp-python-joke2k_faker-timept-inline-dup1000}{0.001}
\DefMacro{exp-python-joke2k_faker-timept-unit-and-inline-dup1000}{0.004}
\DefMacro{exp-python-joke2k_faker-overhead-unit-dup1000}{0.052}
\DefMacro{exp-python-joke2k_faker-overhead-unit-and-inline-dup1000}{0.377}
\DefMacro{exp-python-mitmproxy_mitmproxy-time-vanilla-dup1000}{7.48}
\DefMacro{exp-python-mitmproxy_mitmproxy-time-unit-dup1000}{7.56}
\DefMacro{exp-python-mitmproxy_mitmproxy-time-inline-dup1000}{1.63}
\DefMacro{exp-python-mitmproxy_mitmproxy-time-unit-and-inline-dup1000}{9.19}
\DefMacro{exp-python-mitmproxy_mitmproxy-num-vanilla-dup1000}{1,287}
\DefMacro{exp-python-mitmproxy_mitmproxy-num-unit-dup1000}{1,287}
\DefMacro{exp-python-mitmproxy_mitmproxy-num-inline-dup1000}{1,000}
\DefMacro{exp-python-mitmproxy_mitmproxy-num-unit-and-inline-dup1000}{2,287}
\DefMacro{exp-python-mitmproxy_mitmproxy-timept-vanilla-dup1000}{0.006}
\DefMacro{exp-python-mitmproxy_mitmproxy-timept-unit-dup1000}{0.006}
\DefMacro{exp-python-mitmproxy_mitmproxy-timept-inline-dup1000}{0.002}
\DefMacro{exp-python-mitmproxy_mitmproxy-timept-unit-and-inline-dup1000}{0.004}
\DefMacro{exp-python-mitmproxy_mitmproxy-overhead-unit-dup1000}{0.011}
\DefMacro{exp-python-mitmproxy_mitmproxy-overhead-unit-and-inline-dup1000}{0.229}
\DefMacro{exp-python-numpy_numpy-time-vanilla-dup1000}{146.28}
\DefMacro{exp-python-numpy_numpy-time-unit-dup1000}{146.86}
\DefMacro{exp-python-numpy_numpy-time-inline-dup1000}{3.28}
\DefMacro{exp-python-numpy_numpy-time-unit-and-inline-dup1000}{150.14}
\DefMacro{exp-python-numpy_numpy-num-vanilla-dup1000}{19,644}
\DefMacro{exp-python-numpy_numpy-num-unit-dup1000}{19,644}
\DefMacro{exp-python-numpy_numpy-num-inline-dup1000}{2,000}
\DefMacro{exp-python-numpy_numpy-num-unit-and-inline-dup1000}{21,644}
\DefMacro{exp-python-numpy_numpy-timept-vanilla-dup1000}{0.007}
\DefMacro{exp-python-numpy_numpy-timept-unit-dup1000}{0.007}
\DefMacro{exp-python-numpy_numpy-timept-inline-dup1000}{0.002}
\DefMacro{exp-python-numpy_numpy-timept-unit-and-inline-dup1000}{0.007}
\DefMacro{exp-python-numpy_numpy-overhead-unit-dup1000}{0.004}
\DefMacro{exp-python-numpy_numpy-overhead-unit-and-inline-dup1000}{0.026}
\DefMacro{exp-python-pandas-dev_pandas-time-vanilla-dup1000}{278.08}
\DefMacro{exp-python-pandas-dev_pandas-time-unit-dup1000}{280.19}
\DefMacro{exp-python-pandas-dev_pandas-time-inline-dup1000}{3.68}
\DefMacro{exp-python-pandas-dev_pandas-time-unit-and-inline-dup1000}{283.87}
\DefMacro{exp-python-pandas-dev_pandas-num-vanilla-dup1000}{147,306}
\DefMacro{exp-python-pandas-dev_pandas-num-unit-dup1000}{147,306}
\DefMacro{exp-python-pandas-dev_pandas-num-inline-dup1000}{2,000}
\DefMacro{exp-python-pandas-dev_pandas-num-unit-and-inline-dup1000}{149,306}
\DefMacro{exp-python-pandas-dev_pandas-timept-vanilla-dup1000}{0.002}
\DefMacro{exp-python-pandas-dev_pandas-timept-unit-dup1000}{0.002}
\DefMacro{exp-python-pandas-dev_pandas-timept-inline-dup1000}{0.002}
\DefMacro{exp-python-pandas-dev_pandas-timept-unit-and-inline-dup1000}{0.002}
\DefMacro{exp-python-pandas-dev_pandas-overhead-unit-dup1000}{0.008}
\DefMacro{exp-python-pandas-dev_pandas-overhead-unit-and-inline-dup1000}{0.021}
\DefMacro{exp-python-psf_black-time-vanilla-dup1000}{6.96}
\DefMacro{exp-python-psf_black-time-unit-dup1000}{7.40}
\DefMacro{exp-python-psf_black-time-inline-dup1000}{1.60}
\DefMacro{exp-python-psf_black-time-unit-and-inline-dup1000}{8.99}
\DefMacro{exp-python-psf_black-num-vanilla-dup1000}{236}
\DefMacro{exp-python-psf_black-num-unit-dup1000}{236}
\DefMacro{exp-python-psf_black-num-inline-dup1000}{1,000}
\DefMacro{exp-python-psf_black-num-unit-and-inline-dup1000}{1,236}
\DefMacro{exp-python-psf_black-timept-vanilla-dup1000}{0.029}
\DefMacro{exp-python-psf_black-timept-unit-dup1000}{0.031}
\DefMacro{exp-python-psf_black-timept-inline-dup1000}{0.002}
\DefMacro{exp-python-psf_black-timept-unit-and-inline-dup1000}{0.007}
\DefMacro{exp-python-psf_black-overhead-unit-dup1000}{0.063}
\DefMacro{exp-python-psf_black-overhead-unit-and-inline-dup1000}{0.292}
\DefMacro{exp-python-pypa_pipenv-time-vanilla-dup1000}{3.64}
\DefMacro{exp-python-pypa_pipenv-time-unit-dup1000}{3.66}
\DefMacro{exp-python-pypa_pipenv-time-inline-dup1000}{1.76}
\DefMacro{exp-python-pypa_pipenv-time-unit-and-inline-dup1000}{5.42}
\DefMacro{exp-python-pypa_pipenv-num-vanilla-dup1000}{106}
\DefMacro{exp-python-pypa_pipenv-num-unit-dup1000}{106}
\DefMacro{exp-python-pypa_pipenv-num-inline-dup1000}{1,000}
\DefMacro{exp-python-pypa_pipenv-num-unit-and-inline-dup1000}{1,106}
\DefMacro{exp-python-pypa_pipenv-timept-vanilla-dup1000}{0.034}
\DefMacro{exp-python-pypa_pipenv-timept-unit-dup1000}{0.035}
\DefMacro{exp-python-pypa_pipenv-timept-inline-dup1000}{0.002}
\DefMacro{exp-python-pypa_pipenv-timept-unit-and-inline-dup1000}{0.005}
\DefMacro{exp-python-pypa_pipenv-overhead-unit-dup1000}{0.005}
\DefMacro{exp-python-pypa_pipenv-overhead-unit-and-inline-dup1000}{0.489}
\DefMacro{exp-python-scrapy_scrapy-time-vanilla-dup1000}{130.01}
\DefMacro{exp-python-scrapy_scrapy-time-unit-dup1000}{130.54}
\DefMacro{exp-python-scrapy_scrapy-time-inline-dup1000}{7.01}
\DefMacro{exp-python-scrapy_scrapy-time-unit-and-inline-dup1000}{137.55}
\DefMacro{exp-python-scrapy_scrapy-num-vanilla-dup1000}{2,246}
\DefMacro{exp-python-scrapy_scrapy-num-unit-dup1000}{2,246}
\DefMacro{exp-python-scrapy_scrapy-num-inline-dup1000}{2,000}
\DefMacro{exp-python-scrapy_scrapy-num-unit-and-inline-dup1000}{4,246}
\DefMacro{exp-python-scrapy_scrapy-timept-vanilla-dup1000}{0.058}
\DefMacro{exp-python-scrapy_scrapy-timept-unit-dup1000}{0.058}
\DefMacro{exp-python-scrapy_scrapy-timept-inline-dup1000}{0.004}
\DefMacro{exp-python-scrapy_scrapy-timept-unit-and-inline-dup1000}{0.032}
\DefMacro{exp-python-scrapy_scrapy-overhead-unit-dup1000}{0.004}
\DefMacro{exp-python-scrapy_scrapy-overhead-unit-and-inline-dup1000}{0.058}
\DefMacro{exp-python-time-vanilla-AVG-dup1000}{62.33}
\DefMacro{exp-python-time-vanilla-SUM-dup1000}{872.59}
\DefMacro{exp-python-time-vanilla-MAX-dup1000}{278.08}
\DefMacro{exp-python-time-vanilla-MIN-dup1000}{0.17}
\DefMacro{exp-python-time-vanilla-MEDIAN-dup1000}{7.22}
\DefMacro{exp-python-time-vanilla-STDEV-dup1000}{90.83}
\DefMacro{exp-python-time-vanilla-CNT-dup1000}{14}
\DefMacro{exp-python-time-unit-AVG-dup1000}{63.50}
\DefMacro{exp-python-time-unit-SUM-dup1000}{889.00}
\DefMacro{exp-python-time-unit-MAX-dup1000}{280.19}
\DefMacro{exp-python-time-unit-MIN-dup1000}{0.20}
\DefMacro{exp-python-time-unit-MEDIAN-dup1000}{7.48}
\DefMacro{exp-python-time-unit-STDEV-dup1000}{92.80}
\DefMacro{exp-python-time-unit-CNT-dup1000}{14}
\DefMacro{exp-python-time-inline-AVG-dup1000}{4.29}
\DefMacro{exp-python-time-inline-SUM-dup1000}{60.02}
\DefMacro{exp-python-time-inline-MAX-dup1000}{15.53}
\DefMacro{exp-python-time-inline-MIN-dup1000}{1.60}
\DefMacro{exp-python-time-inline-MEDIAN-dup1000}{3.20}
\DefMacro{exp-python-time-inline-STDEV-dup1000}{3.48}
\DefMacro{exp-python-time-inline-CNT-dup1000}{14}
\DefMacro{exp-python-time-unit-and-inline-AVG-dup1000}{67.79}
\DefMacro{exp-python-time-unit-and-inline-SUM-dup1000}{949.02}
\DefMacro{exp-python-time-unit-and-inline-MAX-dup1000}{283.87}
\DefMacro{exp-python-time-unit-and-inline-MIN-dup1000}{3.41}
\DefMacro{exp-python-time-unit-and-inline-MEDIAN-dup1000}{9.09}
\DefMacro{exp-python-time-unit-and-inline-STDEV-dup1000}{93.07}
\DefMacro{exp-python-time-unit-and-inline-CNT-dup1000}{14}
\DefMacro{exp-python-num-vanilla-AVG-dup1000}{13,046.21}
\DefMacro{exp-python-num-vanilla-SUM-dup1000}{182,647}
\DefMacro{exp-python-num-vanilla-MAX-dup1000}{147,306}
\DefMacro{exp-python-num-vanilla-MIN-dup1000}{1}
\DefMacro{exp-python-num-vanilla-MEDIAN-dup1000}{795.00}
\DefMacro{exp-python-num-vanilla-STDEV-dup1000}{37,594.61}
\DefMacro{exp-python-num-vanilla-CNT-dup1000}{14}
\DefMacro{exp-python-num-unit-AVG-dup1000}{13,046.21}
\DefMacro{exp-python-num-unit-SUM-dup1000}{182,647}
\DefMacro{exp-python-num-unit-MAX-dup1000}{147,306}
\DefMacro{exp-python-num-unit-MIN-dup1000}{1}
\DefMacro{exp-python-num-unit-MEDIAN-dup1000}{795.00}
\DefMacro{exp-python-num-unit-STDEV-dup1000}{37,594.61}
\DefMacro{exp-python-num-unit-CNT-dup1000}{14}
\DefMacro{exp-python-num-inline-AVG-dup1000}{2,428.57}
\DefMacro{exp-python-num-inline-SUM-dup1000}{34,000}
\DefMacro{exp-python-num-inline-MAX-dup1000}{8,000}
\DefMacro{exp-python-num-inline-MIN-dup1000}{1,000}
\DefMacro{exp-python-num-inline-MEDIAN-dup1000}{2,000.00}
\DefMacro{exp-python-num-inline-STDEV-dup1000}{1,840.59}
\DefMacro{exp-python-num-inline-CNT-dup1000}{14}
\DefMacro{exp-python-num-unit-and-inline-AVG-dup1000}{15,474.79}
\DefMacro{exp-python-num-unit-and-inline-SUM-dup1000}{216,647}
\DefMacro{exp-python-num-unit-and-inline-MAX-dup1000}{149,306}
\DefMacro{exp-python-num-unit-and-inline-MIN-dup1000}{1,003}
\DefMacro{exp-python-num-unit-and-inline-MEDIAN-dup1000}{2,984.50}
\DefMacro{exp-python-num-unit-and-inline-STDEV-dup1000}{37,592.11}
\DefMacro{exp-python-num-unit-and-inline-CNT-dup1000}{14}
\DefMacro{exp-python-timept-vanilla-AVG-dup1000}{0.097}
\DefMacro{exp-python-timept-vanilla-SUM-dup1000}{1.364}
\DefMacro{exp-python-timept-vanilla-MAX-dup1000}{0.338}
\DefMacro{exp-python-timept-vanilla-MIN-dup1000}{0.002}
\DefMacro{exp-python-timept-vanilla-MEDIAN-dup1000}{0.032}
\DefMacro{exp-python-timept-vanilla-STDEV-dup1000}{0.118}
\DefMacro{exp-python-timept-vanilla-CNT-dup1000}{14}
\DefMacro{exp-python-timept-unit-AVG-dup1000}{0.100}
\DefMacro{exp-python-timept-unit-SUM-dup1000}{1.404}
\DefMacro{exp-python-timept-unit-MAX-dup1000}{0.335}
\DefMacro{exp-python-timept-unit-MIN-dup1000}{0.002}
\DefMacro{exp-python-timept-unit-MEDIAN-dup1000}{0.033}
\DefMacro{exp-python-timept-unit-STDEV-dup1000}{0.120}
\DefMacro{exp-python-timept-unit-CNT-dup1000}{14}
\DefMacro{exp-python-timept-inline-AVG-dup1000}{0.002}
\DefMacro{exp-python-timept-inline-SUM-dup1000}{0.026}
\DefMacro{exp-python-timept-inline-MAX-dup1000}{0.004}
\DefMacro{exp-python-timept-inline-MIN-dup1000}{0.001}
\DefMacro{exp-python-timept-inline-MEDIAN-dup1000}{0.002}
\DefMacro{exp-python-timept-inline-STDEV-dup1000}{0.001}
\DefMacro{exp-python-timept-inline-CNT-dup1000}{14}
\DefMacro{exp-python-timept-unit-and-inline-AVG-dup1000}{0.011}
\DefMacro{exp-python-timept-unit-and-inline-SUM-dup1000}{0.160}
\DefMacro{exp-python-timept-unit-and-inline-MAX-dup1000}{0.082}
\DefMacro{exp-python-timept-unit-and-inline-MIN-dup1000}{0.001}
\DefMacro{exp-python-timept-unit-and-inline-MEDIAN-dup1000}{0.004}
\DefMacro{exp-python-timept-unit-and-inline-STDEV-dup1000}{0.021}
\DefMacro{exp-python-timept-unit-and-inline-CNT-dup1000}{14}
\DefMacro{exp-python-overhead-unit-AVG-dup1000}{0.025}
\DefMacro{exp-python-overhead-unit-SUM-dup1000}{0.347}
\DefMacro{exp-python-overhead-unit-MAX-dup1000}{0.150}
\DefMacro{exp-python-overhead-unit-MIN-dup1000}{-0.009}
\DefMacro{exp-python-overhead-unit-MEDIAN-dup1000}{0.007}
\DefMacro{exp-python-overhead-unit-STDEV-dup1000}{0.041}
\DefMacro{exp-python-overhead-unit-CNT-dup1000}{14}
\DefMacro{exp-python-overhead-unit-and-inline-AVG-dup1000}{3.596}
\DefMacro{exp-python-overhead-unit-and-inline-SUM-dup1000}{50.341}
\DefMacro{exp-python-overhead-unit-and-inline-MAX-dup1000}{32.877}
\DefMacro{exp-python-overhead-unit-and-inline-MIN-dup1000}{0.021}
\DefMacro{exp-python-overhead-unit-and-inline-MEDIAN-dup1000}{0.341}
\DefMacro{exp-python-overhead-unit-and-inline-STDEV-dup1000}{8.595}
\DefMacro{exp-python-overhead-unit-and-inline-CNT-dup1000}{14}
\DefMacro{exp-python-timept-vanilla-MACROAVG-dup1000}{0.005}
\DefMacro{exp-python-timept-unit-MACROAVG-dup1000}{0.005}
\DefMacro{exp-python-timept-inline-MACROAVG-dup1000}{0.002}
\DefMacro{exp-python-timept-unit-and-inline-MACROAVG-dup1000}{0.004}
\DefMacro{exp-python-overhead-unit-MACROAVG-dup1000}{0.019}
\DefMacro{exp-python-overhead-unit-and-inline-MACROAVG-dup1000}{0.088}

\DefMacro{exp-java-alibaba_fastjson-time-vanilla-dup1}{44.99}
\DefMacro{exp-java-alibaba_fastjson-time-unit-dup1}{44.86}
\DefMacro{exp-java-alibaba_fastjson-time-inline-dup1}{0.73}
\DefMacro{exp-java-alibaba_fastjson-time-unit-and-inline-dup1}{45.59}
\DefMacro{exp-java-alibaba_fastjson-num-vanilla-dup1}{5,022}
\DefMacro{exp-java-alibaba_fastjson-num-unit-dup1}{5,022}
\DefMacro{exp-java-alibaba_fastjson-num-inline-dup1}{2}
\DefMacro{exp-java-alibaba_fastjson-num-unit-and-inline-dup1}{5,024}
\DefMacro{exp-java-alibaba_fastjson-timept-vanilla-dup1}{0.009}
\DefMacro{exp-java-alibaba_fastjson-timept-unit-dup1}{0.009}
\DefMacro{exp-java-alibaba_fastjson-timept-inline-dup1}{0.363}
\DefMacro{exp-java-alibaba_fastjson-timept-unit-and-inline-dup1}{0.009}
\DefMacro{exp-java-alibaba_fastjson-overhead-unit-dup1}{-0.003}
\DefMacro{exp-java-alibaba_fastjson-overhead-unit-and-inline-dup1}{0.013}
\DefMacro{exp-java-alibaba_nacos-time-vanilla-dup1}{249.45}
\DefMacro{exp-java-alibaba_nacos-time-unit-dup1}{249.93}
\DefMacro{exp-java-alibaba_nacos-time-inline-dup1}{0.74}
\DefMacro{exp-java-alibaba_nacos-time-unit-and-inline-dup1}{250.67}
\DefMacro{exp-java-alibaba_nacos-num-vanilla-dup1}{971}
\DefMacro{exp-java-alibaba_nacos-num-unit-dup1}{971}
\DefMacro{exp-java-alibaba_nacos-num-inline-dup1}{1}
\DefMacro{exp-java-alibaba_nacos-num-unit-and-inline-dup1}{972}
\DefMacro{exp-java-alibaba_nacos-timept-vanilla-dup1}{0.257}
\DefMacro{exp-java-alibaba_nacos-timept-unit-dup1}{0.257}
\DefMacro{exp-java-alibaba_nacos-timept-inline-dup1}{0.745}
\DefMacro{exp-java-alibaba_nacos-timept-unit-and-inline-dup1}{0.258}
\DefMacro{exp-java-alibaba_nacos-overhead-unit-dup1}{0.002}
\DefMacro{exp-java-alibaba_nacos-overhead-unit-and-inline-dup1}{0.005}
\DefMacro{exp-java-apache_dubbo-time-vanilla-dup1}{678.86}
\DefMacro{exp-java-apache_dubbo-time-unit-dup1}{679.43}
\DefMacro{exp-java-apache_dubbo-time-inline-dup1}{0.82}
\DefMacro{exp-java-apache_dubbo-time-unit-and-inline-dup1}{680.26}
\DefMacro{exp-java-apache_dubbo-num-vanilla-dup1}{3,180}
\DefMacro{exp-java-apache_dubbo-num-unit-dup1}{3,180}
\DefMacro{exp-java-apache_dubbo-num-inline-dup1}{1}
\DefMacro{exp-java-apache_dubbo-num-unit-and-inline-dup1}{3,181}
\DefMacro{exp-java-apache_dubbo-timept-vanilla-dup1}{0.213}
\DefMacro{exp-java-apache_dubbo-timept-unit-dup1}{0.214}
\DefMacro{exp-java-apache_dubbo-timept-inline-dup1}{0.822}
\DefMacro{exp-java-apache_dubbo-timept-unit-and-inline-dup1}{0.214}
\DefMacro{exp-java-apache_dubbo-overhead-unit-dup1}{0.001}
\DefMacro{exp-java-apache_dubbo-overhead-unit-and-inline-dup1}{0.002}
\DefMacro{exp-java-apache_kafka-time-vanilla-dup1}{9.84}
\DefMacro{exp-java-apache_kafka-time-unit-dup1}{10.09}
\DefMacro{exp-java-apache_kafka-time-inline-dup1}{0.67}
\DefMacro{exp-java-apache_kafka-time-unit-and-inline-dup1}{10.76}
\DefMacro{exp-java-apache_kafka-num-vanilla-dup1}{221}
\DefMacro{exp-java-apache_kafka-num-unit-dup1}{221}
\DefMacro{exp-java-apache_kafka-num-inline-dup1}{1}
\DefMacro{exp-java-apache_kafka-num-unit-and-inline-dup1}{222}
\DefMacro{exp-java-apache_kafka-timept-vanilla-dup1}{0.045}
\DefMacro{exp-java-apache_kafka-timept-unit-dup1}{0.046}
\DefMacro{exp-java-apache_kafka-timept-inline-dup1}{0.670}
\DefMacro{exp-java-apache_kafka-timept-unit-and-inline-dup1}{0.048}
\DefMacro{exp-java-apache_kafka-overhead-unit-dup1}{0.026}
\DefMacro{exp-java-apache_kafka-overhead-unit-and-inline-dup1}{0.094}
\DefMacro{exp-java-apache_shardingsphere-time-vanilla-dup1}{5.03}
\DefMacro{exp-java-apache_shardingsphere-time-unit-dup1}{5.04}
\DefMacro{exp-java-apache_shardingsphere-time-inline-dup1}{0.71}
\DefMacro{exp-java-apache_shardingsphere-time-unit-and-inline-dup1}{5.75}
\DefMacro{exp-java-apache_shardingsphere-num-vanilla-dup1}{44}
\DefMacro{exp-java-apache_shardingsphere-num-unit-dup1}{44}
\DefMacro{exp-java-apache_shardingsphere-num-inline-dup1}{2}
\DefMacro{exp-java-apache_shardingsphere-num-unit-and-inline-dup1}{46}
\DefMacro{exp-java-apache_shardingsphere-timept-vanilla-dup1}{0.114}
\DefMacro{exp-java-apache_shardingsphere-timept-unit-dup1}{0.115}
\DefMacro{exp-java-apache_shardingsphere-timept-inline-dup1}{0.355}
\DefMacro{exp-java-apache_shardingsphere-timept-unit-and-inline-dup1}{0.125}
\DefMacro{exp-java-apache_shardingsphere-overhead-unit-dup1}{0.002}
\DefMacro{exp-java-apache_shardingsphere-overhead-unit-and-inline-dup1}{0.143}
\DefMacro{exp-java-jenkinsci_jenkins-time-vanilla-dup1}{4.67}
\DefMacro{exp-java-jenkinsci_jenkins-time-unit-dup1}{4.64}
\DefMacro{exp-java-jenkinsci_jenkins-time-inline-dup1}{0.65}
\DefMacro{exp-java-jenkinsci_jenkins-time-unit-and-inline-dup1}{5.29}
\DefMacro{exp-java-jenkinsci_jenkins-num-vanilla-dup1}{32}
\DefMacro{exp-java-jenkinsci_jenkins-num-unit-dup1}{32}
\DefMacro{exp-java-jenkinsci_jenkins-num-inline-dup1}{2}
\DefMacro{exp-java-jenkinsci_jenkins-num-unit-and-inline-dup1}{34}
\DefMacro{exp-java-jenkinsci_jenkins-timept-vanilla-dup1}{0.146}
\DefMacro{exp-java-jenkinsci_jenkins-timept-unit-dup1}{0.145}
\DefMacro{exp-java-jenkinsci_jenkins-timept-inline-dup1}{0.325}
\DefMacro{exp-java-jenkinsci_jenkins-timept-unit-and-inline-dup1}{0.156}
\DefMacro{exp-java-jenkinsci_jenkins-overhead-unit-dup1}{-0.007}
\DefMacro{exp-java-jenkinsci_jenkins-overhead-unit-and-inline-dup1}{0.132}
\DefMacro{exp-java-skylot_jadx-time-vanilla-dup1}{66.57}
\DefMacro{exp-java-skylot_jadx-time-unit-dup1}{75.47}
\DefMacro{exp-java-skylot_jadx-time-inline-dup1}{0.74}
\DefMacro{exp-java-skylot_jadx-time-unit-and-inline-dup1}{76.21}
\DefMacro{exp-java-skylot_jadx-num-vanilla-dup1}{709}
\DefMacro{exp-java-skylot_jadx-num-unit-dup1}{709}
\DefMacro{exp-java-skylot_jadx-num-inline-dup1}{1}
\DefMacro{exp-java-skylot_jadx-num-unit-and-inline-dup1}{710}
\DefMacro{exp-java-skylot_jadx-timept-vanilla-dup1}{0.094}
\DefMacro{exp-java-skylot_jadx-timept-unit-dup1}{0.106}
\DefMacro{exp-java-skylot_jadx-timept-inline-dup1}{0.738}
\DefMacro{exp-java-skylot_jadx-timept-unit-and-inline-dup1}{0.107}
\DefMacro{exp-java-skylot_jadx-overhead-unit-dup1}{0.134}
\DefMacro{exp-java-skylot_jadx-overhead-unit-and-inline-dup1}{0.145}
\DefMacro{exp-java-time-vanilla-AVG-dup1}{151.34}
\DefMacro{exp-java-time-vanilla-SUM-dup1}{1,059.41}
\DefMacro{exp-java-time-vanilla-MAX-dup1}{678.86}
\DefMacro{exp-java-time-vanilla-MIN-dup1}{4.67}
\DefMacro{exp-java-time-vanilla-MEDIAN-dup1}{44.99}
\DefMacro{exp-java-time-vanilla-STDEV-dup1}{229.70}
\DefMacro{exp-java-time-vanilla-CNT-dup1}{7}
\DefMacro{exp-java-time-unit-AVG-dup1}{152.78}
\DefMacro{exp-java-time-unit-SUM-dup1}{1,069.47}
\DefMacro{exp-java-time-unit-MAX-dup1}{679.43}
\DefMacro{exp-java-time-unit-MIN-dup1}{4.64}
\DefMacro{exp-java-time-unit-MEDIAN-dup1}{44.86}
\DefMacro{exp-java-time-unit-STDEV-dup1}{229.46}
\DefMacro{exp-java-time-unit-CNT-dup1}{7}
\DefMacro{exp-java-time-inline-AVG-dup1}{0.72}
\DefMacro{exp-java-time-inline-SUM-dup1}{5.06}
\DefMacro{exp-java-time-inline-MAX-dup1}{0.82}
\DefMacro{exp-java-time-inline-MIN-dup1}{0.65}
\DefMacro{exp-java-time-inline-MEDIAN-dup1}{0.73}
\DefMacro{exp-java-time-inline-STDEV-dup1}{0.05}
\DefMacro{exp-java-time-inline-CNT-dup1}{7}
\DefMacro{exp-java-time-unit-and-inline-AVG-dup1}{153.50}
\DefMacro{exp-java-time-unit-and-inline-SUM-dup1}{1,074.53}
\DefMacro{exp-java-time-unit-and-inline-MAX-dup1}{680.26}
\DefMacro{exp-java-time-unit-and-inline-MIN-dup1}{5.29}
\DefMacro{exp-java-time-unit-and-inline-MEDIAN-dup1}{45.59}
\DefMacro{exp-java-time-unit-and-inline-STDEV-dup1}{229.50}
\DefMacro{exp-java-time-unit-and-inline-CNT-dup1}{7}
\DefMacro{exp-java-num-vanilla-AVG-dup1}{1,454.14}
\DefMacro{exp-java-num-vanilla-SUM-dup1}{10,179}
\DefMacro{exp-java-num-vanilla-MAX-dup1}{5,022}
\DefMacro{exp-java-num-vanilla-MIN-dup1}{32}
\DefMacro{exp-java-num-vanilla-MEDIAN-dup1}{709.00}
\DefMacro{exp-java-num-vanilla-STDEV-dup1}{1,773.96}
\DefMacro{exp-java-num-vanilla-CNT-dup1}{7}
\DefMacro{exp-java-num-unit-AVG-dup1}{1,454.14}
\DefMacro{exp-java-num-unit-SUM-dup1}{10,179}
\DefMacro{exp-java-num-unit-MAX-dup1}{5,022}
\DefMacro{exp-java-num-unit-MIN-dup1}{32}
\DefMacro{exp-java-num-unit-MEDIAN-dup1}{709.00}
\DefMacro{exp-java-num-unit-STDEV-dup1}{1,773.96}
\DefMacro{exp-java-num-unit-CNT-dup1}{7}
\DefMacro{exp-java-num-inline-AVG-dup1}{1.43}
\DefMacro{exp-java-num-inline-SUM-dup1}{10}
\DefMacro{exp-java-num-inline-MAX-dup1}{2}
\DefMacro{exp-java-num-inline-MIN-dup1}{1}
\DefMacro{exp-java-num-inline-MEDIAN-dup1}{1.00}
\DefMacro{exp-java-num-inline-STDEV-dup1}{0.49}
\DefMacro{exp-java-num-inline-CNT-dup1}{7}
\DefMacro{exp-java-num-unit-and-inline-AVG-dup1}{1,455.57}
\DefMacro{exp-java-num-unit-and-inline-SUM-dup1}{10,189}
\DefMacro{exp-java-num-unit-and-inline-MAX-dup1}{5,024}
\DefMacro{exp-java-num-unit-and-inline-MIN-dup1}{34}
\DefMacro{exp-java-num-unit-and-inline-MEDIAN-dup1}{710.00}
\DefMacro{exp-java-num-unit-and-inline-STDEV-dup1}{1,774.02}
\DefMacro{exp-java-num-unit-and-inline-CNT-dup1}{7}
\DefMacro{exp-java-timept-vanilla-AVG-dup1}{0.125}
\DefMacro{exp-java-timept-vanilla-SUM-dup1}{0.878}
\DefMacro{exp-java-timept-vanilla-MAX-dup1}{0.257}
\DefMacro{exp-java-timept-vanilla-MIN-dup1}{0.009}
\DefMacro{exp-java-timept-vanilla-MEDIAN-dup1}{0.114}
\DefMacro{exp-java-timept-vanilla-STDEV-dup1}{0.082}
\DefMacro{exp-java-timept-vanilla-CNT-dup1}{7}
\DefMacro{exp-java-timept-unit-AVG-dup1}{0.127}
\DefMacro{exp-java-timept-unit-SUM-dup1}{0.892}
\DefMacro{exp-java-timept-unit-MAX-dup1}{0.257}
\DefMacro{exp-java-timept-unit-MIN-dup1}{0.009}
\DefMacro{exp-java-timept-unit-MEDIAN-dup1}{0.115}
\DefMacro{exp-java-timept-unit-STDEV-dup1}{0.081}
\DefMacro{exp-java-timept-unit-CNT-dup1}{7}
\DefMacro{exp-java-timept-inline-AVG-dup1}{0.574}
\DefMacro{exp-java-timept-inline-SUM-dup1}{4.017}
\DefMacro{exp-java-timept-inline-MAX-dup1}{0.822}
\DefMacro{exp-java-timept-inline-MIN-dup1}{0.325}
\DefMacro{exp-java-timept-inline-MEDIAN-dup1}{0.670}
\DefMacro{exp-java-timept-inline-STDEV-dup1}{0.201}
\DefMacro{exp-java-timept-inline-CNT-dup1}{7}
\DefMacro{exp-java-timept-unit-and-inline-AVG-dup1}{0.131}
\DefMacro{exp-java-timept-unit-and-inline-SUM-dup1}{0.917}
\DefMacro{exp-java-timept-unit-and-inline-MAX-dup1}{0.258}
\DefMacro{exp-java-timept-unit-and-inline-MIN-dup1}{0.009}
\DefMacro{exp-java-timept-unit-and-inline-MEDIAN-dup1}{0.125}
\DefMacro{exp-java-timept-unit-and-inline-STDEV-dup1}{0.081}
\DefMacro{exp-java-timept-unit-and-inline-CNT-dup1}{7}
\DefMacro{exp-java-overhead-unit-AVG-dup1}{0.022}
\DefMacro{exp-java-overhead-unit-SUM-dup1}{0.155}
\DefMacro{exp-java-overhead-unit-MAX-dup1}{0.134}
\DefMacro{exp-java-overhead-unit-MIN-dup1}{-0.007}
\DefMacro{exp-java-overhead-unit-MEDIAN-dup1}{0.002}
\DefMacro{exp-java-overhead-unit-STDEV-dup1}{0.047}
\DefMacro{exp-java-overhead-unit-CNT-dup1}{7}
\DefMacro{exp-java-overhead-unit-and-inline-AVG-dup1}{0.076}
\DefMacro{exp-java-overhead-unit-and-inline-SUM-dup1}{0.535}
\DefMacro{exp-java-overhead-unit-and-inline-MAX-dup1}{0.145}
\DefMacro{exp-java-overhead-unit-and-inline-MIN-dup1}{0.002}
\DefMacro{exp-java-overhead-unit-and-inline-MEDIAN-dup1}{0.094}
\DefMacro{exp-java-overhead-unit-and-inline-STDEV-dup1}{0.062}
\DefMacro{exp-java-overhead-unit-and-inline-CNT-dup1}{7}
\DefMacro{exp-java-timept-vanilla-MACROAVG-dup1}{0.104}
\DefMacro{exp-java-timept-unit-MACROAVG-dup1}{0.105}
\DefMacro{exp-java-timept-inline-MACROAVG-dup1}{0.506}
\DefMacro{exp-java-timept-unit-and-inline-MACROAVG-dup1}{0.105}
\DefMacro{exp-java-overhead-unit-MACROAVG-dup1}{0.009}
\DefMacro{exp-java-overhead-unit-and-inline-MACROAVG-dup1}{0.014}
\DefMacro{exp-java-alibaba_fastjson-time-vanilla-dup10}{44.99}
\DefMacro{exp-java-alibaba_fastjson-time-unit-dup10}{44.73}
\DefMacro{exp-java-alibaba_fastjson-time-inline-dup10}{0.69}
\DefMacro{exp-java-alibaba_fastjson-time-unit-and-inline-dup10}{45.42}
\DefMacro{exp-java-alibaba_fastjson-num-vanilla-dup10}{5,022}
\DefMacro{exp-java-alibaba_fastjson-num-unit-dup10}{5,022}
\DefMacro{exp-java-alibaba_fastjson-num-inline-dup10}{20}
\DefMacro{exp-java-alibaba_fastjson-num-unit-and-inline-dup10}{5,042}
\DefMacro{exp-java-alibaba_fastjson-timept-vanilla-dup10}{0.009}
\DefMacro{exp-java-alibaba_fastjson-timept-unit-dup10}{0.009}
\DefMacro{exp-java-alibaba_fastjson-timept-inline-dup10}{0.035}
\DefMacro{exp-java-alibaba_fastjson-timept-unit-and-inline-dup10}{0.009}
\DefMacro{exp-java-alibaba_fastjson-overhead-unit-dup10}{-0.006}
\DefMacro{exp-java-alibaba_fastjson-overhead-unit-and-inline-dup10}{0.010}
\DefMacro{exp-java-alibaba_nacos-time-vanilla-dup10}{248.65}
\DefMacro{exp-java-alibaba_nacos-time-unit-dup10}{249.20}
\DefMacro{exp-java-alibaba_nacos-time-inline-dup10}{0.67}
\DefMacro{exp-java-alibaba_nacos-time-unit-and-inline-dup10}{249.87}
\DefMacro{exp-java-alibaba_nacos-num-vanilla-dup10}{971}
\DefMacro{exp-java-alibaba_nacos-num-unit-dup10}{971}
\DefMacro{exp-java-alibaba_nacos-num-inline-dup10}{10}
\DefMacro{exp-java-alibaba_nacos-num-unit-and-inline-dup10}{981}
\DefMacro{exp-java-alibaba_nacos-timept-vanilla-dup10}{0.256}
\DefMacro{exp-java-alibaba_nacos-timept-unit-dup10}{0.257}
\DefMacro{exp-java-alibaba_nacos-timept-inline-dup10}{0.067}
\DefMacro{exp-java-alibaba_nacos-timept-unit-and-inline-dup10}{0.255}
\DefMacro{exp-java-alibaba_nacos-overhead-unit-dup10}{0.002}
\DefMacro{exp-java-alibaba_nacos-overhead-unit-and-inline-dup10}{0.005}
\DefMacro{exp-java-apache_dubbo-time-vanilla-dup10}{678.61}
\DefMacro{exp-java-apache_dubbo-time-unit-dup10}{680.85}
\DefMacro{exp-java-apache_dubbo-time-inline-dup10}{0.74}
\DefMacro{exp-java-apache_dubbo-time-unit-and-inline-dup10}{681.59}
\DefMacro{exp-java-apache_dubbo-num-vanilla-dup10}{3,180}
\DefMacro{exp-java-apache_dubbo-num-unit-dup10}{3,180}
\DefMacro{exp-java-apache_dubbo-num-inline-dup10}{10}
\DefMacro{exp-java-apache_dubbo-num-unit-and-inline-dup10}{3,190}
\DefMacro{exp-java-apache_dubbo-timept-vanilla-dup10}{0.213}
\DefMacro{exp-java-apache_dubbo-timept-unit-dup10}{0.214}
\DefMacro{exp-java-apache_dubbo-timept-inline-dup10}{0.074}
\DefMacro{exp-java-apache_dubbo-timept-unit-and-inline-dup10}{0.214}
\DefMacro{exp-java-apache_dubbo-overhead-unit-dup10}{0.003}
\DefMacro{exp-java-apache_dubbo-overhead-unit-and-inline-dup10}{0.004}
\DefMacro{exp-java-apache_kafka-time-vanilla-dup10}{9.91}
\DefMacro{exp-java-apache_kafka-time-unit-dup10}{10.08}
\DefMacro{exp-java-apache_kafka-time-inline-dup10}{0.71}
\DefMacro{exp-java-apache_kafka-time-unit-and-inline-dup10}{10.79}
\DefMacro{exp-java-apache_kafka-num-vanilla-dup10}{221}
\DefMacro{exp-java-apache_kafka-num-unit-dup10}{221}
\DefMacro{exp-java-apache_kafka-num-inline-dup10}{10}
\DefMacro{exp-java-apache_kafka-num-unit-and-inline-dup10}{231}
\DefMacro{exp-java-apache_kafka-timept-vanilla-dup10}{0.045}
\DefMacro{exp-java-apache_kafka-timept-unit-dup10}{0.046}
\DefMacro{exp-java-apache_kafka-timept-inline-dup10}{0.071}
\DefMacro{exp-java-apache_kafka-timept-unit-and-inline-dup10}{0.047}
\DefMacro{exp-java-apache_kafka-overhead-unit-dup10}{0.018}
\DefMacro{exp-java-apache_kafka-overhead-unit-and-inline-dup10}{0.089}
\DefMacro{exp-java-apache_shardingsphere-time-vanilla-dup10}{5.09}
\DefMacro{exp-java-apache_shardingsphere-time-unit-dup10}{5.09}
\DefMacro{exp-java-apache_shardingsphere-time-inline-dup10}{0.68}
\DefMacro{exp-java-apache_shardingsphere-time-unit-and-inline-dup10}{5.76}
\DefMacro{exp-java-apache_shardingsphere-num-vanilla-dup10}{44}
\DefMacro{exp-java-apache_shardingsphere-num-unit-dup10}{44}
\DefMacro{exp-java-apache_shardingsphere-num-inline-dup10}{20}
\DefMacro{exp-java-apache_shardingsphere-num-unit-and-inline-dup10}{64}
\DefMacro{exp-java-apache_shardingsphere-timept-vanilla-dup10}{0.116}
\DefMacro{exp-java-apache_shardingsphere-timept-unit-dup10}{0.116}
\DefMacro{exp-java-apache_shardingsphere-timept-inline-dup10}{0.034}
\DefMacro{exp-java-apache_shardingsphere-timept-unit-and-inline-dup10}{0.090}
\DefMacro{exp-java-apache_shardingsphere-overhead-unit-dup10}{-0.000}
\DefMacro{exp-java-apache_shardingsphere-overhead-unit-and-inline-dup10}{0.133}
\DefMacro{exp-java-jenkinsci_jenkins-time-vanilla-dup10}{4.67}
\DefMacro{exp-java-jenkinsci_jenkins-time-unit-dup10}{4.70}
\DefMacro{exp-java-jenkinsci_jenkins-time-inline-dup10}{0.72}
\DefMacro{exp-java-jenkinsci_jenkins-time-unit-and-inline-dup10}{5.42}
\DefMacro{exp-java-jenkinsci_jenkins-num-vanilla-dup10}{32}
\DefMacro{exp-java-jenkinsci_jenkins-num-unit-dup10}{32}
\DefMacro{exp-java-jenkinsci_jenkins-num-inline-dup10}{20}
\DefMacro{exp-java-jenkinsci_jenkins-num-unit-and-inline-dup10}{52}
\DefMacro{exp-java-jenkinsci_jenkins-timept-vanilla-dup10}{0.146}
\DefMacro{exp-java-jenkinsci_jenkins-timept-unit-dup10}{0.147}
\DefMacro{exp-java-jenkinsci_jenkins-timept-inline-dup10}{0.036}
\DefMacro{exp-java-jenkinsci_jenkins-timept-unit-and-inline-dup10}{0.104}
\DefMacro{exp-java-jenkinsci_jenkins-overhead-unit-dup10}{0.006}
\DefMacro{exp-java-jenkinsci_jenkins-overhead-unit-and-inline-dup10}{0.160}
\DefMacro{exp-java-skylot_jadx-time-vanilla-dup10}{67.43}
\DefMacro{exp-java-skylot_jadx-time-unit-dup10}{65.82}
\DefMacro{exp-java-skylot_jadx-time-inline-dup10}{0.71}
\DefMacro{exp-java-skylot_jadx-time-unit-and-inline-dup10}{66.52}
\DefMacro{exp-java-skylot_jadx-num-vanilla-dup10}{709}
\DefMacro{exp-java-skylot_jadx-num-unit-dup10}{709}
\DefMacro{exp-java-skylot_jadx-num-inline-dup10}{10}
\DefMacro{exp-java-skylot_jadx-num-unit-and-inline-dup10}{719}
\DefMacro{exp-java-skylot_jadx-timept-vanilla-dup10}{0.095}
\DefMacro{exp-java-skylot_jadx-timept-unit-dup10}{0.093}
\DefMacro{exp-java-skylot_jadx-timept-inline-dup10}{0.071}
\DefMacro{exp-java-skylot_jadx-timept-unit-and-inline-dup10}{0.093}
\DefMacro{exp-java-skylot_jadx-overhead-unit-dup10}{-0.024}
\DefMacro{exp-java-skylot_jadx-overhead-unit-and-inline-dup10}{-0.014}
\DefMacro{exp-java-time-vanilla-AVG-dup10}{151.34}
\DefMacro{exp-java-time-vanilla-SUM-dup10}{1,059.36}
\DefMacro{exp-java-time-vanilla-MAX-dup10}{678.61}
\DefMacro{exp-java-time-vanilla-MIN-dup10}{4.67}
\DefMacro{exp-java-time-vanilla-MEDIAN-dup10}{44.99}
\DefMacro{exp-java-time-vanilla-STDEV-dup10}{229.51}
\DefMacro{exp-java-time-vanilla-CNT-dup10}{7}
\DefMacro{exp-java-time-unit-AVG-dup10}{151.50}
\DefMacro{exp-java-time-unit-SUM-dup10}{1,060.47}
\DefMacro{exp-java-time-unit-MAX-dup10}{680.85}
\DefMacro{exp-java-time-unit-MIN-dup10}{4.70}
\DefMacro{exp-java-time-unit-MEDIAN-dup10}{44.73}
\DefMacro{exp-java-time-unit-STDEV-dup10}{230.37}
\DefMacro{exp-java-time-unit-CNT-dup10}{7}
\DefMacro{exp-java-time-inline-AVG-dup10}{0.70}
\DefMacro{exp-java-time-inline-SUM-dup10}{4.91}
\DefMacro{exp-java-time-inline-MAX-dup10}{0.74}
\DefMacro{exp-java-time-inline-MIN-dup10}{0.67}
\DefMacro{exp-java-time-inline-MEDIAN-dup10}{0.71}
\DefMacro{exp-java-time-inline-STDEV-dup10}{0.02}
\DefMacro{exp-java-time-inline-CNT-dup10}{7}
\DefMacro{exp-java-time-unit-and-inline-AVG-dup10}{152.20}
\DefMacro{exp-java-time-unit-and-inline-SUM-dup10}{1,065.38}
\DefMacro{exp-java-time-unit-and-inline-MAX-dup10}{681.59}
\DefMacro{exp-java-time-unit-and-inline-MIN-dup10}{5.42}
\DefMacro{exp-java-time-unit-and-inline-MEDIAN-dup10}{45.42}
\DefMacro{exp-java-time-unit-and-inline-STDEV-dup10}{230.38}
\DefMacro{exp-java-time-unit-and-inline-CNT-dup10}{7}
\DefMacro{exp-java-num-vanilla-AVG-dup10}{1,454.14}
\DefMacro{exp-java-num-vanilla-SUM-dup10}{10,179}
\DefMacro{exp-java-num-vanilla-MAX-dup10}{5,022}
\DefMacro{exp-java-num-vanilla-MIN-dup10}{32}
\DefMacro{exp-java-num-vanilla-MEDIAN-dup10}{709.00}
\DefMacro{exp-java-num-vanilla-STDEV-dup10}{1,773.96}
\DefMacro{exp-java-num-vanilla-CNT-dup10}{7}
\DefMacro{exp-java-num-unit-AVG-dup10}{1,454.14}
\DefMacro{exp-java-num-unit-SUM-dup10}{10,179}
\DefMacro{exp-java-num-unit-MAX-dup10}{5,022}
\DefMacro{exp-java-num-unit-MIN-dup10}{32}
\DefMacro{exp-java-num-unit-MEDIAN-dup10}{709.00}
\DefMacro{exp-java-num-unit-STDEV-dup10}{1,773.96}
\DefMacro{exp-java-num-unit-CNT-dup10}{7}
\DefMacro{exp-java-num-inline-AVG-dup10}{14.29}
\DefMacro{exp-java-num-inline-SUM-dup10}{100}
\DefMacro{exp-java-num-inline-MAX-dup10}{20}
\DefMacro{exp-java-num-inline-MIN-dup10}{10}
\DefMacro{exp-java-num-inline-MEDIAN-dup10}{10.00}
\DefMacro{exp-java-num-inline-STDEV-dup10}{4.95}
\DefMacro{exp-java-num-inline-CNT-dup10}{7}
\DefMacro{exp-java-num-unit-and-inline-AVG-dup10}{1,468.43}
\DefMacro{exp-java-num-unit-and-inline-SUM-dup10}{10,279}
\DefMacro{exp-java-num-unit-and-inline-MAX-dup10}{5,042}
\DefMacro{exp-java-num-unit-and-inline-MIN-dup10}{52}
\DefMacro{exp-java-num-unit-and-inline-MEDIAN-dup10}{719.00}
\DefMacro{exp-java-num-unit-and-inline-STDEV-dup10}{1,774.56}
\DefMacro{exp-java-num-unit-and-inline-CNT-dup10}{7}
\DefMacro{exp-java-timept-vanilla-AVG-dup10}{0.126}
\DefMacro{exp-java-timept-vanilla-SUM-dup10}{0.880}
\DefMacro{exp-java-timept-vanilla-MAX-dup10}{0.256}
\DefMacro{exp-java-timept-vanilla-MIN-dup10}{0.009}
\DefMacro{exp-java-timept-vanilla-MEDIAN-dup10}{0.116}
\DefMacro{exp-java-timept-vanilla-STDEV-dup10}{0.081}
\DefMacro{exp-java-timept-vanilla-CNT-dup10}{7}
\DefMacro{exp-java-timept-unit-AVG-dup10}{0.126}
\DefMacro{exp-java-timept-unit-SUM-dup10}{0.881}
\DefMacro{exp-java-timept-unit-MAX-dup10}{0.257}
\DefMacro{exp-java-timept-unit-MIN-dup10}{0.009}
\DefMacro{exp-java-timept-unit-MEDIAN-dup10}{0.116}
\DefMacro{exp-java-timept-unit-STDEV-dup10}{0.082}
\DefMacro{exp-java-timept-unit-CNT-dup10}{7}
\DefMacro{exp-java-timept-inline-AVG-dup10}{0.055}
\DefMacro{exp-java-timept-inline-SUM-dup10}{0.387}
\DefMacro{exp-java-timept-inline-MAX-dup10}{0.074}
\DefMacro{exp-java-timept-inline-MIN-dup10}{0.034}
\DefMacro{exp-java-timept-inline-MEDIAN-dup10}{0.067}
\DefMacro{exp-java-timept-inline-STDEV-dup10}{0.018}
\DefMacro{exp-java-timept-inline-CNT-dup10}{7}
\DefMacro{exp-java-timept-unit-and-inline-AVG-dup10}{0.116}
\DefMacro{exp-java-timept-unit-and-inline-SUM-dup10}{0.811}
\DefMacro{exp-java-timept-unit-and-inline-MAX-dup10}{0.255}
\DefMacro{exp-java-timept-unit-and-inline-MIN-dup10}{0.009}
\DefMacro{exp-java-timept-unit-and-inline-MEDIAN-dup10}{0.093}
\DefMacro{exp-java-timept-unit-and-inline-STDEV-dup10}{0.081}
\DefMacro{exp-java-timept-unit-and-inline-CNT-dup10}{7}
\DefMacro{exp-java-overhead-unit-AVG-dup10}{-0.000}
\DefMacro{exp-java-overhead-unit-SUM-dup10}{-0.000}
\DefMacro{exp-java-overhead-unit-MAX-dup10}{0.018}
\DefMacro{exp-java-overhead-unit-MIN-dup10}{-0.024}
\DefMacro{exp-java-overhead-unit-MEDIAN-dup10}{0.002}
\DefMacro{exp-java-overhead-unit-STDEV-dup10}{0.012}
\DefMacro{exp-java-overhead-unit-CNT-dup10}{7}
\DefMacro{exp-java-overhead-unit-and-inline-AVG-dup10}{0.055}
\DefMacro{exp-java-overhead-unit-and-inline-SUM-dup10}{0.388}
\DefMacro{exp-java-overhead-unit-and-inline-MAX-dup10}{0.160}
\DefMacro{exp-java-overhead-unit-and-inline-MIN-dup10}{-0.014}
\DefMacro{exp-java-overhead-unit-and-inline-MEDIAN-dup10}{0.010}
\DefMacro{exp-java-overhead-unit-and-inline-STDEV-dup10}{0.066}
\DefMacro{exp-java-overhead-unit-and-inline-CNT-dup10}{7}
\DefMacro{exp-java-timept-vanilla-MACROAVG-dup10}{0.104}
\DefMacro{exp-java-timept-unit-MACROAVG-dup10}{0.104}
\DefMacro{exp-java-timept-inline-MACROAVG-dup10}{0.049}
\DefMacro{exp-java-timept-unit-and-inline-MACROAVG-dup10}{0.104}
\DefMacro{exp-java-overhead-unit-MACROAVG-dup10}{0.001}
\DefMacro{exp-java-overhead-unit-and-inline-MACROAVG-dup10}{0.006}
\DefMacro{exp-java-alibaba_fastjson-time-vanilla-dup100}{44.86}
\DefMacro{exp-java-alibaba_fastjson-time-unit-dup100}{52.40}
\DefMacro{exp-java-alibaba_fastjson-time-inline-dup100}{0.72}
\DefMacro{exp-java-alibaba_fastjson-time-unit-and-inline-dup100}{53.13}
\DefMacro{exp-java-alibaba_fastjson-num-vanilla-dup100}{5,022}
\DefMacro{exp-java-alibaba_fastjson-num-unit-dup100}{5,022}
\DefMacro{exp-java-alibaba_fastjson-num-inline-dup100}{200}
\DefMacro{exp-java-alibaba_fastjson-num-unit-and-inline-dup100}{5,222}
\DefMacro{exp-java-alibaba_fastjson-timept-vanilla-dup100}{0.009}
\DefMacro{exp-java-alibaba_fastjson-timept-unit-dup100}{0.010}
\DefMacro{exp-java-alibaba_fastjson-timept-inline-dup100}{0.004}
\DefMacro{exp-java-alibaba_fastjson-timept-unit-and-inline-dup100}{0.010}
\DefMacro{exp-java-alibaba_fastjson-overhead-unit-dup100}{0.168}
\DefMacro{exp-java-alibaba_fastjson-overhead-unit-and-inline-dup100}{0.184}
\DefMacro{exp-java-alibaba_nacos-time-vanilla-dup100}{249.59}
\DefMacro{exp-java-alibaba_nacos-time-unit-dup100}{249.86}
\DefMacro{exp-java-alibaba_nacos-time-inline-dup100}{0.69}
\DefMacro{exp-java-alibaba_nacos-time-unit-and-inline-dup100}{250.54}
\DefMacro{exp-java-alibaba_nacos-num-vanilla-dup100}{971}
\DefMacro{exp-java-alibaba_nacos-num-unit-dup100}{971}
\DefMacro{exp-java-alibaba_nacos-num-inline-dup100}{100}
\DefMacro{exp-java-alibaba_nacos-num-unit-and-inline-dup100}{1,071}
\DefMacro{exp-java-alibaba_nacos-timept-vanilla-dup100}{0.257}
\DefMacro{exp-java-alibaba_nacos-timept-unit-dup100}{0.257}
\DefMacro{exp-java-alibaba_nacos-timept-inline-dup100}{0.007}
\DefMacro{exp-java-alibaba_nacos-timept-unit-and-inline-dup100}{0.234}
\DefMacro{exp-java-alibaba_nacos-overhead-unit-dup100}{0.001}
\DefMacro{exp-java-alibaba_nacos-overhead-unit-and-inline-dup100}{0.004}
\DefMacro{exp-java-apache_dubbo-time-vanilla-dup100}{678.24}
\DefMacro{exp-java-apache_dubbo-time-unit-dup100}{680.08}
\DefMacro{exp-java-apache_dubbo-time-inline-dup100}{0.74}
\DefMacro{exp-java-apache_dubbo-time-unit-and-inline-dup100}{680.81}
\DefMacro{exp-java-apache_dubbo-num-vanilla-dup100}{3,180}
\DefMacro{exp-java-apache_dubbo-num-unit-dup100}{3,180}
\DefMacro{exp-java-apache_dubbo-num-inline-dup100}{100}
\DefMacro{exp-java-apache_dubbo-num-unit-and-inline-dup100}{3,280}
\DefMacro{exp-java-apache_dubbo-timept-vanilla-dup100}{0.213}
\DefMacro{exp-java-apache_dubbo-timept-unit-dup100}{0.214}
\DefMacro{exp-java-apache_dubbo-timept-inline-dup100}{0.007}
\DefMacro{exp-java-apache_dubbo-timept-unit-and-inline-dup100}{0.208}
\DefMacro{exp-java-apache_dubbo-overhead-unit-dup100}{0.003}
\DefMacro{exp-java-apache_dubbo-overhead-unit-and-inline-dup100}{0.004}
\DefMacro{exp-java-apache_kafka-time-vanilla-dup100}{9.93}
\DefMacro{exp-java-apache_kafka-time-unit-dup100}{10.13}
\DefMacro{exp-java-apache_kafka-time-inline-dup100}{0.72}
\DefMacro{exp-java-apache_kafka-time-unit-and-inline-dup100}{10.86}
\DefMacro{exp-java-apache_kafka-num-vanilla-dup100}{221}
\DefMacro{exp-java-apache_kafka-num-unit-dup100}{221}
\DefMacro{exp-java-apache_kafka-num-inline-dup100}{100}
\DefMacro{exp-java-apache_kafka-num-unit-and-inline-dup100}{321}
\DefMacro{exp-java-apache_kafka-timept-vanilla-dup100}{0.045}
\DefMacro{exp-java-apache_kafka-timept-unit-dup100}{0.046}
\DefMacro{exp-java-apache_kafka-timept-inline-dup100}{0.007}
\DefMacro{exp-java-apache_kafka-timept-unit-and-inline-dup100}{0.034}
\DefMacro{exp-java-apache_kafka-overhead-unit-dup100}{0.021}
\DefMacro{exp-java-apache_kafka-overhead-unit-and-inline-dup100}{0.094}
\DefMacro{exp-java-apache_shardingsphere-time-vanilla-dup100}{5.08}
\DefMacro{exp-java-apache_shardingsphere-time-unit-dup100}{5.12}
\DefMacro{exp-java-apache_shardingsphere-time-inline-dup100}{0.72}
\DefMacro{exp-java-apache_shardingsphere-time-unit-and-inline-dup100}{5.84}
\DefMacro{exp-java-apache_shardingsphere-num-vanilla-dup100}{44}
\DefMacro{exp-java-apache_shardingsphere-num-unit-dup100}{44}
\DefMacro{exp-java-apache_shardingsphere-num-inline-dup100}{200}
\DefMacro{exp-java-apache_shardingsphere-num-unit-and-inline-dup100}{244}
\DefMacro{exp-java-apache_shardingsphere-timept-vanilla-dup100}{0.116}
\DefMacro{exp-java-apache_shardingsphere-timept-unit-dup100}{0.116}
\DefMacro{exp-java-apache_shardingsphere-timept-inline-dup100}{0.004}
\DefMacro{exp-java-apache_shardingsphere-timept-unit-and-inline-dup100}{0.024}
\DefMacro{exp-java-apache_shardingsphere-overhead-unit-dup100}{0.007}
\DefMacro{exp-java-apache_shardingsphere-overhead-unit-and-inline-dup100}{0.149}
\DefMacro{exp-java-jenkinsci_jenkins-time-vanilla-dup100}{4.65}
\DefMacro{exp-java-jenkinsci_jenkins-time-unit-dup100}{4.68}
\DefMacro{exp-java-jenkinsci_jenkins-time-inline-dup100}{0.72}
\DefMacro{exp-java-jenkinsci_jenkins-time-unit-and-inline-dup100}{5.40}
\DefMacro{exp-java-jenkinsci_jenkins-num-vanilla-dup100}{32}
\DefMacro{exp-java-jenkinsci_jenkins-num-unit-dup100}{32}
\DefMacro{exp-java-jenkinsci_jenkins-num-inline-dup100}{200}
\DefMacro{exp-java-jenkinsci_jenkins-num-unit-and-inline-dup100}{232}
\DefMacro{exp-java-jenkinsci_jenkins-timept-vanilla-dup100}{0.145}
\DefMacro{exp-java-jenkinsci_jenkins-timept-unit-dup100}{0.146}
\DefMacro{exp-java-jenkinsci_jenkins-timept-inline-dup100}{0.004}
\DefMacro{exp-java-jenkinsci_jenkins-timept-unit-and-inline-dup100}{0.023}
\DefMacro{exp-java-jenkinsci_jenkins-overhead-unit-dup100}{0.006}
\DefMacro{exp-java-jenkinsci_jenkins-overhead-unit-and-inline-dup100}{0.162}
\DefMacro{exp-java-skylot_jadx-time-vanilla-dup100}{66.76}
\DefMacro{exp-java-skylot_jadx-time-unit-dup100}{66.17}
\DefMacro{exp-java-skylot_jadx-time-inline-dup100}{0.74}
\DefMacro{exp-java-skylot_jadx-time-unit-and-inline-dup100}{66.91}
\DefMacro{exp-java-skylot_jadx-num-vanilla-dup100}{709}
\DefMacro{exp-java-skylot_jadx-num-unit-dup100}{709}
\DefMacro{exp-java-skylot_jadx-num-inline-dup100}{100}
\DefMacro{exp-java-skylot_jadx-num-unit-and-inline-dup100}{809}
\DefMacro{exp-java-skylot_jadx-timept-vanilla-dup100}{0.094}
\DefMacro{exp-java-skylot_jadx-timept-unit-dup100}{0.093}
\DefMacro{exp-java-skylot_jadx-timept-inline-dup100}{0.007}
\DefMacro{exp-java-skylot_jadx-timept-unit-and-inline-dup100}{0.083}
\DefMacro{exp-java-skylot_jadx-overhead-unit-dup100}{-0.009}
\DefMacro{exp-java-skylot_jadx-overhead-unit-and-inline-dup100}{0.002}
\DefMacro{exp-java-time-vanilla-AVG-dup100}{151.30}
\DefMacro{exp-java-time-vanilla-SUM-dup100}{1,059.11}
\DefMacro{exp-java-time-vanilla-MAX-dup100}{678.24}
\DefMacro{exp-java-time-vanilla-MIN-dup100}{4.65}
\DefMacro{exp-java-time-vanilla-MEDIAN-dup100}{44.86}
\DefMacro{exp-java-time-vanilla-STDEV-dup100}{229.49}
\DefMacro{exp-java-time-vanilla-CNT-dup100}{7}
\DefMacro{exp-java-time-unit-AVG-dup100}{152.63}
\DefMacro{exp-java-time-unit-SUM-dup100}{1,068.44}
\DefMacro{exp-java-time-unit-MAX-dup100}{680.08}
\DefMacro{exp-java-time-unit-MIN-dup100}{4.68}
\DefMacro{exp-java-time-unit-MEDIAN-dup100}{52.40}
\DefMacro{exp-java-time-unit-STDEV-dup100}{229.64}
\DefMacro{exp-java-time-unit-CNT-dup100}{7}
\DefMacro{exp-java-time-inline-AVG-dup100}{0.72}
\DefMacro{exp-java-time-inline-SUM-dup100}{5.06}
\DefMacro{exp-java-time-inline-MAX-dup100}{0.74}
\DefMacro{exp-java-time-inline-MIN-dup100}{0.69}
\DefMacro{exp-java-time-inline-MEDIAN-dup100}{0.72}
\DefMacro{exp-java-time-inline-STDEV-dup100}{0.01}
\DefMacro{exp-java-time-inline-CNT-dup100}{7}
\DefMacro{exp-java-time-unit-and-inline-AVG-dup100}{153.36}
\DefMacro{exp-java-time-unit-and-inline-SUM-dup100}{1,073.50}
\DefMacro{exp-java-time-unit-and-inline-MAX-dup100}{680.81}
\DefMacro{exp-java-time-unit-and-inline-MIN-dup100}{5.40}
\DefMacro{exp-java-time-unit-and-inline-MEDIAN-dup100}{53.13}
\DefMacro{exp-java-time-unit-and-inline-STDEV-dup100}{229.64}
\DefMacro{exp-java-time-unit-and-inline-CNT-dup100}{7}
\DefMacro{exp-java-num-vanilla-AVG-dup100}{1,454.14}
\DefMacro{exp-java-num-vanilla-SUM-dup100}{10,179}
\DefMacro{exp-java-num-vanilla-MAX-dup100}{5,022}
\DefMacro{exp-java-num-vanilla-MIN-dup100}{32}
\DefMacro{exp-java-num-vanilla-MEDIAN-dup100}{709.00}
\DefMacro{exp-java-num-vanilla-STDEV-dup100}{1,773.96}
\DefMacro{exp-java-num-vanilla-CNT-dup100}{7}
\DefMacro{exp-java-num-unit-AVG-dup100}{1,454.14}
\DefMacro{exp-java-num-unit-SUM-dup100}{10,179}
\DefMacro{exp-java-num-unit-MAX-dup100}{5,022}
\DefMacro{exp-java-num-unit-MIN-dup100}{32}
\DefMacro{exp-java-num-unit-MEDIAN-dup100}{709.00}
\DefMacro{exp-java-num-unit-STDEV-dup100}{1,773.96}
\DefMacro{exp-java-num-unit-CNT-dup100}{7}
\DefMacro{exp-java-num-inline-AVG-dup100}{142.86}
\DefMacro{exp-java-num-inline-SUM-dup100}{1,000}
\DefMacro{exp-java-num-inline-MAX-dup100}{200}
\DefMacro{exp-java-num-inline-MIN-dup100}{100}
\DefMacro{exp-java-num-inline-MEDIAN-dup100}{100.00}
\DefMacro{exp-java-num-inline-STDEV-dup100}{49.49}
\DefMacro{exp-java-num-inline-CNT-dup100}{7}
\DefMacro{exp-java-num-unit-and-inline-AVG-dup100}{1,597.00}
\DefMacro{exp-java-num-unit-and-inline-SUM-dup100}{11,179}
\DefMacro{exp-java-num-unit-and-inline-MAX-dup100}{5,222}
\DefMacro{exp-java-num-unit-and-inline-MIN-dup100}{232}
\DefMacro{exp-java-num-unit-and-inline-MEDIAN-dup100}{809.00}
\DefMacro{exp-java-num-unit-and-inline-STDEV-dup100}{1,780.56}
\DefMacro{exp-java-num-unit-and-inline-CNT-dup100}{7}
\DefMacro{exp-java-timept-vanilla-AVG-dup100}{0.126}
\DefMacro{exp-java-timept-vanilla-SUM-dup100}{0.879}
\DefMacro{exp-java-timept-vanilla-MAX-dup100}{0.257}
\DefMacro{exp-java-timept-vanilla-MIN-dup100}{0.009}
\DefMacro{exp-java-timept-vanilla-MEDIAN-dup100}{0.116}
\DefMacro{exp-java-timept-vanilla-STDEV-dup100}{0.082}
\DefMacro{exp-java-timept-vanilla-CNT-dup100}{7}
\DefMacro{exp-java-timept-unit-AVG-dup100}{0.126}
\DefMacro{exp-java-timept-unit-SUM-dup100}{0.883}
\DefMacro{exp-java-timept-unit-MAX-dup100}{0.257}
\DefMacro{exp-java-timept-unit-MIN-dup100}{0.010}
\DefMacro{exp-java-timept-unit-MEDIAN-dup100}{0.116}
\DefMacro{exp-java-timept-unit-STDEV-dup100}{0.081}
\DefMacro{exp-java-timept-unit-CNT-dup100}{7}
\DefMacro{exp-java-timept-inline-AVG-dup100}{0.006}
\DefMacro{exp-java-timept-inline-SUM-dup100}{0.040}
\DefMacro{exp-java-timept-inline-MAX-dup100}{0.007}
\DefMacro{exp-java-timept-inline-MIN-dup100}{0.004}
\DefMacro{exp-java-timept-inline-MEDIAN-dup100}{0.007}
\DefMacro{exp-java-timept-inline-STDEV-dup100}{0.002}
\DefMacro{exp-java-timept-inline-CNT-dup100}{7}
\DefMacro{exp-java-timept-unit-and-inline-AVG-dup100}{0.088}
\DefMacro{exp-java-timept-unit-and-inline-SUM-dup100}{0.615}
\DefMacro{exp-java-timept-unit-and-inline-MAX-dup100}{0.234}
\DefMacro{exp-java-timept-unit-and-inline-MIN-dup100}{0.010}
\DefMacro{exp-java-timept-unit-and-inline-MEDIAN-dup100}{0.034}
\DefMacro{exp-java-timept-unit-and-inline-STDEV-dup100}{0.087}
\DefMacro{exp-java-timept-unit-and-inline-CNT-dup100}{7}
\DefMacro{exp-java-overhead-unit-AVG-dup100}{0.028}
\DefMacro{exp-java-overhead-unit-SUM-dup100}{0.197}
\DefMacro{exp-java-overhead-unit-MAX-dup100}{0.168}
\DefMacro{exp-java-overhead-unit-MIN-dup100}{-0.009}
\DefMacro{exp-java-overhead-unit-MEDIAN-dup100}{0.006}
\DefMacro{exp-java-overhead-unit-STDEV-dup100}{0.058}
\DefMacro{exp-java-overhead-unit-CNT-dup100}{7}
\DefMacro{exp-java-overhead-unit-and-inline-AVG-dup100}{0.086}
\DefMacro{exp-java-overhead-unit-and-inline-SUM-dup100}{0.599}
\DefMacro{exp-java-overhead-unit-and-inline-MAX-dup100}{0.184}
\DefMacro{exp-java-overhead-unit-and-inline-MIN-dup100}{0.002}
\DefMacro{exp-java-overhead-unit-and-inline-MEDIAN-dup100}{0.094}
\DefMacro{exp-java-overhead-unit-and-inline-STDEV-dup100}{0.076}
\DefMacro{exp-java-overhead-unit-and-inline-CNT-dup100}{7}
\DefMacro{exp-java-timept-vanilla-MACROAVG-dup100}{0.104}
\DefMacro{exp-java-timept-unit-MACROAVG-dup100}{0.105}
\DefMacro{exp-java-timept-inline-MACROAVG-dup100}{0.005}
\DefMacro{exp-java-timept-unit-and-inline-MACROAVG-dup100}{0.096}
\DefMacro{exp-java-overhead-unit-MACROAVG-dup100}{0.009}
\DefMacro{exp-java-overhead-unit-and-inline-MACROAVG-dup100}{0.014}
\DefMacro{exp-java-alibaba_fastjson-time-vanilla-dup1000}{44.82}
\DefMacro{exp-java-alibaba_fastjson-time-unit-dup1000}{ERROR}
\DefMacro{exp-java-alibaba_fastjson-time-inline-dup1000}{ERROR}
\DefMacro{exp-java-alibaba_fastjson-time-unit-and-inline-dup1000}{ERROR}
\DefMacro{exp-java-alibaba_fastjson-num-vanilla-dup1000}{5,022}
\DefMacro{exp-java-alibaba_fastjson-num-unit-dup1000}{ERROR}
\DefMacro{exp-java-alibaba_fastjson-num-inline-dup1000}{ERROR}
\DefMacro{exp-java-alibaba_fastjson-num-unit-and-inline-dup1000}{ERROR}
\DefMacro{exp-java-alibaba_fastjson-timept-vanilla-dup1000}{ERROR}
\DefMacro{exp-java-alibaba_fastjson-timept-unit-dup1000}{ERROR}
\DefMacro{exp-java-alibaba_fastjson-timept-inline-dup1000}{ERROR}
\DefMacro{exp-java-alibaba_fastjson-timept-unit-and-inline-dup1000}{ERROR}
\DefMacro{exp-java-alibaba_fastjson-overhead-unit-dup1000}{ERROR}
\DefMacro{exp-java-alibaba_fastjson-overhead-unit-and-inline-dup1000}{ERROR}
\DefMacro{exp-java-alibaba_nacos-time-vanilla-dup1000}{249.34}
\DefMacro{exp-java-alibaba_nacos-time-unit-dup1000}{249.96}
\DefMacro{exp-java-alibaba_nacos-time-inline-dup1000}{0.71}
\DefMacro{exp-java-alibaba_nacos-time-unit-and-inline-dup1000}{250.67}
\DefMacro{exp-java-alibaba_nacos-num-vanilla-dup1000}{971}
\DefMacro{exp-java-alibaba_nacos-num-unit-dup1000}{971}
\DefMacro{exp-java-alibaba_nacos-num-inline-dup1000}{1,000}
\DefMacro{exp-java-alibaba_nacos-num-unit-and-inline-dup1000}{1,971}
\DefMacro{exp-java-alibaba_nacos-timept-vanilla-dup1000}{0.257}
\DefMacro{exp-java-alibaba_nacos-timept-unit-dup1000}{0.257}
\DefMacro{exp-java-alibaba_nacos-timept-inline-dup1000}{0.001}
\DefMacro{exp-java-alibaba_nacos-timept-unit-and-inline-dup1000}{0.127}
\DefMacro{exp-java-alibaba_nacos-overhead-unit-dup1000}{0.003}
\DefMacro{exp-java-alibaba_nacos-overhead-unit-and-inline-dup1000}{0.005}
\DefMacro{exp-java-apache_dubbo-time-vanilla-dup1000}{678.03}
\DefMacro{exp-java-apache_dubbo-time-unit-dup1000}{679.09}
\DefMacro{exp-java-apache_dubbo-time-inline-dup1000}{0.71}
\DefMacro{exp-java-apache_dubbo-time-unit-and-inline-dup1000}{679.79}
\DefMacro{exp-java-apache_dubbo-num-vanilla-dup1000}{3,180}
\DefMacro{exp-java-apache_dubbo-num-unit-dup1000}{3,180}
\DefMacro{exp-java-apache_dubbo-num-inline-dup1000}{1,000}
\DefMacro{exp-java-apache_dubbo-num-unit-and-inline-dup1000}{4,180}
\DefMacro{exp-java-apache_dubbo-timept-vanilla-dup1000}{0.213}
\DefMacro{exp-java-apache_dubbo-timept-unit-dup1000}{0.214}
\DefMacro{exp-java-apache_dubbo-timept-inline-dup1000}{0.001}
\DefMacro{exp-java-apache_dubbo-timept-unit-and-inline-dup1000}{0.163}
\DefMacro{exp-java-apache_dubbo-overhead-unit-dup1000}{0.002}
\DefMacro{exp-java-apache_dubbo-overhead-unit-and-inline-dup1000}{0.003}
\DefMacro{exp-java-apache_kafka-time-vanilla-dup1000}{9.92}
\DefMacro{exp-java-apache_kafka-time-unit-dup1000}{ERROR}
\DefMacro{exp-java-apache_kafka-time-inline-dup1000}{ERROR}
\DefMacro{exp-java-apache_kafka-time-unit-and-inline-dup1000}{ERROR}
\DefMacro{exp-java-apache_kafka-num-vanilla-dup1000}{221}
\DefMacro{exp-java-apache_kafka-num-unit-dup1000}{ERROR}
\DefMacro{exp-java-apache_kafka-num-inline-dup1000}{ERROR}
\DefMacro{exp-java-apache_kafka-num-unit-and-inline-dup1000}{ERROR}
\DefMacro{exp-java-apache_kafka-timept-vanilla-dup1000}{ERROR}
\DefMacro{exp-java-apache_kafka-timept-unit-dup1000}{ERROR}
\DefMacro{exp-java-apache_kafka-timept-inline-dup1000}{ERROR}
\DefMacro{exp-java-apache_kafka-timept-unit-and-inline-dup1000}{ERROR}
\DefMacro{exp-java-apache_kafka-overhead-unit-dup1000}{ERROR}
\DefMacro{exp-java-apache_kafka-overhead-unit-and-inline-dup1000}{ERROR}
\DefMacro{exp-java-apache_shardingsphere-time-vanilla-dup1000}{5.06}
\DefMacro{exp-java-apache_shardingsphere-time-unit-dup1000}{5.09}
\DefMacro{exp-java-apache_shardingsphere-time-inline-dup1000}{0.78}
\DefMacro{exp-java-apache_shardingsphere-time-unit-and-inline-dup1000}{5.87}
\DefMacro{exp-java-apache_shardingsphere-num-vanilla-dup1000}{44}
\DefMacro{exp-java-apache_shardingsphere-num-unit-dup1000}{44}
\DefMacro{exp-java-apache_shardingsphere-num-inline-dup1000}{2,000}
\DefMacro{exp-java-apache_shardingsphere-num-unit-and-inline-dup1000}{2,044}
\DefMacro{exp-java-apache_shardingsphere-timept-vanilla-dup1000}{0.115}
\DefMacro{exp-java-apache_shardingsphere-timept-unit-dup1000}{0.116}
\DefMacro{exp-java-apache_shardingsphere-timept-inline-dup1000}{0.000}
\DefMacro{exp-java-apache_shardingsphere-timept-unit-and-inline-dup1000}{0.003}
\DefMacro{exp-java-apache_shardingsphere-overhead-unit-dup1000}{0.005}
\DefMacro{exp-java-apache_shardingsphere-overhead-unit-and-inline-dup1000}{0.160}
\DefMacro{exp-java-jenkinsci_jenkins-time-vanilla-dup1000}{4.70}
\DefMacro{exp-java-jenkinsci_jenkins-time-unit-dup1000}{4.67}
\DefMacro{exp-java-jenkinsci_jenkins-time-inline-dup1000}{0.71}
\DefMacro{exp-java-jenkinsci_jenkins-time-unit-and-inline-dup1000}{5.38}
\DefMacro{exp-java-jenkinsci_jenkins-num-vanilla-dup1000}{32}
\DefMacro{exp-java-jenkinsci_jenkins-num-unit-dup1000}{32}
\DefMacro{exp-java-jenkinsci_jenkins-num-inline-dup1000}{2,000}
\DefMacro{exp-java-jenkinsci_jenkins-num-unit-and-inline-dup1000}{2,032}
\DefMacro{exp-java-jenkinsci_jenkins-timept-vanilla-dup1000}{0.147}
\DefMacro{exp-java-jenkinsci_jenkins-timept-unit-dup1000}{0.146}
\DefMacro{exp-java-jenkinsci_jenkins-timept-inline-dup1000}{0.000}
\DefMacro{exp-java-jenkinsci_jenkins-timept-unit-and-inline-dup1000}{0.003}
\DefMacro{exp-java-jenkinsci_jenkins-overhead-unit-dup1000}{-0.006}
\DefMacro{exp-java-jenkinsci_jenkins-overhead-unit-and-inline-dup1000}{0.145}
\DefMacro{exp-java-skylot_jadx-time-vanilla-dup1000}{67.11}
\DefMacro{exp-java-skylot_jadx-time-unit-dup1000}{69.73}
\DefMacro{exp-java-skylot_jadx-time-inline-dup1000}{0.70}
\DefMacro{exp-java-skylot_jadx-time-unit-and-inline-dup1000}{70.44}
\DefMacro{exp-java-skylot_jadx-num-vanilla-dup1000}{709}
\DefMacro{exp-java-skylot_jadx-num-unit-dup1000}{709}
\DefMacro{exp-java-skylot_jadx-num-inline-dup1000}{1,000}
\DefMacro{exp-java-skylot_jadx-num-unit-and-inline-dup1000}{1,709}
\DefMacro{exp-java-skylot_jadx-timept-vanilla-dup1000}{0.095}
\DefMacro{exp-java-skylot_jadx-timept-unit-dup1000}{0.098}
\DefMacro{exp-java-skylot_jadx-timept-inline-dup1000}{0.001}
\DefMacro{exp-java-skylot_jadx-timept-unit-and-inline-dup1000}{0.041}
\DefMacro{exp-java-skylot_jadx-overhead-unit-dup1000}{0.039}
\DefMacro{exp-java-skylot_jadx-overhead-unit-and-inline-dup1000}{0.050}
\DefMacro{exp-java-time-vanilla-AVG-dup1000}{200.85}
\DefMacro{exp-java-time-vanilla-SUM-dup1000}{1,004.24}
\DefMacro{exp-java-time-vanilla-MAX-dup1000}{678.03}
\DefMacro{exp-java-time-vanilla-MIN-dup1000}{4.70}
\DefMacro{exp-java-time-vanilla-MEDIAN-dup1000}{67.11}
\DefMacro{exp-java-time-vanilla-STDEV-dup1000}{254.85}
\DefMacro{exp-java-time-vanilla-CNT-dup1000}{5}
\DefMacro{exp-java-time-unit-AVG-dup1000}{201.71}
\DefMacro{exp-java-time-unit-SUM-dup1000}{1,008.55}
\DefMacro{exp-java-time-unit-MAX-dup1000}{679.09}
\DefMacro{exp-java-time-unit-MIN-dup1000}{4.67}
\DefMacro{exp-java-time-unit-MEDIAN-dup1000}{69.73}
\DefMacro{exp-java-time-unit-STDEV-dup1000}{255.00}
\DefMacro{exp-java-time-unit-CNT-dup1000}{5}
\DefMacro{exp-java-time-inline-AVG-dup1000}{0.72}
\DefMacro{exp-java-time-inline-SUM-dup1000}{3.61}
\DefMacro{exp-java-time-inline-MAX-dup1000}{0.78}
\DefMacro{exp-java-time-inline-MIN-dup1000}{0.70}
\DefMacro{exp-java-time-inline-MEDIAN-dup1000}{0.71}
\DefMacro{exp-java-time-inline-STDEV-dup1000}{0.03}
\DefMacro{exp-java-time-inline-CNT-dup1000}{5}
\DefMacro{exp-java-time-unit-and-inline-AVG-dup1000}{202.43}
\DefMacro{exp-java-time-unit-and-inline-SUM-dup1000}{1,012.16}
\DefMacro{exp-java-time-unit-and-inline-MAX-dup1000}{679.79}
\DefMacro{exp-java-time-unit-and-inline-MIN-dup1000}{5.38}
\DefMacro{exp-java-time-unit-and-inline-MEDIAN-dup1000}{70.44}
\DefMacro{exp-java-time-unit-and-inline-STDEV-dup1000}{254.99}
\DefMacro{exp-java-time-unit-and-inline-CNT-dup1000}{5}
\DefMacro{exp-java-num-vanilla-AVG-dup1000}{987.20}
\DefMacro{exp-java-num-vanilla-SUM-dup1000}{4,936}
\DefMacro{exp-java-num-vanilla-MAX-dup1000}{3,180}
\DefMacro{exp-java-num-vanilla-MIN-dup1000}{32}
\DefMacro{exp-java-num-vanilla-MEDIAN-dup1000}{709.00}
\DefMacro{exp-java-num-vanilla-STDEV-dup1000}{1,156.55}
\DefMacro{exp-java-num-vanilla-CNT-dup1000}{5}
\DefMacro{exp-java-num-unit-AVG-dup1000}{987.20}
\DefMacro{exp-java-num-unit-SUM-dup1000}{4,936}
\DefMacro{exp-java-num-unit-MAX-dup1000}{3,180}
\DefMacro{exp-java-num-unit-MIN-dup1000}{32}
\DefMacro{exp-java-num-unit-MEDIAN-dup1000}{709.00}
\DefMacro{exp-java-num-unit-STDEV-dup1000}{1,156.55}
\DefMacro{exp-java-num-unit-CNT-dup1000}{5}
\DefMacro{exp-java-num-inline-AVG-dup1000}{1,400.00}
\DefMacro{exp-java-num-inline-SUM-dup1000}{7,000}
\DefMacro{exp-java-num-inline-MAX-dup1000}{2,000}
\DefMacro{exp-java-num-inline-MIN-dup1000}{1,000}
\DefMacro{exp-java-num-inline-MEDIAN-dup1000}{1,000.00}
\DefMacro{exp-java-num-inline-STDEV-dup1000}{489.90}
\DefMacro{exp-java-num-inline-CNT-dup1000}{5}
\DefMacro{exp-java-num-unit-and-inline-AVG-dup1000}{2,387.20}
\DefMacro{exp-java-num-unit-and-inline-SUM-dup1000}{11,936}
\DefMacro{exp-java-num-unit-and-inline-MAX-dup1000}{4,180}
\DefMacro{exp-java-num-unit-and-inline-MIN-dup1000}{1,709}
\DefMacro{exp-java-num-unit-and-inline-MEDIAN-dup1000}{2,032.00}
\DefMacro{exp-java-num-unit-and-inline-STDEV-dup1000}{904.57}
\DefMacro{exp-java-num-unit-and-inline-CNT-dup1000}{5}
\DefMacro{exp-java-timept-vanilla-AVG-dup1000}{0.165}
\DefMacro{exp-java-timept-vanilla-SUM-dup1000}{0.827}
\DefMacro{exp-java-timept-vanilla-MAX-dup1000}{0.257}
\DefMacro{exp-java-timept-vanilla-MIN-dup1000}{0.095}
\DefMacro{exp-java-timept-vanilla-MEDIAN-dup1000}{0.147}
\DefMacro{exp-java-timept-vanilla-STDEV-dup1000}{0.061}
\DefMacro{exp-java-timept-vanilla-CNT-dup1000}{5}
\DefMacro{exp-java-timept-unit-AVG-dup1000}{0.166}
\DefMacro{exp-java-timept-unit-SUM-dup1000}{0.831}
\DefMacro{exp-java-timept-unit-MAX-dup1000}{0.257}
\DefMacro{exp-java-timept-unit-MIN-dup1000}{0.098}
\DefMacro{exp-java-timept-unit-MEDIAN-dup1000}{0.146}
\DefMacro{exp-java-timept-unit-STDEV-dup1000}{0.060}
\DefMacro{exp-java-timept-unit-CNT-dup1000}{5}
\DefMacro{exp-java-timept-inline-AVG-dup1000}{0.001}
\DefMacro{exp-java-timept-inline-SUM-dup1000}{0.003}
\DefMacro{exp-java-timept-inline-MAX-dup1000}{0.001}
\DefMacro{exp-java-timept-inline-MIN-dup1000}{0.000}
\DefMacro{exp-java-timept-inline-MEDIAN-dup1000}{0.001}
\DefMacro{exp-java-timept-inline-STDEV-dup1000}{0.000}
\DefMacro{exp-java-timept-inline-CNT-dup1000}{5}
\DefMacro{exp-java-timept-unit-and-inline-AVG-dup1000}{0.067}
\DefMacro{exp-java-timept-unit-and-inline-SUM-dup1000}{0.337}
\DefMacro{exp-java-timept-unit-and-inline-MAX-dup1000}{0.163}
\DefMacro{exp-java-timept-unit-and-inline-MIN-dup1000}{0.003}
\DefMacro{exp-java-timept-unit-and-inline-MEDIAN-dup1000}{0.041}
\DefMacro{exp-java-timept-unit-and-inline-STDEV-dup1000}{0.066}
\DefMacro{exp-java-timept-unit-and-inline-CNT-dup1000}{5}
\DefMacro{exp-java-overhead-unit-AVG-dup1000}{0.009}
\DefMacro{exp-java-overhead-unit-SUM-dup1000}{0.043}
\DefMacro{exp-java-overhead-unit-MAX-dup1000}{0.039}
\DefMacro{exp-java-overhead-unit-MIN-dup1000}{-0.006}
\DefMacro{exp-java-overhead-unit-MEDIAN-dup1000}{0.003}
\DefMacro{exp-java-overhead-unit-STDEV-dup1000}{0.016}
\DefMacro{exp-java-overhead-unit-CNT-dup1000}{5}
\DefMacro{exp-java-overhead-unit-and-inline-AVG-dup1000}{0.073}
\DefMacro{exp-java-overhead-unit-and-inline-SUM-dup1000}{0.363}
\DefMacro{exp-java-overhead-unit-and-inline-MAX-dup1000}{0.160}
\DefMacro{exp-java-overhead-unit-and-inline-MIN-dup1000}{0.003}
\DefMacro{exp-java-overhead-unit-and-inline-MEDIAN-dup1000}{0.050}
\DefMacro{exp-java-overhead-unit-and-inline-STDEV-dup1000}{0.068}
\DefMacro{exp-java-overhead-unit-and-inline-CNT-dup1000}{5}
\DefMacro{exp-java-timept-vanilla-MACROAVG-dup1000}{0.203}
\DefMacro{exp-java-timept-unit-MACROAVG-dup1000}{0.204}
\DefMacro{exp-java-timept-inline-MACROAVG-dup1000}{0.001}
\DefMacro{exp-java-timept-unit-and-inline-MACROAVG-dup1000}{0.085}
\DefMacro{exp-java-overhead-unit-MACROAVG-dup1000}{0.004}
\DefMacro{exp-java-overhead-unit-and-inline-MACROAVG-dup1000}{0.008}

\DefMacro{NumTotalProjs}{68}
\DefMacro{NumTotalFiles}{100}
\DefMacro{NumTotalInlineTests}{152}
\DefMacro{NumTotalTestedStmts}{144}
\DefMacro{exp-total-proj-can-run-unit-tests}{21}
\DefMacro{exp-java-skipped-all}{30}
\DefMacro{exp-total-skipped-all}{47}
\DefMacro{exp-python-skipped-all}{17}

\newcommand{\IncludingComment}[1]{\vspace{2pt}\noindent\hspace{5pt}#1\vspace{2pt}}

\copyrightyear{2022}
\acmYear{2022}
\setcopyright{acmlicensed}\acmConference[ASE '22]{37th IEEE/ACM International Conference on Automated Software Engineering}{October 10--14, 2022}{Rochester, MI, USA}
\acmBooktitle{37th IEEE/ACM International Conference on Automated Software Engineering (ASE '22), October 10--14, 2022, Rochester, MI, USA}
\acmPrice{15.00}
\acmDOI{10.1145/3551349.3556952}
\acmISBN{978-1-4503-9475-8/22/10}

\begin{document}

\title{\Title}

\author{Yu Liu}
\affiliation{
  \mbox{
    \institution{UT Austin}
    \country{USA}
  }
}
\email{yuki.liu@utexas.edu}

\author{Pengyu Nie}
\affiliation{
  \mbox{
    \institution{UT Austin}
    \country{USA}
  }
}
\email{pynie@utexas.edu}

\author{Owolabi Legunsen}
\affiliation{
  \mbox{
    \institution{Cornell University}
    \country{USA}
  }
}
\email{legunsen@cornell.edu}

\author{Milos Gligoric}
\affiliation{
  \mbox{
    \institution{UT Austin}
    \country{USA}
  }
}
\email{gligoric@utexas.edu}

\begin{abstract}
  Unit tests are widely used to check source code quality, but they
  can be too coarse-grained or ill-suited for testing individual
  program statements. We introduce \emph{\inlinetests} to make it
  easier to check for faults in statements.
  We motivate \inlinetests through several language features and a
  common testing scenario in which \inlinetests could be beneficial.  For
  example, \inlinetests can allow a developer to test a regular expression in
  place. We also define language-agnostic \desiderata for
  \inlinetesting frameworks.  Lastly, we implement \InlineTest, the
  first \inlinetesting framework. \InlineTest works for Python and
  Java, and it satisfies most of the \desiderata.
  We evaluate \InlineTest on open-source projects by using it to test
  \UseMacro{NumTotalTestedStmts} statements in \UseMacro{NumPythonProjs} Python programs and
  \UseMacro{NumJavaProjs} Java programs.  We also perform a user study. All
  \UseMacro{NumUserStudyValidResponses} user study participants say that \inlinetests are easy
  to write and that \inlinetesting is beneficial. The cost of running
  \inlinetests is negligible, at \UseMacro{exp-python-overhead-unit-and-inline-MACROAVG-dup1}x--\UseMacro{exp-java-overhead-unit-and-inline-MACROAVG-dup1}x, and our \inlinetests helped
  find two faults that have been fixed by the developers.

\end{abstract}

\begin{CCSXML}
  <ccs2012>
     <concept>
         <concept_id>10011007.10011074.10011099.10011102.10011103</concept_id>
         <concept_desc>Software and its engineering~Software testing and debugging</concept_desc>
         <concept_significance>500</concept_significance>
         </concept>
   </ccs2012>
\end{CCSXML}
  
\ccsdesc[500]{Software and its engineering~Software testing and debugging}

\keywords{inline tests, software testing}

\maketitle

\section{Introduction}
\label{sec:intro}

\begin{figure}[t]
  \includegraphics[width=0.7\columnwidth]{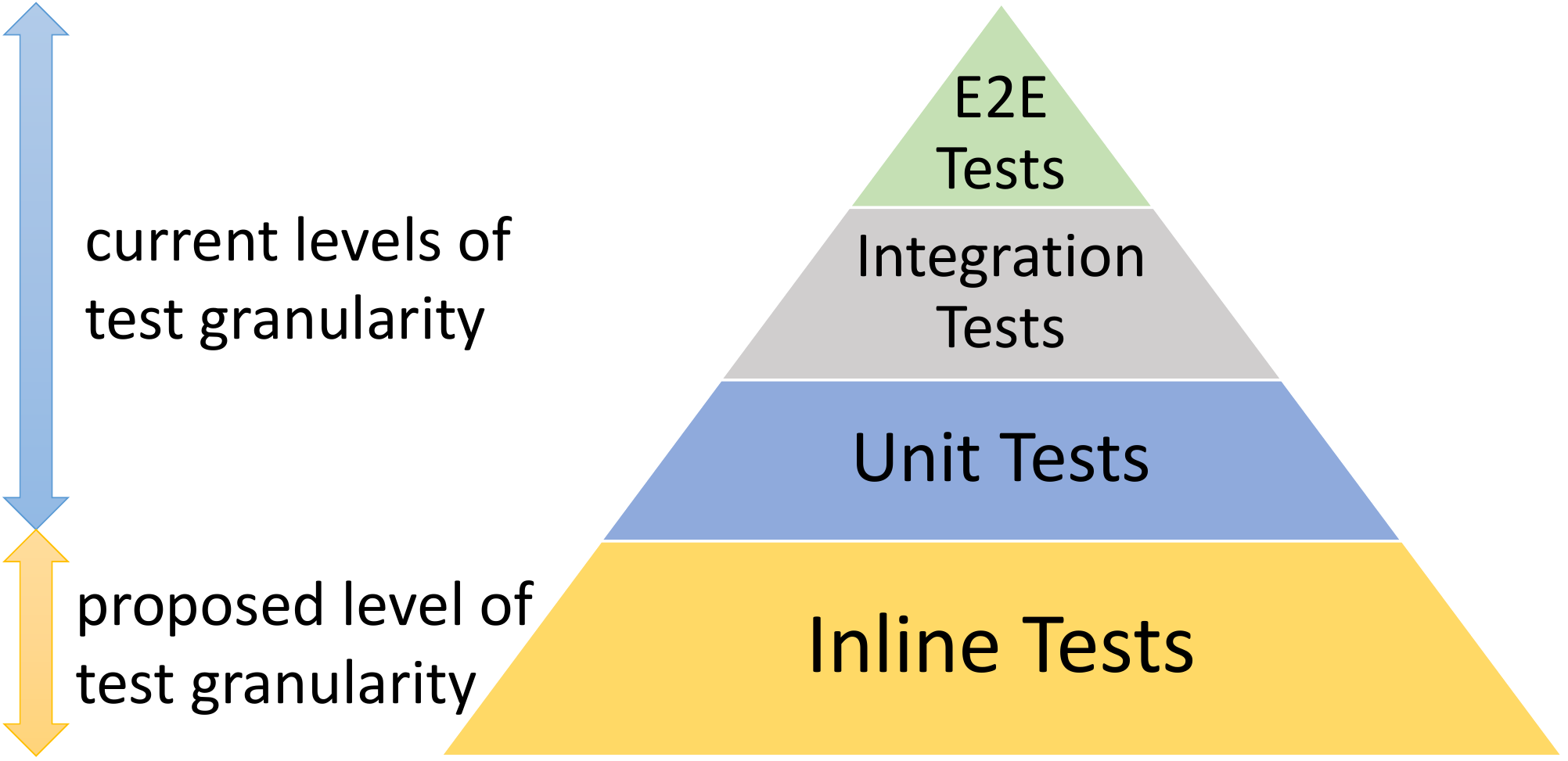}
  \vspace{-8px}
  \caption{Testing pyramid.}
  \label{intro:fig:test-hierarchy}
  \vspace{-10px}
\end{figure}

Testing is essential for checking code quality during software development.
Today, testing frameworks only support three levels of test granularity---unit
testing, integration testing and end-to-end testing. These levels,
shown in the top three layers of Figure~\ref{intro:fig:test-hierarchy} (known as the test pyramid), reflect
developer testing needs. Developers write unit tests to check the correctness
of logical units of functionality, e.g., methods or
functions~\cite{runeson2006survey, daka2014survey}. Integration tests are used to check that logical units interact
correctly~\cite{grechanik2019generating, leung1990study,
  orso1998integration, trautsch2019analysis}.
Developers use end-to-end tests to check if code runs
correctly in its operating environment, and if functional and
non-functional requirements are being met~\cite{tsaiEndToEnd, yandrapally2021mutation}.

Unfortunately, there is little support for developer testing needs
below the unit-test level. Yet, developers may want to test
individual statements for at least four reasons:

\begin{enumerate}[topsep=.2ex,itemsep=.2ex,leftmargin=1.5em]

\item Single-statement bugs occur
  frequently~\cite{karampatsis2020often, kamienski2021pysstubs}, but
  unit tests rarely fail on commits that introduce single-statement
  bugs~\cite{latendresse2021effective}.

\item The statement to be checked, i.e., the \emph{target statement},
  may be buried deeply inside complicated program logic.

\item Developers may want to check and better comprehend
  harder-to-understand traditional programming language features like
  regular expressions (regexes)~\cite{michael2019regexes,
    davis2019testing, davis2018impact, ZhongETAL18NL2Regex,
    KushmanAndBarzilay13GenRegex}, bit manipulation~\cite{bae2019bit,
    lee2016verifying}, and string manipulation~\cite{eghbali2020no,
    larson2003high, ploski2007research}.

\item Recent language features, e.g., Java's stream
  API~\cite{JavaStreamAPI}, allow writing complex program logic in one
  statement where one would previously have written a method that can be unit tested.

\end{enumerate}

Due to the lack of direct support for statement-level testing, developers often resort to
wasteful or \emph{ad hoc} manual approaches. We briefly mention three
of them here and describe them and others in
Section~\ref{sec:example}. First, in the commonly-practiced
``\texttt{printf}
debugging''~\cite{HowDevelopersDebugBellerEtAlICSE2018,
banken2018debugging, perscheid2017studying, ida2012performance,
li2015medic, grabner1995debugging}, developers wastefully add and then
remove print statements to visually check correctness at specific
program points. Second, if the target statement is in privately
accessible code, some developers violate core software engineering
principles to enable checking them with unit tests. For example,
\CodeIn{google/guava}~\cite{guava2009} developers use the
``@VisibleForTesting'' annotation to expose non-public variables or
methods for unit testing~\cite{BadPracticesOfTesting,
CaseForVisibleForTesting}. Lastly, developers lose productivity when
they repeatedly use any of the many third-party websites \cite{regex101Link,
debuggexLink, regexTesterLink} or in-IDE pop-ups like the one in
IntelliJ~\cite{intellijideaRegex} to test regexes.

We argue that there is a need for specialized support to allow testing
individual statements ``in place''. A simple approach is to first
extract the target statement into a method by itself and then write a
unit test for the extracted method. Doing so would not be effective
for three reasons. First, to correctly set up the right state for
testing, developers may have to duplicate code from the method that contains the target
statement in the test for the extracted method. Second, if there are
many target statements, extracting each one can devolve into a
hard-to-maintain ``one unit test per statement'' scenario. Finally,
programs may become harder to comprehend if one has to look up method
bodies to understand individual statements.

We introduce \emph{\inlinetests}, a new kind of tests that makes it easier to
check individual program statements. An \inlinetest is a
statement that allows to provide arbitrary inputs and test oracles
for checking the immediately preceding statement that is not an \inlinetest.
\Inlinetests can be viewed as a way to bring the power of unit tests to the
statement level. Structurally, \inlinetests add a new level of granularity
below unit tests to the testing pyramid in
Figure~\ref{intro:fig:test-hierarchy}.

\Inlinetests could provide software development
benefits beyond testing. For example, prior work showed that tests and code do not
usually co-evolve gracefully~\cite{beller2015and}. Unlike
unit tests, \inlinetests are co-located in the same file as target statements. So, \inlinetests could be easier to co-evolve with code. Prior work also showed
that test coverage can stay stable over time because existing
tests cover newly-added code~\cite{marinescu2014covrig}. \Inlinetests
can help find faults in newly-added code. The
inputs and expected outputs in \inlinetests are a form of
documentation and they could improve code comprehension. Also,
\inlinetests could improve developer productivity by being more
durable and less wasteful than ``\texttt{printf} debugging''.

\Inlinetests are different from the \texttt{assert} construct that
many programming languages provide, e.g.,~\cite{JavaAssert,
  PythonAssert}. \texttt{Assert} statements can enable
\emph{production-time enforcement} of conditions on program state at
given code locations without requiring developer-provided inputs. For
example, an \texttt{assert} can be used to ensure that a variable is
in range, or that a method's return value is not \texttt{null}.
Differently, \inlinetests require developer-provided inputs and
oracles, and they only enable \emph{test-time checking} of individual statements.

We implement \InlineTest, the first \inlinetesting framework. Our
starting point is to define language-agnostic \desiderata for
\inlinetesting frameworks (Section~\ref{sec:desiderata}). For example, it should \emph{not} be
possible to use inline tests in place of unit tests or debuggers. The
\desiderata that we define provide a basis for \InlineTest and they
can provide guidance for the development of future \inlinetesting
frameworks. Our current \InlineTest implementation supports
\inlinetesting for Python and Java, and it satisfies most of the
\desiderata.

We evaluate \InlineTest on open-source projects by using it to test
\UseMacro{NumTotalTestedStmts} statements in \UseMacro{NumPythonProjs}
Python programs and \UseMacro{NumJavaProjs} Java programs. We perform
a user study to assess how easy it is to write \inlinetests, and to
obtain feedback about \inlinetesting. Lastly, we measure the runtime
cost of \inlinetests.  All \UseMacro{NumUserStudyValidResponses} user
participants who completed the study say that \inlinetests are easy to
write, needing an average of \UseMacro{AVG_time_per_test} minutes to
write each \inlinetest, and that \inlinetesting is beneficial.
\Inlinetests incur negligible cost, at
\UseMacro{exp-python-overhead-unit-and-inline-MACROAVG-dup1}x for
Python and \UseMacro{exp-java-overhead-unit-and-inline-MACROAVG-dup1}x
for Java on average, and our \inlinetests helped find two new faults
that have been fixed by developers after we reported the bugs. These
results show the promise of \inlinetests.

\vspace{2pt}
\noindent
The main contributions of this paper include:

\begin{itemize}[topsep=.2ex,itemsep=2pt,leftmargin=0.8em]

  \item[\Contrib{}]\textbf{Idea.} We introduce \inlinetests, the
    benefits that they provide, and \desiderata for testing frameworks
    that support them.

  \item[\Contrib{}]\textbf{Framework.} We implement \InlineTest, the
    first \inlinetesting framework. \InlineTest works for Python and
    Java.

  \item[\Contrib{}]\textbf{User study.} We evaluate programmer
  perceptions about \inlinetesting, and obtain feedback about
  their \inlinetesting needs.

\item[\Contrib{}]\textbf{Performance evaluation.} We measure runtime
costs of \InlineTest using \UseMacro{NumTotalInlineTests} \inlinetests
that we write in \UseMacro{NumTotalProjs} open-source projects.

\end{itemize}

\noindent{Our code and data is publicly available at\\\url{\UseMacro{repo}}}.

\section{Motivation and Examples}
\label{sec:example}

\begin{figure}[t]
  \centering
\begin{lstlisting}[language=python-pretty]
def parse_diff(diff: str) -> Diff:
 ...
 nm = re.match(r'^--- (?:(?:/dev/null)|(?:a/(.*)))$', line) (*@\label{fig:example:regex:start}@*)
 (*@{\color{darkblue}Here().given(line, \textquotesingle-{}-{}-~a/python/regex.py\textquotesingle).check\_true(nm).check\_eq(nm.groups(), (\textquotesingle python/regex.py\textquotesingle,)) }@*) (*@\label{fig:example:regex:itest}@*)
 if nm:
  name, = nm.groups()
\end{lstlisting}
  \vspace{-10px}
  \caption{Regex in Python code, and an \inlinetest in {\color{darkblue} blue}.}
  \label{fig:example:regular-expression}
\end{figure}

We motivate \inlinetests by showing examples of some programming language (PL)
features and one common testing scenario for which \inlinetests could be
beneficial.  For each, we discuss problems that developers face
due to the lack of direct support for statement-level testing, and
show example \inlinetests that can help.

\subsection{An Example \InlineTTest}
\label{sec:example:inline-test-structure}

We start by illustrating what \inlinetests look like
because we show several of them in this section, before the
\InlineTest API is described (Section~\ref{sec:tech:api}). Consider
this \inlinetest that we write for a target statement in \CodeIn{apprenticeharper/DeDRM\_tools}~\cite{dedrmtools};
that target statement is shown and described in
Figure~\ref{fig:example:bit-manipulation}:

\vspace{1ex}
\IncludingComment{
\tikz[na] \coordinate(a1b);Here()\tikz[na] \coordinate(a1e);
.\tikz[na] \coordinate(t1b);given(dt, (1980, 1, 25, 17, 13, 14))\tikz[na] \coordinate(t1e);
.\tikz[na] \coordinate(c1b);check\_eq(dosdate, 57)}\tikz[na] \coordinate(c1e);

\begin{tikzpicture}[overlay]
  \draw (a1b) to node[below] {\color{darkblue} Declare} (a1e);
  \draw (t1b) to node[below] {\color{darkblue} Assign} (t1e);
  \draw (c1b) to node[below] {\color{darkblue} Assert} (c1e);
\end{tikzpicture}
\vspace{1ex}

The ``Declare'' portion tells the \inlinetesting framework to process
the statement as an \inlinetest. The ``Assign'' portion allows the
developer to provide test inputs to the \inlinetest. In this case,
\CodeIn{(1980, 1, 25, 17, 13, 14)} is to be used as the value of the
\CodeIn{dt} variable that is in the target statement. Finally,
the ``Assert'' portion allows the developer to specify a test oracle. In
this case, given the test input for \CodeIn{dt}, the \CodeIn{dosdate}
variable that is being computed in the target statement should
equal 57 for the \inlinetest to pass.

\begin{figure*}[t]
  \centering
  \begin{subfigure}[b]{.5\textwidth}
    \centering
    \includegraphics[width=0.95\columnwidth]{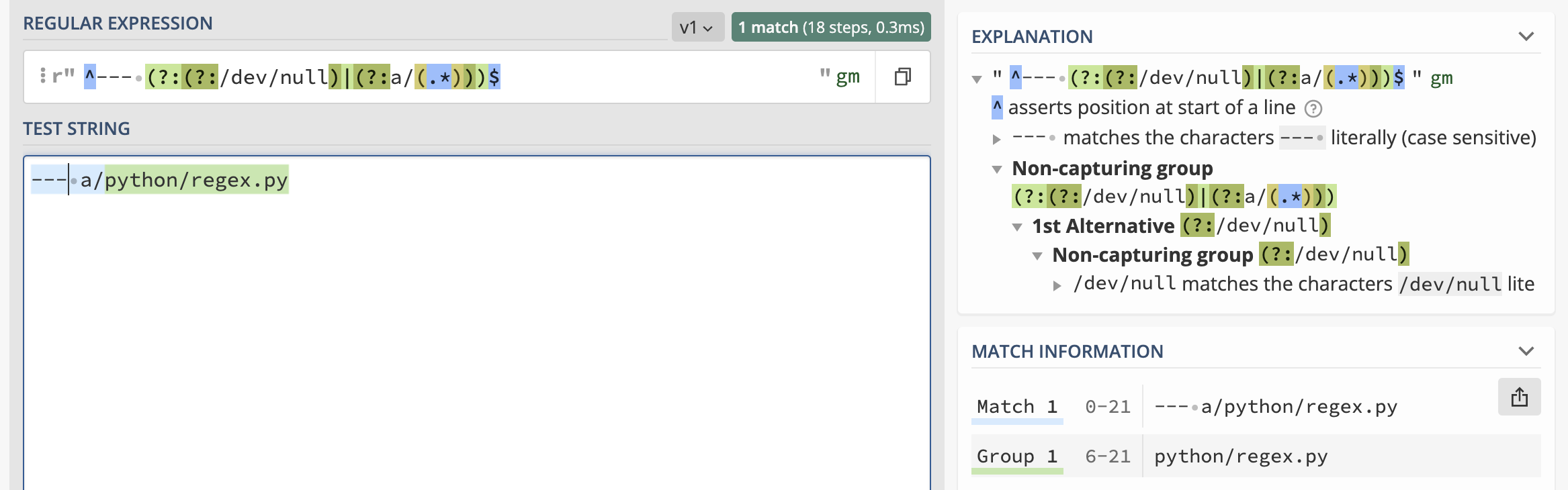}
    \subcaption{A regex-checking website~\cite{regex101Link}}
    \label{fig:example:online-regex-debugger}
  \end{subfigure}%
  \begin{subfigure}[b]{.5\textwidth}
    \centering
    \includegraphics[width=0.95\columnwidth]{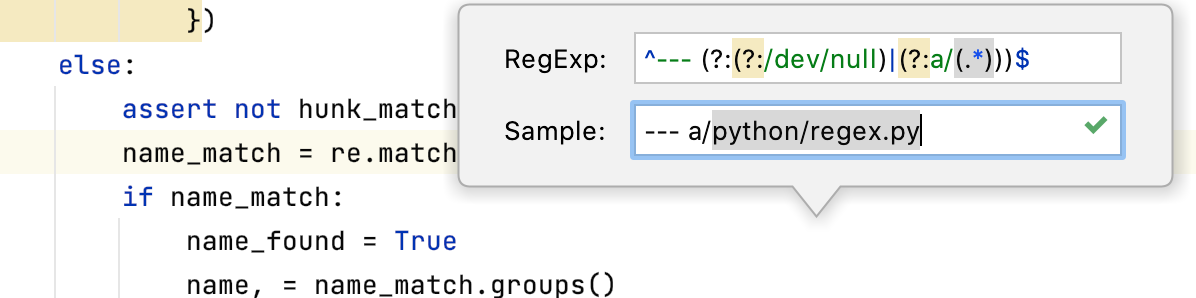}
    \subcaption{IntelliJ pop-up for checking regexes~\cite{intellijideaRegex}}
    \label{fig:example:intellij-regex}
  \end{subfigure}
  \vspace{-18pt}
  \caption{Screenshots of an online website and an in-IDE pop-up for checking regexes.}
\end{figure*}

\subsection{PL Features That \InlineTests Help Check}
\label{sec:example:language}

\MyPara{Regular expressions (regexes)} Prior work showed that regexes
are widely used, but they are difficult for developers to understand
and to use correctly~\cite{chapman2016exploring, michael2019regexes,
davis2019testing, davis2018impact}. So, \inlinetests can allow
developers to check what regexes do, and to test them in place.
Consider the Python code fragment in
Figure~\ref{fig:example:regular-expression}, which is simplified from
\CodeIn{pytorch/pytorch}~\cite{pytorch2016}. The regex on
line~\ref{fig:example:regex:start} is a search pattern that starts
with ``\nobreakdash-\nobreakdash-\nobreakdash-\ '' and ends with the
non-capturing group ``/dev/null'' or ``a/(.*)''. A matched string is
assigned to the \CodeIn{name} variable.

Directly checking what the regex on line~\ref{fig:example:regex:start}
matches, or testing that it is correct, is difficult without support
for statement-level testing. Three unit
tests check \CodeIn{parse\_diff} (written in a different file and
executed using \pytest~\cite{pytestLink}), but they mock the
\CodeIn{parse\_diff} inputs and do not directly test the regex.  In
fact, we are not aware of an easy way to
directly unit-test the regex on line~\ref{fig:example:regex:start}
with \pytest.

In practice, a main way of checking regexes is to use regex-checking
websites~\cite{regex101Link, debuggexLink, regexTesterLink}.
Figure~\ref{fig:example:online-regex-debugger} shows one such website.
One could also use in-IDE pop-ups like the one in
Figure~\ref{fig:example:intellij-regex} for
IntelliJ~\cite{intellijideaRegex}.  These websites and in-IDE pop-ups
strengthen our argument for statement-level
testing in four ways. First, the existence and usage of these websites
or pop-ups show that developers have a need to directly test
regexes. Second, these websites
and pop-ups are not connected to the target statement(s), so
developers cannot easily specify where in the code the checks should
be performed, what kind of oracles should be used, and what the
expected outcome should be.   Third, each time developers leave their development
environment to use websites or pop-ups, they mentally switch context
and may lose productivity as a result~\cite{latoza2006maintaining,
latoza2011designing}. Lastly, knowledge gained from using websites and pop-ups may not be documented, so (other)
developers in the same organization may later wastefully re-check the
same regex.

Line~\ref{fig:example:regex:itest} in
Figure~\ref{fig:example:regular-expression} shows how \inlinetests can
be used to directly test a regex. There, a developer specifies an
input and an expected output. Then a framework like \InlineTest can
run the \inlinetest to provide feedback on
what the regex does. Using \inlinetests as shown in
Figure~\ref{fig:example:regular-expression} mitigates the
aforementioned problems of using regex-checking websites and in-IDE
pop-ups: developers have more control to specify how to test the
target statement, they do not have to leave their development
environment to perform checks, and \inlinetests self-document
knowledge about regexes.
    
\begin{figure}[t]
  \centering
\begin{lstlisting}[language=java-diff]
 orig = os.path.splitext(os.path.basename(infile))[0]
 if (re.match('^B[A-Z0-9]{9}(_EBOK|_EBSP|_sample)?$', orig) or
-(*@\label{fig:example:fix-regex:bug}@*)re.match('^{0-9A-F-}{36}$', orig)
+(*@\label{fig:example:fix-regex:fix}@*)re.match('^[0-9A-F-]{36}$', orig)
 ):
   # Kindle for PC / Mac / Android / Fire / iOS(*@\label{fig:example:regex:comment}@*)
   (*@{\color{darkblue}Here().given(orig, \textquotesingle 0123456789ABCDEF0123456789ABCDEF0123\textquotesingle).check\_true(Group(1)) }@*)(*@\label{fig:example:fix-regex:itest}@*)
   clean_title = cleanup_name(book.getBookTitle())
\end{lstlisting}
  \vspace{-10px}
  \caption{Fix for faulty regex that an \inlinetest helped find.}
  \label{fig:example:regex_pull_request}
\end{figure}

We showcase an additional benefit of using \inlinetests to check
regexes: it helped us find a fault.
Figure~\ref{fig:example:regex_pull_request} shows a fix that we report
to developers of a project in our evaluation, who have since accepted
our pull
request\footnote{\url{https://github.com/noDRM/DeDRM_tools/commit/012ff533ab6ba6920813284a4eb7}}.
The goal of the faulty regex on line~\ref{fig:example:fix-regex:bug}
is to match valid string representations of 36-digit hexadecimal
numbers or ``-'', but it wrongly matches ``\{0-9A-F-'' followed by 36
repetitions of ``\}''. The \inlinetest on
line~\ref{fig:example:fix-regex:itest} helped us find this fault.  The
\inlinetest input (provided using \InlineTest's \texttt{given} function) is a
string that represents a 36-digit hexadecimal number. \texttt{Group}
is an \InlineTest{} construct for automatically matching conditional
expressions in \texttt{if} or \texttt{while} statement headers; it
accepts a zero-based index that represents the position of a condition
in the header. So, \texttt{Group(1)} matches the second conditional
expression in the \texttt{if} statement in
Figure~\ref{fig:example:regex_pull_request}, i.e.,
\texttt{re.match(\textquotesingle
\^{}\{0-9A-F-\}\{36\}\$\textquotesingle, orig)}. We expected the matched
condition to be \texttt{True}, but it was \texttt{False} and the
\inlinetest failed. Our fix is on
line~\ref{fig:example:fix-regex:fix}.
 In sum, an \inlinetest was useful
for reducing the burden of setting up and writing a unit test for this
regex without the need to first perform some throw-away refactoring to
extract the regex from the conditional expression.

\MyPara{Bit manipulation} Figure~\ref{fig:example:bit-manipulation}
shows a simplified code fragment from
\CodeIn{apprenticeharper/DeDRM\_tools}~\cite{dedrmtools}.
 Line~\ref{fig:example:bit-manipulation:dosdate} parses
the year, month, and day into a 32-bit DOS date.
Line~\ref{fig:example:bit-manipulation:dostime} uses the hour, minute,
and second to compute a 32-bit DOS time. The \CodeIn{FileHeader}
function that contains the fragment in
Figure~\ref{fig:example:bit-manipulation} has many other statements
that we elide, and it can be unit tested to check that it constructs
correct headers. However, it is hard to directly test
lines~\ref{fig:example:bit-manipulation:dosdate}
and~\ref{fig:example:bit-manipulation:dostime} without first
extracting these statements into separate functions. Also, bit
manipulation is fast but it may be hard to understand. With the
\inlinetests on lines~\ref{fig:example:bit-manipulation:dosdate:itest}
and~\ref{fig:example:bit-manipulation:dostime:itest}, we are able to
directly check the code, and the inputs and expected outputs in those
\inlinetests document what the target statements compute.

\begin{figure}[t]
  \centering
\begin{lstlisting}[language=python-pretty]
def FileHeader(self):
 dt = self.date_time
 dosdate = (dt[0] - 1980) << 9 | dt[1] << 5 | dt[2](*@\label{fig:example:bit-manipulation:dosdate}@*)
 (*@{\color{darkblue} Here().given(dt, (1980, 1, 25, 17, 13, 14)).check\_eq(dosdate, 57) }@*)(*@\label{fig:example:bit-manipulation:dosdate:itest}@*)
 dostime = dt[3] << 11 | dt[4] << 5 | (dt[5] // 2)(*@\label{fig:example:bit-manipulation:dostime}@*)
 (*@{\color{darkblue} Here().given(dt, (1980, 1, 25, 17, 13, 14)).check\_eq(dostime, 35239) }@*)(*@\label{fig:example:bit-manipulation:dostime:itest}@*)
 if self.flag_bits & 0x08:
  # Set these to zero because we write them after the file data
  CRC = compress_size = file_size = 0
\end{lstlisting}
  \vspace{-10px}
  \caption{\scalebox{0.95}{Bit manipulation in Python, and \inlinetests in {\color{darkblue} blue}.}}
  \label{fig:example:bit-manipulation}
\end{figure}

\vspace{10pt}
 \MyPara{String manipulation}
Figure~\ref{fig:example:string-manipulation} shows simplified code in
a method from \CodeIn{greenrobot/GreenDAO}~\cite{greenDAO}.
Line~\ref{fig:example:string-operation:line-split} uses a regex to
tokenize a string. The result of
line~\ref{fig:example:string-operation:line-split} is subsequently
used to query a database on
line~\ref{fig:example:string-operation:query-database}, so a developer
may want to check that the split is correct. Although there is a unit
test for this function, it only indirectly checks
line~\ref{fig:example:string-operation:line-split} together with the
logic that is implemented in
lines~\ref{fig:example:string-operation:complex:begin}
to~\ref{fig:example:string-operation:complex:end}. The \inlinetest on
line~\ref{fig:example:string-operation:itest} directly tests
line~\ref{fig:example:string-operation:line-split}.

\begin{figure}[t]
  \centering
\begin{lstlisting}[language=java-pretty]
public static int executeSqlScript(Context context, Database db, String assetFilename, boolean transactional)
        throws IOException {
    byte[] bytes = readAsset(context, assetFilename);
    String sql = new String(bytes, "UTF-8");
    String[] lines = sql.split(";(\\s)*[\n\r]"); (*@\label{fig:example:string-operation:line-split}@*)
    (*@{\color{darkblue}new Here().given(sql, "CREATE TABLE MINIMAL\_ENTITY (\_id INTEGER PRIMARY KEY);\textbackslash nINSERT INTO MINIMAL\_ENTITY VALUES (1);\textbackslash nINSERT INTO MINIMAL\_ENTITY \textbackslash nVALUES (2);"}@*) (*@\label{fig:example:string-operation:itest}@*)
   (*@{\color{darkblue}.checkEq(lines.length, 3);}@*)
    int count;(*@\label{fig:example:string-operation:complex:begin}@*)
    if (transactional) {
        count = executeSqlStatementsInTx(db, lines); (*@\label{fig:example:string-operation:query-database}@*)
    } 
    ...(*@\label{fig:example:string-operation:complex:end}@*)
    return count;
}
\end{lstlisting}
  \vspace{-10px}
  \caption{String manipulation in Java, and \inlinetest in {\color{darkblue} blue}.}
  \label{fig:example:string-manipulation}
\end{figure}

Using an \inlinetest to check statements that manipulate strings also
helped us find a fault, which we show together with the fix in
Figure~\ref{fig:example:string_manipulation_pull_request}.
Specifically, the condition on
line~\ref{fig:patch:string-manipulation:target} is faulty because it
directly compares a string with an integer. So, the \inlinetest on
line~\ref{fig:patch:string-manipulation:itest-old} fails with the
message, ``TypeError: ord() expected string of length 1, but int
found''.  Changing the condition to be as shown on
line~\ref{fig:patch:string-manipulation:fix} fixes the fault and the
developers have accepted our pull request\footnote{\url{https://github.com/python/cpython/commit/5535f3f745761e53a6ff941b8ef74b5ce}}.
Line~\ref{fig:patch:string-manipulation:itest-new} is our updated
\inlinetest after our fix. No unit test covers this function, but other functions can call it in production.

\begin{figure}[t]
  \centering
\begin{lstlisting}[language=java-diff]
  ...
-(*@\label{fig:patch:string-manipulation:target}@*) elif ch < ' ' or ch == 0x7F:
+(*@\label{fig:patch:string-manipulation:fix}@*) elif ch < ' ' or ord(ch) == 0x7F:
   out.write('\\x')
   out.write(hexdigits[(ord(ch) >> 4) & 0x000F])
-(*@\label{fig:patch:string-manipulation:itest-old}@*)  Here().given(ch, 0x7F).check_eq((ord(ch)>>4)&0x000F, 0x07)
+(*@\label{fig:patch:string-manipulation:itest-new}@*)  Here().given(ch, chr(0x7F)).check_eq((ord(ch)>>4)&0x000F, 0x07)
   out.write(hexdigits[ord(ch) & 0x000F])
\end{lstlisting}
  \vspace{-10px}
  \caption{\Inlinetest helped find string manipulation fault.}
  \label{fig:example:string_manipulation_pull_request}
\end{figure}

\vspace{10pt} \MyPara{Stream} The target statement on
lines~\ref{fig:example:stream:line-start}
to~\ref{fig:example:stream:line-end} in
Figure~\ref{fig:example:stream} uses Java's stream API; it is from
\CodeIn{apache/flink}~\cite{flink2011} and it extracts the values of
an expression's children to a list. Using unit tests to check whether
the \CodeIn{aliases} variable is computed correctly will require using
sophisticated Java features like reflection~\cite{JavaReflection} (the
target statement is in a private method). Moreover, a unit test cannot
help to directly check \CodeIn{aliases}; only the value computed on
line~\ref{fig:example:stream:return} is returned. Lastly,
the \CodeIn{unwrapFromAlias} method is not directly tested by any unit
test but it is called by methods in other classes.
The \inlinetest on line~\ref{fig:example:stream:itest} directly tests
the target statement. Also, given the complexity of the statement on
lines~\ref{fig:example:stream:line-start}
to~\ref{fig:example:stream:line-end}, a developer who is new to
\CodeIn{apache/flink} is likely to be better able to understand the
code with the \inlinetest than they would do without it.

\begin{figure}[t]
  \centering
\begin{lstlisting}[language=java-pretty]
private CalculatedQueryOperation unwrapFromAlias(CallExpression call) {
 List<Expression> children = call.getChildren();
 List<String> aliases = (*@\label{fig:example:stream:line-start}@*)
  children.subList(1, children.size())
  .stream()
  .map(alias -> ExpressionUtils.extractValue(alias, String.class)
   .orElseThrow(() ->  new ValidationException("Unexpected: " + alias)))
  .collect(toList()); (*@\label{fig:example:stream:line-end}@*)
  (*@{\color{darkblue}new Here().given(children, Arrays.asList(new Expression[]\{new SqlCallExpression("SELECT MIN(Price) AS SmallestPrice FROM Products; "), new SqlCallExpression("SELECT COUNT(ProductID) FROM Products;")\})).checkEq(aliases, Arrays.asList("SELECT COUNT(ProductID) FROM Products;")); }@*) (*@\label{fig:example:stream:itest}@*)
 CallExpression tc = (CallExpression) children.get(0);
 return createFunctionCall(tc, aliases, tc.getResolvedChildren());  (*@\label{fig:example:stream:return}@*)
}
\end{lstlisting}
  \vspace{-10px}
  \caption{Java code using stream, and an \inlinetest in {\color{darkblue} blue}.}
  \label{fig:example:stream}
\end{figure}

\subsection{A Common Scenario: \printfdebugging}

Developers commonly perform \printfdebugging, in which they
temporarily add print statements so that they can visually check
whether correct values are being computed at the target
statement. Then, after some time, they remove these print statements.

One indication of \printfdebugging popularity can be seen by searching
for ``remove debug'' on GitHub or by going
to~\cite{RemoveDebugGitHub}. (We found 3,344,094 matching commits in
May 2022, but we did not look through them all to see if they are all
about \printfdebugging.) GitHub commits likely underestimate
\printfdebugging popularity; developers may clean the print statements
before committing their code. Dedicated utilities like
\lstinline{git-remove-debug}~\cite{git-remove-debug} and others~\cite{DebugPurge, grunt-groundskeeper, DebugStatementsFixers}
clean up after \printfdebugging.
Figure~\ref{fig:printf:debugging}
shows a GitHub commit\footnote{\url{https://github.com/redis/redis-om-spring/commit/f808c9b3a0c72d22c14221e37228a389a3ff139d}} that cleaned up after \printfdebugging a complex
statement in a private method.
Researchers found many reasons why developers do \printfdebugging:
lack of familiarity with
debuggers~\cite{HowDevelopersDebugBellerEtAlICSE2018}, lack of
platform-specific debuggers~\cite{banken2018debugging,
ida2012performance}, perceived speed~\cite{perscheid2017studying} and
simplicity~\cite{li2015medic} of \printfdebugging, the inability of
debuggers to handle parallel PL
constructs~\cite{grabner1995debugging}, etc.

We do not claim that \inlinetests could replace \printfdebugging. The
many reasons for the longevity and popularity of \printfdebugging
suggests that there is no silver bullet. However, \inlinetests can
help to reduce some of the wastefulness of adding and then removing print
statements during \printfdebugging. Specifically developers could use \inlinetests to persist
knowledge that they gain during \printfdebugging.  For example,
line~\ref{fig:printf:debugging:inline-start} to
line~\ref{fig:printf:debugging:inline-end} in
Figure~\ref{fig:printf:debugging} shows how one could migrate the
print statements from \printfdebugging into \inlinetests.

\begin{figure}[t]
  \centering
\begin{lstlisting}[language=java-diff]
private List<Field> getNestedField(...) {
 if (subField.isAnnotationPresent(Indexed.class)) {
- System.out.println(">>> Found Indexed SUBFIELD....");
  boolean sfIsTagField = ((subField
   .isAnnotationPresent(Indexed.class)
   && ((CharSequence.class.isAssignableFrom(subField.getType())
    || (subField.getType() == Boolean.class)
     || (maybeCollectionType.isPresent()
      && (CharSequence.class
       .isAssignableFrom(maybeCollectionType.get())
      || (maybeCollectionType.get() == Boolean.class)))))));
- System.out.println(">>> sfIsTagField ==> " + sfIsTagField);
  (*@{\color{darkblue}new Here().given(subField, new Object() \{@Indexed CharSequence f;\}}@*) (*@\label{fig:printf:debugging:inline-start}@*)
    (*@{\color{darkblue}.getClass().getDeclaredField("f"))}@*)
   (*@{\color{darkblue}.checkEq(sfIsTagField, true);}@*) (*@\label{fig:printf:debugging:inline-end}@*)
   ...
}}}
\end{lstlisting}
  \vspace{-10px}
  \caption{How \inlinetests can help Java \printfdebugging.}
  \label{fig:printf:debugging}
\end{figure}

\section{The \InlineTest Framework}
\label{sec:tech}

We start with a list of language-agnostic
\desiderata for \inlinetesting frameworks. Then, we give an overview of
\InlineTest{}, the \inlinetesting framework that we implement in this
paper. Lastly, we introduce \InlineTest{}'s API, and describe our current
implementation.

\subsection{\InlineTesting Framework Requirements}
\label{sec:desiderata}

Section~\ref{sec:example} motivated the need for \inlinetests.
We now turn to the question, \emph{what are
  the requirements for \inlinetesting frameworks?}  Answering it helps to
(1)~distinguish \inlinetesting from existing forms of testing,
(2)~provide a road map for \inlinetesting development, and
(3)~provide a basis for evaluating \InlineTest.
\Inlinetesting frameworks should meet this minimum set of requirements:

\begin{enumerate}[topsep=1ex,itemsep=2pt,leftmargin=1.6em]
  
\item \Inlinetests are \emph{not} replacements for unit tests
  or debuggers. \cmark
  
\item An \inlinetest should only check one target statement. \cmark

\item Multiple \inlinetests can check the same target statement. \cmark

\item An \inlinetest should allow developers to provide multiple
  values for a variable in the target statement. \cmark
  
\item \Inlinetests should be easy for developers to write and run
  using similar idioms as those they already use, to ease adoption. \pmark

\item \Inlinetesting frameworks should be easy to integrate with
  testing frameworks and IDEs that developers use. \pmark

\item To aid readability, when integrated with IDEs, \inlinetesting
  frameworks should hide \inlinetests by default, and allow developers
  to hide or view \inlinetests as needed. \xmark \label{requirement:readability}

\item It should be possible to enable \inlinetests during testing and
  to disable them in production. \cmark

\item When \emph{enabled}, the runtime cost of \inlinetests should be low. \cmark

\item When \emph{disabled}, \inlinetests should have negligible
  overhead. \cmark

\item It should be possible for developers to run subsets of all
  \inlinetests---developers often
  perform manual test selection~\cite{gligoric2014empirical}. \pmark \label{requirement:subset}

\item It should be possible to run \inlinetests in parallel. \xmark

\item It should be possible to write \inlinetests for target
  statements that invoke methods or functions whose arguments need
  initialization. \xmark \label{requirement:arguments}

\item It should be possible to write \inlinetests for expressions in
  branch conditions, without requiring developers to copy those expressions into
  the \inlinetest. \pmark \label{requirement:expressions}

\end{enumerate}

\noindent
\InlineTest currently meets \desiderata marked as \cmark; it only
partially supports those marked \pmark and it does not support those
marked \xmark. The \pmark{} in requirements~\ref{requirement:subset} and~\ref{requirement:expressions} means that our current Python implementation satisfies the requirement but our current Java implementation does not.  Other \pmark{} marks mean that we partially satisfy the requirement in Python and Java.

These \desiderata that we enumerate are \emph{initial},
based on our understanding so far, and they are likely incomplete. Our
goal for providing them is to bootstrap the development of
\inlinetesting and to aid better community understanding of
\inlinetesting.

\subsection{Overview of the \InlineTest Framework}
\label{sec:tech:overview}

\begin{figure}[t]
  \includegraphics[width=0.65\columnwidth]{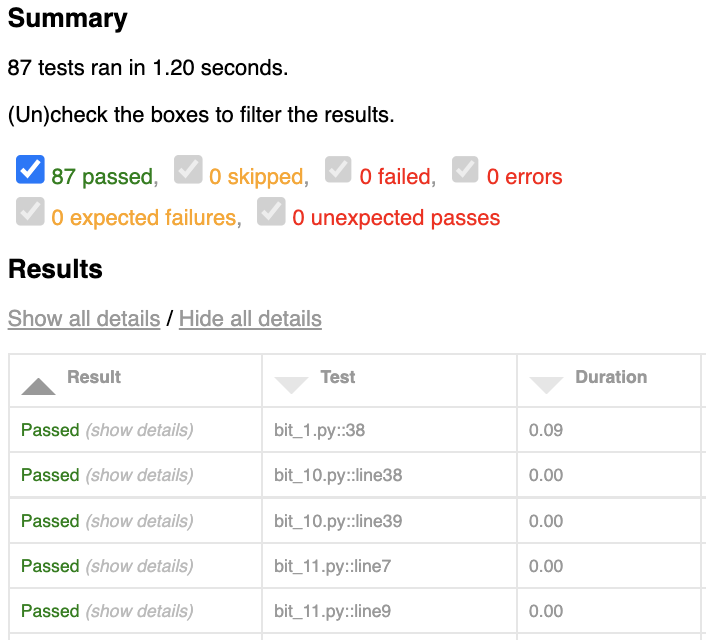}
  \vspace{-8px}
  \caption{Test report in HTML format.}
  \label{framework:fig:test-report}
  \vspace{-10px}
\end{figure}

\InlineTest{} is our \inlinetesting framework that provides developers
with support for statement-level testing. \InlineTest's API provides
three kinds of methods that allow developers to (1)~declare an
\inlinetest, (2)~provide input values that should be assigned to the
variables in the target statement during testing, and (3)~specify test
oracles. If developers write multiple \inlinetests, they can run the
\inlinetests separately or in a batch. We started integrating
\InlineTest with two popular unit testing frameworks---\CodeIn{JUnit}
and \CodeIn{pytest}---and it also generates test reports.
Figure~\ref{framework:fig:test-report} is an example test
report generated by \InlineTest, based on the \texttt{pytest-html}
plugin~\cite{pytest-html} that it uses.
\Inlinetests must be the next statements after a
target statement that is being tested. Since \inlinetests are
co-located with code, \InlineTest provides facilities for turning off
the execution of \inlinetests in production environments. When
\inlinetesting is turned off, the \inlinetests are still in the code
but running the code should incur negligible runtime overhead.

\subsection{\InlineTest Development Process and API}
\label{sec:tech:design}

To ground \InlineTest in likely developer needs, we focus our
current implementation on selected kinds of statements from
open-source projects. Based on our own programming experience, these
kinds of statements could benefit from \inlinetesting. We described
some of these kinds of statements in
Section~\ref{sec:example:language}, but we focus our implementation on
five of them: regexes, string manipulation, bit manipulation, stream API usage, and
collection handling code.

One challenge is to better understand the API that \InlineTest{}
should provide to support statement-level testing for the kinds of
statements that we focus on. To address this challenge, we collect
examples of these kinds of statements from open-source, manually
inspect them, and iteratively refine our \InlineTest
API. Specifically, we first collect Java and Python projects from
GitHub. Then, we filter out projects that do not contain the kinds of
statements that we focus on. Lastly, we find examples from those that
remain and we use them to guide our API design. We next describe our example collection process,
and provide more details on the current API.

\subsubsection{Example Collection Process}
\label{sec:tech:data}

We are interested in target statements that are in possibly
complicated code blocks, such that the target statement may be
difficult to test directly with unit tests. (See
Section~\ref{sec:intro} for a discussion of the pitfalls of extracting
individual statements into methods or functions for the sole purpose
of enabling unit testing.) We look for Java and Python statements with
regular expressions, as well as those that manipulate strings and
bits. We also look for statements that use the stream API in Java and
those that manipulate collections in Python.

We perform keyword
search (such as ``re.match'' and ``re.split'' for Python regular
expressions) among the 100 top-starred Java and Python projects on
GitHub (a total of 200 projects). All keywords that we use for each
language and the number of matches that we find are provided in the
data package for this paper. We manually inspect metadata for these projects
and remove those that are about tutorials, e.g., interview questions.
We then use the
remaining 83 Java projects and 91 Python projects.
For each project that
remains, we select examples and manually inspect them for suitability to
help guide our API design.

To make our manual check easier, we make our keyword search return five lines
of leading and trailing context for each match. We then manually check
whether the matched lines are for the kinds of target statements that
we focus on.  We filter out cases where keywords only appear in
comments or in which we deem the code too simple to warrant an
\inlinetest, e.g., for keyword ``split'' we find 
\CodeIn{String[]~errorMessageSplit = e.getMessage().split("~");}.
We also filter out
keyword searches that yield false positives. For example, we search
for $>>$ as the right shift operator in bit manipulation but sometimes
match the closing tag of a parameterized generic type, e.g.,
\lstinline{<String, Box<Integer>>}. Among the rest, for each kind of
target statement per project, we extract an example which is the first
snippet with a target statement that can be tested at the statement
level. Finally, based on randomly extracted \UseMacro{NumPythonFiles}
examples of Python and \UseMacro{NumJavaFiles} examples of Java, we
design the \InlineTest API.

\subsubsection{Corpus}
\label{sec:tech:corpus}

Data about the selected examples that we base our design of
\InlineTest API on are shown in
Table~\ref{table:basic-stat}. For Python, we write
\UseMacro{NumPythonInlineTests} \inlinetests for
\UseMacro{NumPythonTestedStmts} statements in
\UseMacro{NumPythonFiles} examples from \UseMacro{NumPythonProjs}
projects. For Java, we write \UseMacro{NumJavaInlineTests}
\inlinetests for \UseMacro{NumJavaTestedStmts} statements in
\UseMacro{NumJavaFiles} examples from \UseMacro{NumJavaProjs}
projects. There are sometimes multiple target statements in some
examples, and we sometimes write multiple inline tests for a target
statement.

\begin{table}[t]
  \centering
  \vspace{-10px}
  \caption{\CaptionExpBasicStats}
  \vspace{-10px}
  \scalebox{0.85}{
  \begin{tabular}{|l|c|c|c|c|}
    \hline
    \HeaderPL & \HeaderProjs    & \HeaderFiles    & \HeaderTestedStmts   & \HeaderInlineTests    \\ \hline
    Python    & \UseMacro{NumPythonProjs} & \UseMacro{NumPythonFiles} & \UseMacro{NumPythonTestedStmts} & \UseMacro{NumPythonInlineTests} \\ \hline
    Java      & \UseMacro{NumJavaProjs}   & \UseMacro{NumJavaFiles}   & \UseMacro{NumJavaTestedStmts}   & \UseMacro{NumJavaInlineTests}   \\ \hline
  \end{tabular}
  }
  \label{table:basic-stat}
\end{table}

Table~\ref{table:category-stat} shows a breakdown of the number of
\inlinetests that we write for each kind of target statement.  Columns
represent the kind of target statement, the PL, the number of
projects, the number of examples, the number of target statements, and
the number of \inlinetests. We write at least one \inlinetest per
target statement. There are fewer numbers in the ``Collection'' row
because although operations on collections, like list comprehension or
sorting, look complicated, some developers may want to test them and
others may not. Our user study proves this variation in preferences
(Section~\ref{sec:userstudy}).

\begin{table}[t]
  \begin{center}
\vspace{-10px}    
\caption{\CaptionInlineTestsCategory}
\vspace{-10px}
\scalebox{0.85}{
\begin{tabular}{|l|l|c|c|c|c|}
\hline
\HeaderKind
& \HeaderPL
& \HeaderProjs
& \HeaderFiles
& \HeaderTestedStmts
& \HeaderInlineTests
\\ \hline
\multirow{2}{*}{Regex}
 & Python
 & \UseMacro{NumPythonRegexProjs}
 & \UseMacro{NumPythonRegexFiles}
 & \UseMacro{NumPythonRegexTestedStmts}
 & \UseMacro{NumPythonRegexInlineTests}
\\ \cline{2-6}
 & Java
 & \UseMacro{NumJavaRegexProjs}
 & \UseMacro{NumJavaRegexFiles}
 & \UseMacro{NumJavaRegexTestedStmts}
 & \UseMacro{NumJavaRegexInlineTests}
\\ \hline
\multirow{2}{*}{String}
 & Python
 & \UseMacro{NumPythonStringProjs}
 & \UseMacro{NumPythonStringFiles}
 & \UseMacro{NumPythonStringTestedStmts}
 & \UseMacro{NumPythonStringInlineTests}
\\ \cline{2-6}
 & Java
 & \UseMacro{NumJavaStringProjs}
 & \UseMacro{NumJavaStringFiles}
 & \UseMacro{NumJavaStringTestedStmts}
 & \UseMacro{NumJavaStringInlineTests}
\\ \hline
\multirow{2}{*}{Bit}
 & Python
 & \UseMacro{NumPythonBitProjs}
 & \UseMacro{NumPythonBitFiles}
 & \UseMacro{NumPythonBitTestedStmts}
 & \UseMacro{NumPythonBitInlineTests}
\\ \cline{2-6}
 & Java
 & \UseMacro{NumJavaBitProjs}
 & \UseMacro{NumJavaBitFiles}
 & \UseMacro{NumJavaBitTestedStmts}
 & \UseMacro{NumJavaBitInlineTests}
\\ \hline
Collection
 & Python
 & \UseMacro{NumPythonCollectionProjs}
 & \UseMacro{NumPythonCollectionFiles}
 & \UseMacro{NumPythonCollectionTestedStmts}
 & \UseMacro{NumPythonCollectionInlineTests}
\\ \hline
Stream
 & Java
 & \UseMacro{NumJavaStreamProjs}
 & \UseMacro{NumJavaStreamFiles}
 & \UseMacro{NumJavaStreamTestedStmts}
 & \UseMacro{NumJavaStreamInlineTests}
\\ \hline
\end{tabular}
}
\label{table:category-stat}
\vspace{-10px}
\end{center}
\end{table}

\subsection{The \InlineTest API}
\label{sec:tech:api}

We design the \InlineTest API to
have three components, based on what they allow developers to do:

\MyPara{(1)~Declare and initialize an \inlinetest} This API component
signals to the \InlineTest framework to process a statement as an
\inlinetest and allows users to optionally specify a name for the
\inlinetest. If a test name is not specified, \InlineTest defaults to
using a name which is the concatenation of the current file name and
the line number of the \inlinetest. This component comprises the
\CodeIn{Here()} and \CodeIn{Here(test\_name = ``")} functions in Python
and the \CodeIn{Here()} and \Code{Here(testName)} methods in Java.
With these \CodeIn{Here()} functions or methods, users can also
provide optional parameters for customizing \inlinetest execution.
These parameters include those that (1)~set the number of times to
re-rerun an \inlinetest; (2)~disable the \inlinetest so that it is not
executed (similar to the \CodeIn{@Ignore} annotation in
\CodeIn{JUnit}); (3)~indicate that sets of values can be used to parameterize an
\inlinetest; and (4)~tag \inlinetests so that users can filter out those that they do not want to run (similar to the \CodeIn{@Tag}
annotation in \CodeIn{JUnit}~\cite{junit}).
We plan to implement support for other features in this
\InlineTest API component, including allowing users to specify that an
\inlinetest should be run conditionally.

\MyPara{(2)~Provide test inputs} Developers can use this API component
to initialize variables in the target statement to desired test input
values. The rationale is that, to directly test a target statement,
\InlineTest has to be able to re-initialize the variables in that
statement to the values that should be used for testing. In Python and
Java, this API component is the \CodeIn{given(variable, value)}
function or method. \InlineTest assigns \CodeIn{value} to
\CodeIn{variable} only while running the \inlinetest. Two
input-related needs may arise during inline testing: a target
statement may have multiple variables, or a developer may want to test
a target statement using multiple values of the same variable. To
address the first need, \InlineTest allows chaining
\CodeIn{given(\ldots)} calls. To address the second need,  \InlineTest allows to
provide a list of values in each \CodeIn{given(\ldots)} call if
\CodeIn{Here(parameterized=True)} is used. This feature is similar to
parameterized unit tests~\cite{tillmann2005parameterized,
  tillmann2006unit}.

\MyPara{(3)~Specify test oracles} This API component allows developers
to make assertions on the results of running the \inlinetest. Driven
by the examples that we base our design on, \InlineTest supports
checking equality of two expressions with \CodeIn{check\_eq(expr1,
expr2)}, checking whether a condition holds or not with
\CodeIn{check\_true(expr)} and \CodeIn{check\_false(expr)}. The last
two are for convenience; they are equivalent to
\CodeIn{check\_eq(expr, True)} and \CodeIn{check\_eq(expr, False)},
respectively. In
Java, we support oracles with the same functionality but they have camel-case naming.
Unit testing frameworks typically support more kinds of
assertions. As \InlineTest grows, we may
need to add more kinds of assertions. These three suffice to check the target statements in our
corpus (Section~\ref{sec:tech:corpus}).

Even though we base our design on selected examples from
open source, we are encouraged that our API design resulted in
components that should be familiar to developers who already know how
to write unit tests.  The API is also the same for
Java and Python. Even if small tweaks are needed to support other
programming languages, current evidence suggests that the same
\inlinetesting API components may be useful more broadly.

\begin{figure}[t]
  \centering
  \includegraphics[width=\linewidth]{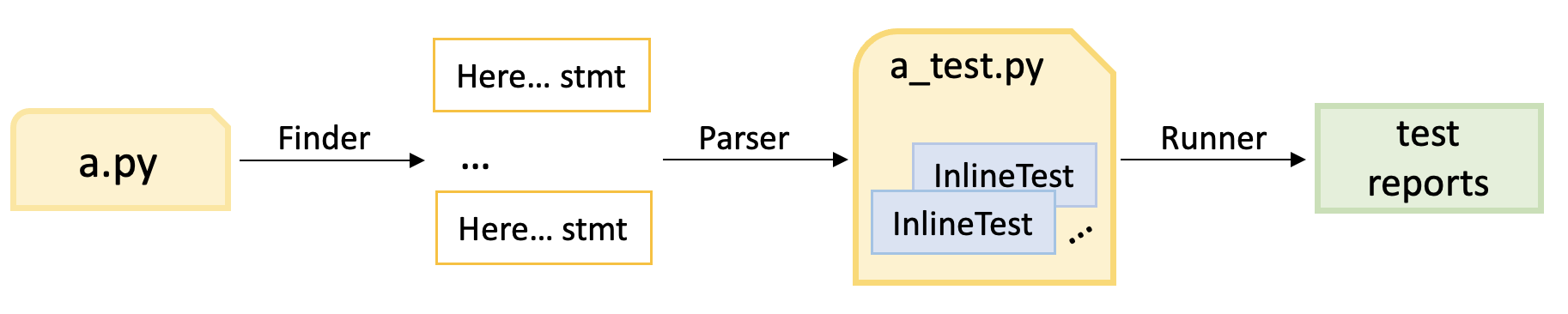}
  \vspace{-25pt}
  \caption{Workflow of \InlineTest{} for Python.}
  \label{fig:framework:work-flow}
  \vspace{-15px}
\end{figure}

\subsection{\InlineTest Implementation}
\label{sec:tech:impl}

Figure~\ref{fig:framework:work-flow} shows the workflow of
\InlineTest{} for Python; it is similar for Java. Given a source file, \CodeIn{Finder}
searches for statements that start with \CodeIn{Here} calls.
\CodeIn{Parser} traverses the AST of the source file to discover the
target statement. \CodeIn{Parser} also uses the output of
\CodeIn{Finder} to reconstruct assignments and assertions and to collect
each \inlinetest into a new source file that can be executed.  
Moreover, \CodeIn{Parser} copies the import statements used by the
target statement and the \inlinetest to the new source file; thus,
execution of this new source file only requires the packages used by
the target statement and the \inlinetest.
Finally, \Code{Runner} executes the \inlinetest files and generates test reports like the one shown in Figure~\ref{framework:fig:test-report}.

\MyPara{Python} We implement \InlineTest{} as a standalone Python
library, which can be run from the command line; we also integrate
\InlineTest into \pytest.
\InlineTest{} uses the Python
AST library~\cite{PythonASTLibrary} to
parse the source code, extract the tested statement, process the
input assignments and assertions, compose an executable test, and execute the
\inlinetest in the name space of the module in which tested statement
exists. More precisely, \InlineTest{} uses the visitor design pattern to
detect \inlinetest initialization and to find target statements.
Oracles are implemented on top of the \CodeIn{assert} construct in
Python. If an assertion fails, the resulting error message shows the
line number of the failing \inlinetest, and its observed and
expected outputs.
We integrate \InlineTest{} as a plugin into \pytest to reuse the
various testing options that \pytest provides and to generate test
reports.

\MyPara{Java} We use Javaparser~\cite{JavaASTLibrary} to manipulate
Java AST.  Java \InlineTest{} additionally infers variable types in
\Code{given} calls using a symbol table that it maintains.  For
example, in \CodeIn{given(a, 1)}, it looks up the type that \CodeIn{a}
was declared with in the program.
We support two compilation modes for Java \inlinetests.  The first
(guard mode) keeps the \inlinetest in the resulting bytecode and uses
a flag to skip or run the \inlinetest. The second (delete mode) discards the
\inlinetests from the bytecode.
We also support two ways to run \inlinetests{} in Java. The first
generates an \emph{ad hoc} class for each source file, where each
\inlinetest is converted to a method and a main method is added to run
all the \inlinetests.  The second produces a JUnit test class for
the given file, where each \inlinetest is converted to a test method,
which can be executed using the JUnit runner.

\section{Performance Evaluation for \InlineTest}
\label{sec:eval}

We answer these research questions to assess \inlinetesting costs:

\begin{enumerate}

\item[\textbf{\RQCostAlone:}] How long does it take to run \inlinetests?
  
\item[\textbf{\RQCostWithUnitEnabled:}] What is the runtime overhead when \inlinetests
  are \emph{enabled} during the execution of existing unit tests?

\item[\textbf{\RQCostWithUnitDisabled:}] What is the runtime overhead when \inlinetests
  are \emph{disabled} during the execution of existing unit tests?
  
\end{enumerate}

We measure the times for answering these questions using the
\inlinetests from the \UseMacro{NumTotalFiles} examples that we write
(Section~\ref{sec:tech:corpus}). We also duplicate each of these
\inlinetests 10, 100, and 1000 times, so that we can simulate the
costs as the number of \inlinetests grows.  We evaluate
\RQCostWithUnitEnabled and \RQCostWithUnitDisabled on
\UseMacro{exp-total-proj-can-run-unit-tests} projects in our corpus
where we could run the unit tests.

\subsection{Experimental Setup}
\label{sec:eval:setup}

\MyPara{Standalone experiments}
To run the \inlinetests in an example, \InlineTest does not need all
code elements (class, method, or field) in that example. Rather, it
only needs code elements used by the target statement and the
\inlinetest.  For example, the code fragment in
Figure~\ref{fig:example:string-manipulation} has classes
\CodeIn{Context} and \CodeIn{Database} in the method signature. But, the
\inlinetest there does not need these classes; it only needs the
\CodeIn{String} class from the standard library and the \CodeIn{Here}
class in \InlineTest.  On the contrary, running a unit test
for the same example requires loading all the classes.
So, \InlineTest can run all \UseMacro{NumTotalInlineTests}
\inlinetests under the standalone mode without setting up the
environments needed to run unit tests.
For Python, we run the \inlinetests in each example using the
\InlineTest plugin that we integrate into \pytest.  For Java, we run the
\inlinetests in each example by using \InlineTest to produce an ad-hoc
class and then invoke its main method.

\MyPara{Integrated experiments}
To measure the runtime overhead of \inlinetests, we need to run them
together with unit tests using the runtime environment specified by
each project.
We write \inlinetests directly in the projects from which we extract the examples.
But, we face difficulties in setting up some runtime environments or
in running unit tests. So, we perform the experiments for answering
\RQCostWithUnitEnabled and \RQCostWithUnitDisabled on a subset of
\UseMacro{exp-total-proj-can-run-unit-tests} projects.
Below, we discuss the difficulties that we face for
Python and Java, respectively.

\InlineTest for Python relies
on \pytest to run \inlinetests.  Of
\UseMacro{NumPythonProjs} Python projects in our corpus, we could not
setup the appropriate \pytest runtime environment for
\UseMacro{exp-python-skipped-import-error}:
\UseMacro{proj-keras-team_keras} uses the bazel build system which requires additional
time to setup; and \UseMacro{proj-kovidgoyal_kitty} mixes C++ with
Python code, leading to problems with importing C++ code into \pytest
using a \CodeIn{pyi} interface file. Of the other
\UseMacro{exp-python-proj-can-run} projects,
\UseMacro{exp-python-skipped-unit-tests-no-tests} have no unit tests.
We confirm absence of unit tests by (1)~checking the README.md and
CONTRIBUTING.md files which contain instructions for setting up the
projects; (2)~inspecting the Continuous Integration logs, if any; and
(3)~searching for \CodeIn{*test*.py} in the repositories.
\UseMacro{exp-python-skipped-unit-tests-does-not-use-pytest} projects
do not use \pytest to run unit tests. Lastly, another
\UseMacro{exp-python-skipped-unit-tests-too-many-test-failures}
projects have many unit tests that consistently fail. If a project
manifests less than 10 flaky unit
tests~\cite{luo2014empirical,lam2020study,lam2019root,BellETALICSE2018Deflaker,
  ShiETAL16NonDex} that can be skipped without causing more failures,
we run the remaining unit tests in that project.
We run \inlinetests and unit tests for the remaining
\UseMacro{exp-python-proj-can-run-unit-tests} projects
(first column of Table~\ref{tab:exp-ut-and-it-python}).  

For Java, we use \InlineTest to generate ad-hoc classes for the
integrated examples, and compile the generated classes together with
the other source code in the project.  Of \UseMacro{NumJavaProjs}
projects in our corpus,
\UseMacro{exp-java-skipped-compilation-failure} have compilation
failures (before integrating any \inlinetest) and
\UseMacro{exp-java-skipped-unit-tests-no-tests} have no unit tests.
We confirm that these projects have no unit tests and handle flaky
tests similarly as we did for Python. If
running unit tests across a multi-module project fails, we retry
running only the unit tests in the sub-modules that we write
\inlinetests for (and refrain from using the project in our
experiments if there are still too many failures).   We run
\inlinetests and unit tests for the remaining
\UseMacro{exp-java-proj-can-run-unit-tests} projects, shown in the
first column of Table~\ref{tab:exp-ut-and-it-java}.

\MyPara{Duplicating \inlinetests} Since we are the first to explore
\inlinetests, the number of \inlinetests we have written for each
project is often not as much as the number of unit tests that a
project typically has. In the future, about equal or even
more \inlinetests than unit tests may be written. To simulate the
performance of \InlineTest in such scenario with the corpus we
currently have, we experiment with duplicating each \inlinetest 10,
100, and 1000 times.  When duplicating \inlinetests 1000 times, two
Java projects (\UseMacro{proj-alibaba_fastjson} and
\UseMacro{proj-apache_kafka}) do not compile because the size of the
bytecode in the method containing the target statement exceeded the
allowable limit in Java~\cite{JavaMethodLimit}. So, we exclude these
two projects (only when
duplicating 1000 times).

\MyPara{Experimental procedure and environment} We run \inlinetests
and unit tests four times. The first run is for warm-up, and we
average the times for the last three runs.  We run experiments on a
machine with Intel Core i7-11700K @ 3.60GHz (8 cores, 16 threads) CPU,
64 GB RAM, and Ubuntu 20.04.  We use Java 8 and Python 3.9 in the
standalone experiments, and use the software versions required by each
project in the integrated experiments.

\subsection{Results}
\label{sec:eval:results}

\begin{table}[t]
  \begin{small}
    \centering
  \caption{\CaptionExpItStandalone}
  \vspace{-10pt}
  \begin{minipage}[t]{.23\textwidth}
    \centering
    \subcaption{\CaptionExpItStandalonePython}
    \vspace{-4pt}
    
%% Automatically generated by pyutil.latex 

\begin{tabular}{|l@{\hspace{2pt}}|@{\hspace{2pt}}r|r|r|}
\hline
\HeaderDup & \HeaderNumITStandalone & \HeaderTimeITStandalone & \HeaderTimePerTestITStandalone \\
\hline
\UseMacro{dup-1}
 & \UseMacro{exp-python-num-SUM-dup1-its}
 & \UseMacro{exp-python-time-SUM-dup1-its}
 & \UseMacro{exp-python-timept-MACROAVG-dup1-its}
 \\
\UseMacro{dup-10}
 & \UseMacro{exp-python-num-SUM-dup10-its}
 & \UseMacro{exp-python-time-SUM-dup10-its}
 & \UseMacro{exp-python-timept-MACROAVG-dup10-its}
 \\
\UseMacro{dup-100}
 & \UseMacro{exp-python-num-SUM-dup100-its}
 & \UseMacro{exp-python-time-SUM-dup100-its}
 & \UseMacro{exp-python-timept-MACROAVG-dup100-its}
 \\
\UseMacro{dup-1000}
 & \UseMacro{exp-python-num-SUM-dup1000-its}
 & \UseMacro{exp-python-time-SUM-dup1000-its}
 & \UseMacro{exp-python-timept-MACROAVG-dup1000-its}
 \\
\hline
\end{tabular}

  \end{minipage}
  \hspace{3pt}
  \begin{minipage}[t]{.23\textwidth}
    \centering
    \subcaption{\CaptionExpItStandaloneJava}
    \vspace{-4pt}
    
%% Automatically generated by pyutil.latex 

\begin{tabular}{|l@{\hspace{2pt}}|@{\hspace{2pt}}r|r|r|}
\hline
\HeaderDup & \HeaderNumITStandalone & \HeaderTimeITStandalone & \HeaderTimePerTestITStandalone \\
\hline
\UseMacro{dup-1}
 & \UseMacro{exp-java-num-SUM-dup1-its}
 & \UseMacro{exp-java-time-SUM-dup1-its}
 & \UseMacro{exp-java-timept-MACROAVG-dup1-its}
 \\
\UseMacro{dup-10}
 & \UseMacro{exp-java-num-SUM-dup10-its}
 & \UseMacro{exp-java-time-SUM-dup10-its}
 & \UseMacro{exp-java-timept-MACROAVG-dup10-its}
 \\
\UseMacro{dup-100}
 & \UseMacro{exp-java-num-SUM-dup100-its}
 & \UseMacro{exp-java-time-SUM-dup100-its}
 & \UseMacro{exp-java-timept-MACROAVG-dup100-its}
 \\
\UseMacro{dup-1000}
 & \UseMacro{exp-java-num-SUM-dup1000-its}
 & \UseMacro{exp-java-time-SUM-dup1000-its}
 & \UseMacro{exp-java-timept-MACROAVG-dup1000-its}
 \\
\hline
\end{tabular}

  \end{minipage}
  \vspace{-5pt}
  \end{small}
\end{table}

\begin{figure}[t]
  \begin{small}
  \centering
  \begin{minipage}[t]{.23\textwidth}
    \centering
    \includegraphics[width=\textwidth]{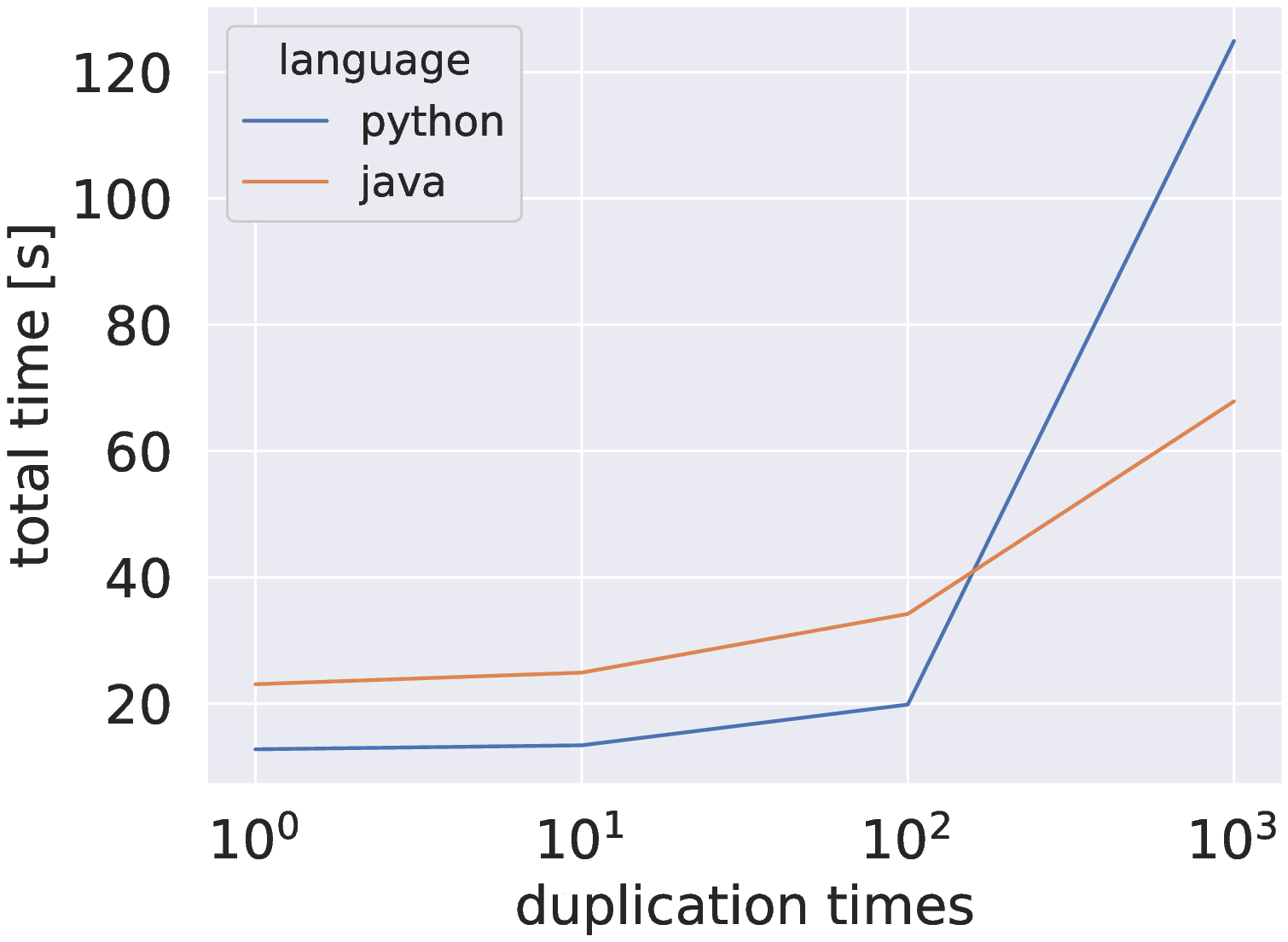}
    \vspace{-15pt}
    \subcaption{total time\label{fig:exp-it-standalone:total}}
  \end{minipage}
  \begin{minipage}[t]{.23\textwidth}
    \centering
    \includegraphics[width=\textwidth]{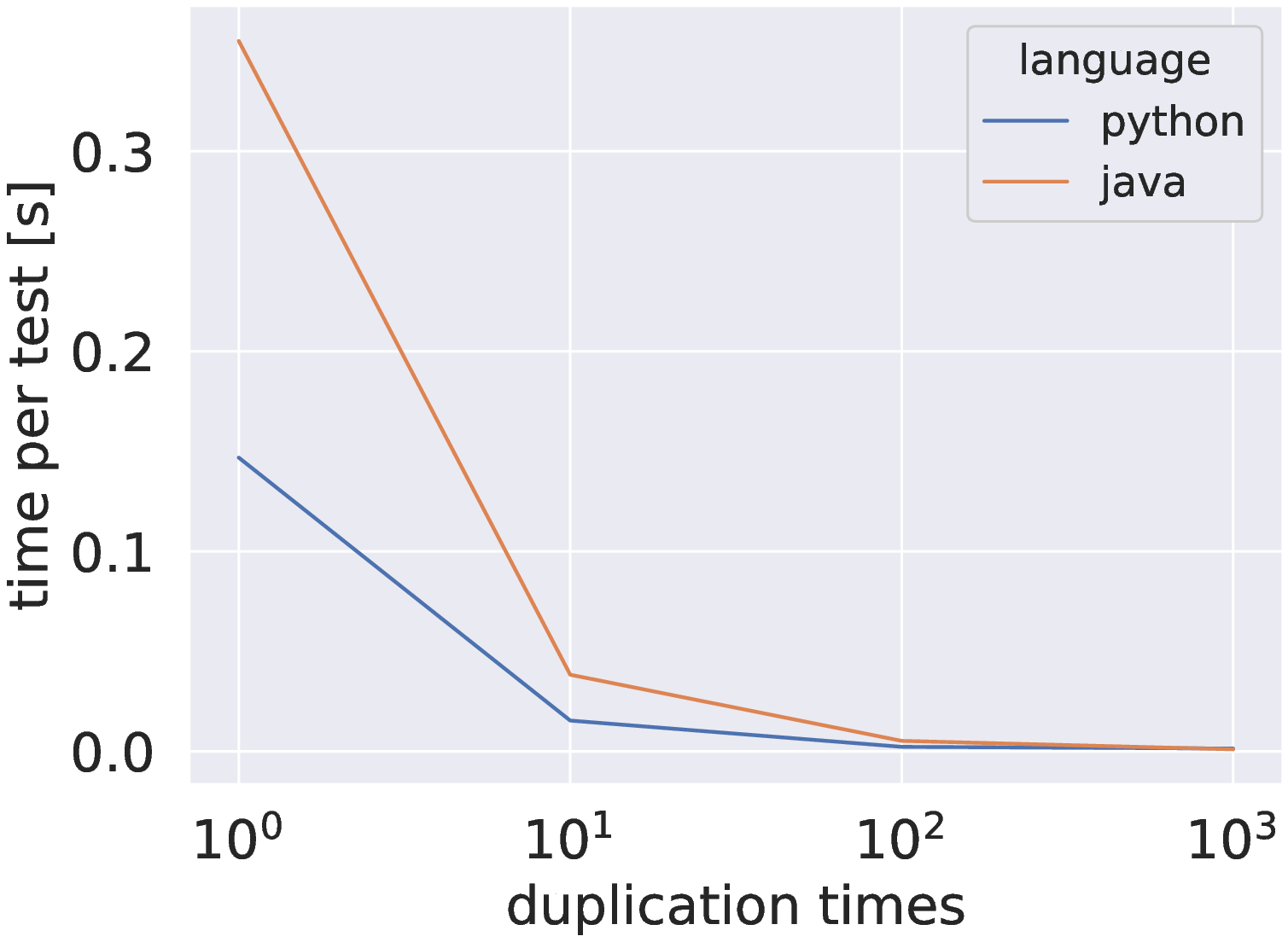}
    \vspace{-15pt}
    \subcaption{per-test time\label{fig:exp-it-standalone:per-test}}
  \end{minipage}
  \vspace{-10pt}
  \caption{Line plots of duplication times vs. total/per-test time when running \inlinetests standalone.\label{fig:exp-it-standalone}}
  \end{small}
\end{figure}

\begin{table*}[t]
  \begin{small}
  \centering
  \caption{\CaptionExpUtAndIt}
  \vspace{-8pt}
  \begin{minipage}[t]{\textwidth}
    \centering
    \subcaption{\CaptionExpUtAndItPython}
    \vspace{-4pt}
    
%% Automatically generated by pyutil.latex 

\begin{tabular}{|l|r|r||r|r||r|r|r||r|r|r|}
\hline
\HeaderProj & \HeaderNumVanilla & \HeaderNumITOnly & \HeaderTimeVanilla & \HeaderTimePerTestVanilla & \HeaderTimeITEnabled & \HeaderTimePerTestITEnabled & \HeaderOverheadITEnabled & \HeaderTimeITDisabled & \HeaderTimePerTestITDisabled & \HeaderOverheadITDisabled \\
\hline
\UseMacro{proj-RaRe-Technologies_gensim}
 & \UseMacro{exp-python-RaRe-Technologies_gensim-num-vanilla-dup1}
 & \UseMacro{exp-python-RaRe-Technologies_gensim-num-inline-dup1}
 & \UseMacro{exp-python-RaRe-Technologies_gensim-time-vanilla-dup1}
 & \UseMacro{exp-python-RaRe-Technologies_gensim-timept-vanilla-dup1}
 & \UseMacro{exp-python-RaRe-Technologies_gensim-time-unit-and-inline-dup1}
 & \UseMacro{exp-python-RaRe-Technologies_gensim-timept-unit-and-inline-dup1}
 & \UseMacro{exp-python-RaRe-Technologies_gensim-overhead-unit-and-inline-dup1}
 & \UseMacro{exp-python-RaRe-Technologies_gensim-time-unit-dup1}
 & \UseMacro{exp-python-RaRe-Technologies_gensim-timept-unit-dup1}
 & \UseMacro{exp-python-RaRe-Technologies_gensim-overhead-unit-dup1}
 \\
\UseMacro{proj-Textualize_rich}
 & \UseMacro{exp-python-Textualize_rich-num-vanilla-dup1}
 & \UseMacro{exp-python-Textualize_rich-num-inline-dup1}
 & \UseMacro{exp-python-Textualize_rich-time-vanilla-dup1}
 & \UseMacro{exp-python-Textualize_rich-timept-vanilla-dup1}
 & \UseMacro{exp-python-Textualize_rich-time-unit-and-inline-dup1}
 & \UseMacro{exp-python-Textualize_rich-timept-unit-and-inline-dup1}
 & \UseMacro{exp-python-Textualize_rich-overhead-unit-and-inline-dup1}
 & \UseMacro{exp-python-Textualize_rich-time-unit-dup1}
 & \UseMacro{exp-python-Textualize_rich-timept-unit-dup1}
 & \UseMacro{exp-python-Textualize_rich-overhead-unit-dup1}
 \\
\UseMacro{proj-bokeh_bokeh}
 & \UseMacro{exp-python-bokeh_bokeh-num-vanilla-dup1}
 & \UseMacro{exp-python-bokeh_bokeh-num-inline-dup1}
 & \UseMacro{exp-python-bokeh_bokeh-time-vanilla-dup1}
 & \UseMacro{exp-python-bokeh_bokeh-timept-vanilla-dup1}
 & \UseMacro{exp-python-bokeh_bokeh-time-unit-and-inline-dup1}
 & \UseMacro{exp-python-bokeh_bokeh-timept-unit-and-inline-dup1}
 & \UseMacro{exp-python-bokeh_bokeh-overhead-unit-and-inline-dup1}
 & \UseMacro{exp-python-bokeh_bokeh-time-unit-dup1}
 & \UseMacro{exp-python-bokeh_bokeh-timept-unit-dup1}
 & \UseMacro{exp-python-bokeh_bokeh-overhead-unit-dup1}
 \\
\UseMacro{proj-chubin_cheat.sh}
 & \UseMacro{exp-python-chubin_cheat.sh-num-vanilla-dup1}
 & \UseMacro{exp-python-chubin_cheat.sh-num-inline-dup1}
 & \UseMacro{exp-python-chubin_cheat.sh-time-vanilla-dup1}
 & \UseMacro{exp-python-chubin_cheat.sh-timept-vanilla-dup1}
 & \UseMacro{exp-python-chubin_cheat.sh-time-unit-and-inline-dup1}
 & \UseMacro{exp-python-chubin_cheat.sh-timept-unit-and-inline-dup1}
 & \UseMacro{exp-python-chubin_cheat.sh-overhead-unit-and-inline-dup1}
 & \UseMacro{exp-python-chubin_cheat.sh-time-unit-dup1}
 & \UseMacro{exp-python-chubin_cheat.sh-timept-unit-dup1}
 & \UseMacro{exp-python-chubin_cheat.sh-overhead-unit-dup1}
 \\
\UseMacro{proj-davidsandberg_facenet}
 & \UseMacro{exp-python-davidsandberg_facenet-num-vanilla-dup1}
 & \UseMacro{exp-python-davidsandberg_facenet-num-inline-dup1}
 & \UseMacro{exp-python-davidsandberg_facenet-time-vanilla-dup1}
 & \UseMacro{exp-python-davidsandberg_facenet-timept-vanilla-dup1}
 & \UseMacro{exp-python-davidsandberg_facenet-time-unit-and-inline-dup1}
 & \UseMacro{exp-python-davidsandberg_facenet-timept-unit-and-inline-dup1}
 & \UseMacro{exp-python-davidsandberg_facenet-overhead-unit-and-inline-dup1}
 & \UseMacro{exp-python-davidsandberg_facenet-time-unit-dup1}
 & \UseMacro{exp-python-davidsandberg_facenet-timept-unit-dup1}
 & \UseMacro{exp-python-davidsandberg_facenet-overhead-unit-dup1}
 \\
\UseMacro{proj-geekcomputers_Python}
 & \UseMacro{exp-python-geekcomputers_Python-num-vanilla-dup1}
 & \UseMacro{exp-python-geekcomputers_Python-num-inline-dup1}
 & \UseMacro{exp-python-geekcomputers_Python-time-vanilla-dup1}
 & \UseMacro{exp-python-geekcomputers_Python-timept-vanilla-dup1}
 & \UseMacro{exp-python-geekcomputers_Python-time-unit-and-inline-dup1}
 & \UseMacro{exp-python-geekcomputers_Python-timept-unit-and-inline-dup1}
 & \UseMacro{exp-python-geekcomputers_Python-overhead-unit-and-inline-dup1}
 & \UseMacro{exp-python-geekcomputers_Python-time-unit-dup1}
 & \UseMacro{exp-python-geekcomputers_Python-timept-unit-dup1}
 & \UseMacro{exp-python-geekcomputers_Python-overhead-unit-dup1}
 \\
\UseMacro{proj-google-research_bert}
 & \UseMacro{exp-python-google-research_bert-num-vanilla-dup1}
 & \UseMacro{exp-python-google-research_bert-num-inline-dup1}
 & \UseMacro{exp-python-google-research_bert-time-vanilla-dup1}
 & \UseMacro{exp-python-google-research_bert-timept-vanilla-dup1}
 & \UseMacro{exp-python-google-research_bert-time-unit-and-inline-dup1}
 & \UseMacro{exp-python-google-research_bert-timept-unit-and-inline-dup1}
 & \UseMacro{exp-python-google-research_bert-overhead-unit-and-inline-dup1}
 & \UseMacro{exp-python-google-research_bert-time-unit-dup1}
 & \UseMacro{exp-python-google-research_bert-timept-unit-dup1}
 & \UseMacro{exp-python-google-research_bert-overhead-unit-dup1}
 \\
\UseMacro{proj-joke2k_faker}
 & \UseMacro{exp-python-joke2k_faker-num-vanilla-dup1}
 & \UseMacro{exp-python-joke2k_faker-num-inline-dup1}
 & \UseMacro{exp-python-joke2k_faker-time-vanilla-dup1}
 & \UseMacro{exp-python-joke2k_faker-timept-vanilla-dup1}
 & \UseMacro{exp-python-joke2k_faker-time-unit-and-inline-dup1}
 & \UseMacro{exp-python-joke2k_faker-timept-unit-and-inline-dup1}
 & \UseMacro{exp-python-joke2k_faker-overhead-unit-and-inline-dup1}
 & \UseMacro{exp-python-joke2k_faker-time-unit-dup1}
 & \UseMacro{exp-python-joke2k_faker-timept-unit-dup1}
 & \UseMacro{exp-python-joke2k_faker-overhead-unit-dup1}
 \\
\UseMacro{proj-mitmproxy_mitmproxy}
 & \UseMacro{exp-python-mitmproxy_mitmproxy-num-vanilla-dup1}
 & \UseMacro{exp-python-mitmproxy_mitmproxy-num-inline-dup1}
 & \UseMacro{exp-python-mitmproxy_mitmproxy-time-vanilla-dup1}
 & \UseMacro{exp-python-mitmproxy_mitmproxy-timept-vanilla-dup1}
 & \UseMacro{exp-python-mitmproxy_mitmproxy-time-unit-and-inline-dup1}
 & \UseMacro{exp-python-mitmproxy_mitmproxy-timept-unit-and-inline-dup1}
 & \UseMacro{exp-python-mitmproxy_mitmproxy-overhead-unit-and-inline-dup1}
 & \UseMacro{exp-python-mitmproxy_mitmproxy-time-unit-dup1}
 & \UseMacro{exp-python-mitmproxy_mitmproxy-timept-unit-dup1}
 & \UseMacro{exp-python-mitmproxy_mitmproxy-overhead-unit-dup1}
 \\
\UseMacro{proj-numpy_numpy}
 & \UseMacro{exp-python-numpy_numpy-num-vanilla-dup1}
 & \UseMacro{exp-python-numpy_numpy-num-inline-dup1}
 & \UseMacro{exp-python-numpy_numpy-time-vanilla-dup1}
 & \UseMacro{exp-python-numpy_numpy-timept-vanilla-dup1}
 & \UseMacro{exp-python-numpy_numpy-time-unit-and-inline-dup1}
 & \UseMacro{exp-python-numpy_numpy-timept-unit-and-inline-dup1}
 & \UseMacro{exp-python-numpy_numpy-overhead-unit-and-inline-dup1}
 & \UseMacro{exp-python-numpy_numpy-time-unit-dup1}
 & \UseMacro{exp-python-numpy_numpy-timept-unit-dup1}
 & \UseMacro{exp-python-numpy_numpy-overhead-unit-dup1}
 \\
\UseMacro{proj-pandas-dev_pandas}
 & \UseMacro{exp-python-pandas-dev_pandas-num-vanilla-dup1}
 & \UseMacro{exp-python-pandas-dev_pandas-num-inline-dup1}
 & \UseMacro{exp-python-pandas-dev_pandas-time-vanilla-dup1}
 & \UseMacro{exp-python-pandas-dev_pandas-timept-vanilla-dup1}
 & \UseMacro{exp-python-pandas-dev_pandas-time-unit-and-inline-dup1}
 & \UseMacro{exp-python-pandas-dev_pandas-timept-unit-and-inline-dup1}
 & \UseMacro{exp-python-pandas-dev_pandas-overhead-unit-and-inline-dup1}
 & \UseMacro{exp-python-pandas-dev_pandas-time-unit-dup1}
 & \UseMacro{exp-python-pandas-dev_pandas-timept-unit-dup1}
 & \UseMacro{exp-python-pandas-dev_pandas-overhead-unit-dup1}
 \\
\UseMacro{proj-psf_black}
 & \UseMacro{exp-python-psf_black-num-vanilla-dup1}
 & \UseMacro{exp-python-psf_black-num-inline-dup1}
 & \UseMacro{exp-python-psf_black-time-vanilla-dup1}
 & \UseMacro{exp-python-psf_black-timept-vanilla-dup1}
 & \UseMacro{exp-python-psf_black-time-unit-and-inline-dup1}
 & \UseMacro{exp-python-psf_black-timept-unit-and-inline-dup1}
 & \UseMacro{exp-python-psf_black-overhead-unit-and-inline-dup1}
 & \UseMacro{exp-python-psf_black-time-unit-dup1}
 & \UseMacro{exp-python-psf_black-timept-unit-dup1}
 & \UseMacro{exp-python-psf_black-overhead-unit-dup1}
 \\
\UseMacro{proj-pypa_pipenv}
 & \UseMacro{exp-python-pypa_pipenv-num-vanilla-dup1}
 & \UseMacro{exp-python-pypa_pipenv-num-inline-dup1}
 & \UseMacro{exp-python-pypa_pipenv-time-vanilla-dup1}
 & \UseMacro{exp-python-pypa_pipenv-timept-vanilla-dup1}
 & \UseMacro{exp-python-pypa_pipenv-time-unit-and-inline-dup1}
 & \UseMacro{exp-python-pypa_pipenv-timept-unit-and-inline-dup1}
 & \UseMacro{exp-python-pypa_pipenv-overhead-unit-and-inline-dup1}
 & \UseMacro{exp-python-pypa_pipenv-time-unit-dup1}
 & \UseMacro{exp-python-pypa_pipenv-timept-unit-dup1}
 & \UseMacro{exp-python-pypa_pipenv-overhead-unit-dup1}
 \\
\UseMacro{proj-scrapy_scrapy}
 & \UseMacro{exp-python-scrapy_scrapy-num-vanilla-dup1}
 & \UseMacro{exp-python-scrapy_scrapy-num-inline-dup1}
 & \UseMacro{exp-python-scrapy_scrapy-time-vanilla-dup1}
 & \UseMacro{exp-python-scrapy_scrapy-timept-vanilla-dup1}
 & \UseMacro{exp-python-scrapy_scrapy-time-unit-and-inline-dup1}
 & \UseMacro{exp-python-scrapy_scrapy-timept-unit-and-inline-dup1}
 & \UseMacro{exp-python-scrapy_scrapy-overhead-unit-and-inline-dup1}
 & \UseMacro{exp-python-scrapy_scrapy-time-unit-dup1}
 & \UseMacro{exp-python-scrapy_scrapy-timept-unit-dup1}
 & \UseMacro{exp-python-scrapy_scrapy-overhead-unit-dup1}
 \\
\hline
\HeaderAvg
 & \UseMacro{exp-python-num-vanilla-AVG-dup1}
 & \UseMacro{exp-python-num-inline-AVG-dup1}
 & \UseMacro{exp-python-time-vanilla-AVG-dup1}
 & \UseMacro{exp-python-timept-vanilla-MACROAVG-dup1}
 & \UseMacro{exp-python-time-unit-and-inline-AVG-dup1}
 & \UseMacro{exp-python-timept-unit-and-inline-MACROAVG-dup1}
 & \UseMacro{exp-python-overhead-unit-and-inline-MACROAVG-dup1}
 & \UseMacro{exp-python-time-unit-AVG-dup1}
 & \UseMacro{exp-python-timept-unit-MACROAVG-dup1}
 & \UseMacro{exp-python-overhead-unit-MACROAVG-dup1}
 \\
\HeaderSum
 & \UseMacro{exp-python-num-vanilla-SUM-dup1}
 & \UseMacro{exp-python-num-inline-SUM-dup1}
 & \UseMacro{exp-python-time-vanilla-SUM-dup1}
 & \HeaderNA
 & \UseMacro{exp-python-time-unit-and-inline-SUM-dup1}
 & \HeaderNA
 & \HeaderNA
 & \UseMacro{exp-python-time-unit-SUM-dup1}
 & \HeaderNA
 & \HeaderNA
 \\
\hline
\end{tabular}

    \vspace{4pt}
  \end{minipage}
  \begin{minipage}[t]{\textwidth}
    \centering
    \subcaption{\CaptionExpUtAndItDupsPython}
    \vspace{-4pt}
    
%% Automatically generated by pyutil.latex 

\begin{tabular}{|l|r|r||r|r||r|r|r||r|r|r|}
\hline
\HeaderDup & \HeaderNumVanilla & \HeaderNumITOnly & \HeaderTimeVanilla & \HeaderTimePerTestVanilla & \HeaderTimeITEnabled & \HeaderTimePerTestITEnabled & \HeaderOverheadITEnabled & \HeaderTimeITDisabled & \HeaderTimePerTestITDisabled & \HeaderOverheadITDisabled \\
\hline
\UseMacro{dup-1}
 & \UseMacro{exp-python-num-vanilla-SUM-dup1}
 & \UseMacro{exp-python-num-inline-SUM-dup1}
 & \UseMacro{exp-python-time-vanilla-SUM-dup1}
 & \UseMacro{exp-python-timept-vanilla-MACROAVG-dup1}
 & \UseMacro{exp-python-time-unit-and-inline-SUM-dup1}
 & \UseMacro{exp-python-timept-unit-and-inline-MACROAVG-dup1}
 & \UseMacro{exp-python-overhead-unit-and-inline-MACROAVG-dup1}
 & \UseMacro{exp-python-time-unit-SUM-dup1}
 & \UseMacro{exp-python-timept-unit-MACROAVG-dup1}
 & \UseMacro{exp-python-overhead-unit-MACROAVG-dup1}
 \\
\UseMacro{dup-10}
 & \UseMacro{exp-python-num-vanilla-SUM-dup10}
 & \UseMacro{exp-python-num-inline-SUM-dup10}
 & \UseMacro{exp-python-time-vanilla-SUM-dup10}
 & \UseMacro{exp-python-timept-vanilla-MACROAVG-dup10}
 & \UseMacro{exp-python-time-unit-and-inline-SUM-dup10}
 & \UseMacro{exp-python-timept-unit-and-inline-MACROAVG-dup10}
 & \UseMacro{exp-python-overhead-unit-and-inline-MACROAVG-dup10}
 & \UseMacro{exp-python-time-unit-SUM-dup10}
 & \UseMacro{exp-python-timept-unit-MACROAVG-dup10}
 & \UseMacro{exp-python-overhead-unit-MACROAVG-dup10}
 \\
\UseMacro{dup-100}
 & \UseMacro{exp-python-num-vanilla-SUM-dup100}
 & \UseMacro{exp-python-num-inline-SUM-dup100}
 & \UseMacro{exp-python-time-vanilla-SUM-dup100}
 & \UseMacro{exp-python-timept-vanilla-MACROAVG-dup100}
 & \UseMacro{exp-python-time-unit-and-inline-SUM-dup100}
 & \UseMacro{exp-python-timept-unit-and-inline-MACROAVG-dup100}
 & \UseMacro{exp-python-overhead-unit-and-inline-MACROAVG-dup100}
 & \UseMacro{exp-python-time-unit-SUM-dup100}
 & \UseMacro{exp-python-timept-unit-MACROAVG-dup100}
 & \UseMacro{exp-python-overhead-unit-MACROAVG-dup100}
 \\
\UseMacro{dup-1000}
 & \UseMacro{exp-python-num-vanilla-SUM-dup1000}
 & \UseMacro{exp-python-num-inline-SUM-dup1000}
 & \UseMacro{exp-python-time-vanilla-SUM-dup1000}
 & \UseMacro{exp-python-timept-vanilla-MACROAVG-dup1000}
 & \UseMacro{exp-python-time-unit-and-inline-SUM-dup1000}
 & \UseMacro{exp-python-timept-unit-and-inline-MACROAVG-dup1000}
 & \UseMacro{exp-python-overhead-unit-and-inline-MACROAVG-dup1000}
 & \UseMacro{exp-python-time-unit-SUM-dup1000}
 & \UseMacro{exp-python-timept-unit-MACROAVG-dup1000}
 & \UseMacro{exp-python-overhead-unit-MACROAVG-dup1000}
 \\
\hline
\end{tabular}

    \vspace{4pt}
  \end{minipage}
  \begin{minipage}[t]{\textwidth}
    \centering
    \subcaption{\CaptionExpUtAndItJava}
    \vspace{-4pt}
    
%% Automatically generated by pyutil.latex 

\begin{tabular}{|l|r|r||r|r||r|r|r||r|r|r|}
\hline
\HeaderProj & \HeaderNumVanilla & \HeaderNumITOnly & \HeaderTimeVanilla & \HeaderTimePerTestVanilla & \HeaderTimeITEnabled & \HeaderTimePerTestITEnabled & \HeaderOverheadITEnabled & \HeaderTimeITDisabled & \HeaderTimePerTestITDisabled & \HeaderOverheadITDisabled \\
\hline
\UseMacro{proj-alibaba_fastjson}
 & \UseMacro{exp-java-alibaba_fastjson-num-vanilla-dup1}
 & \UseMacro{exp-java-alibaba_fastjson-num-inline-dup1}
 & \UseMacro{exp-java-alibaba_fastjson-time-vanilla-dup1}
 & \UseMacro{exp-java-alibaba_fastjson-timept-vanilla-dup1}
 & \UseMacro{exp-java-alibaba_fastjson-time-unit-and-inline-dup1}
 & \UseMacro{exp-java-alibaba_fastjson-timept-unit-and-inline-dup1}
 & \UseMacro{exp-java-alibaba_fastjson-overhead-unit-and-inline-dup1}
 & \UseMacro{exp-java-alibaba_fastjson-time-unit-dup1}
 & \UseMacro{exp-java-alibaba_fastjson-timept-unit-dup1}
 & \UseMacro{exp-java-alibaba_fastjson-overhead-unit-dup1}
 \\
\UseMacro{proj-alibaba_nacos}
 & \UseMacro{exp-java-alibaba_nacos-num-vanilla-dup1}
 & \UseMacro{exp-java-alibaba_nacos-num-inline-dup1}
 & \UseMacro{exp-java-alibaba_nacos-time-vanilla-dup1}
 & \UseMacro{exp-java-alibaba_nacos-timept-vanilla-dup1}
 & \UseMacro{exp-java-alibaba_nacos-time-unit-and-inline-dup1}
 & \UseMacro{exp-java-alibaba_nacos-timept-unit-and-inline-dup1}
 & \UseMacro{exp-java-alibaba_nacos-overhead-unit-and-inline-dup1}
 & \UseMacro{exp-java-alibaba_nacos-time-unit-dup1}
 & \UseMacro{exp-java-alibaba_nacos-timept-unit-dup1}
 & \UseMacro{exp-java-alibaba_nacos-overhead-unit-dup1}
 \\
\UseMacro{proj-apache_dubbo}
 & \UseMacro{exp-java-apache_dubbo-num-vanilla-dup1}
 & \UseMacro{exp-java-apache_dubbo-num-inline-dup1}
 & \UseMacro{exp-java-apache_dubbo-time-vanilla-dup1}
 & \UseMacro{exp-java-apache_dubbo-timept-vanilla-dup1}
 & \UseMacro{exp-java-apache_dubbo-time-unit-and-inline-dup1}
 & \UseMacro{exp-java-apache_dubbo-timept-unit-and-inline-dup1}
 & \UseMacro{exp-java-apache_dubbo-overhead-unit-and-inline-dup1}
 & \UseMacro{exp-java-apache_dubbo-time-unit-dup1}
 & \UseMacro{exp-java-apache_dubbo-timept-unit-dup1}
 & \UseMacro{exp-java-apache_dubbo-overhead-unit-dup1}
 \\
\UseMacro{proj-apache_kafka}
 & \UseMacro{exp-java-apache_kafka-num-vanilla-dup1}
 & \UseMacro{exp-java-apache_kafka-num-inline-dup1}
 & \UseMacro{exp-java-apache_kafka-time-vanilla-dup1}
 & \UseMacro{exp-java-apache_kafka-timept-vanilla-dup1}
 & \UseMacro{exp-java-apache_kafka-time-unit-and-inline-dup1}
 & \UseMacro{exp-java-apache_kafka-timept-unit-and-inline-dup1}
 & \UseMacro{exp-java-apache_kafka-overhead-unit-and-inline-dup1}
 & \UseMacro{exp-java-apache_kafka-time-unit-dup1}
 & \UseMacro{exp-java-apache_kafka-timept-unit-dup1}
 & \UseMacro{exp-java-apache_kafka-overhead-unit-dup1}
 \\
\UseMacro{proj-apache_shardingsphere}
 & \UseMacro{exp-java-apache_shardingsphere-num-vanilla-dup1}
 & \UseMacro{exp-java-apache_shardingsphere-num-inline-dup1}
 & \UseMacro{exp-java-apache_shardingsphere-time-vanilla-dup1}
 & \UseMacro{exp-java-apache_shardingsphere-timept-vanilla-dup1}
 & \UseMacro{exp-java-apache_shardingsphere-time-unit-and-inline-dup1}
 & \UseMacro{exp-java-apache_shardingsphere-timept-unit-and-inline-dup1}
 & \UseMacro{exp-java-apache_shardingsphere-overhead-unit-and-inline-dup1}
 & \UseMacro{exp-java-apache_shardingsphere-time-unit-dup1}
 & \UseMacro{exp-java-apache_shardingsphere-timept-unit-dup1}
 & \UseMacro{exp-java-apache_shardingsphere-overhead-unit-dup1}
 \\
\UseMacro{proj-jenkinsci_jenkins}
 & \UseMacro{exp-java-jenkinsci_jenkins-num-vanilla-dup1}
 & \UseMacro{exp-java-jenkinsci_jenkins-num-inline-dup1}
 & \UseMacro{exp-java-jenkinsci_jenkins-time-vanilla-dup1}
 & \UseMacro{exp-java-jenkinsci_jenkins-timept-vanilla-dup1}
 & \UseMacro{exp-java-jenkinsci_jenkins-time-unit-and-inline-dup1}
 & \UseMacro{exp-java-jenkinsci_jenkins-timept-unit-and-inline-dup1}
 & \UseMacro{exp-java-jenkinsci_jenkins-overhead-unit-and-inline-dup1}
 & \UseMacro{exp-java-jenkinsci_jenkins-time-unit-dup1}
 & \UseMacro{exp-java-jenkinsci_jenkins-timept-unit-dup1}
 & \UseMacro{exp-java-jenkinsci_jenkins-overhead-unit-dup1}
 \\
\UseMacro{proj-skylot_jadx}
 & \UseMacro{exp-java-skylot_jadx-num-vanilla-dup1}
 & \UseMacro{exp-java-skylot_jadx-num-inline-dup1}
 & \UseMacro{exp-java-skylot_jadx-time-vanilla-dup1}
 & \UseMacro{exp-java-skylot_jadx-timept-vanilla-dup1}
 & \UseMacro{exp-java-skylot_jadx-time-unit-and-inline-dup1}
 & \UseMacro{exp-java-skylot_jadx-timept-unit-and-inline-dup1}
 & \UseMacro{exp-java-skylot_jadx-overhead-unit-and-inline-dup1}
 & \UseMacro{exp-java-skylot_jadx-time-unit-dup1}
 & \UseMacro{exp-java-skylot_jadx-timept-unit-dup1}
 & \UseMacro{exp-java-skylot_jadx-overhead-unit-dup1}
 \\
\hline
\HeaderAvg
 & \UseMacro{exp-java-num-vanilla-AVG-dup1}
 & \UseMacro{exp-java-num-inline-AVG-dup1}
 & \UseMacro{exp-java-time-vanilla-AVG-dup1}
 & \UseMacro{exp-java-timept-vanilla-MACROAVG-dup1}
 & \UseMacro{exp-java-time-unit-and-inline-AVG-dup1}
 & \UseMacro{exp-java-timept-unit-and-inline-MACROAVG-dup1}
 & \UseMacro{exp-java-overhead-unit-and-inline-MACROAVG-dup1}
 & \UseMacro{exp-java-time-unit-AVG-dup1}
 & \UseMacro{exp-java-timept-unit-MACROAVG-dup1}
 & \UseMacro{exp-java-overhead-unit-MACROAVG-dup1}
 \\
\HeaderSum
 & \UseMacro{exp-java-num-vanilla-SUM-dup1}
 & \UseMacro{exp-java-num-inline-SUM-dup1}
 & \UseMacro{exp-java-time-vanilla-SUM-dup1}
 & \HeaderNA
 & \UseMacro{exp-java-time-unit-and-inline-SUM-dup1}
 & \HeaderNA
 & \HeaderNA
 & \UseMacro{exp-java-time-unit-SUM-dup1}
 & \HeaderNA
 & \HeaderNA
 \\
\hline
\end{tabular}

    \vspace{4pt}
  \end{minipage}
  \begin{minipage}[t]{\textwidth}
    \centering
    \subcaption{\CaptionExpUtAndItDupsJava}
    \vspace{-4pt}
    
%% Automatically generated by pyutil.latex 

\begin{tabular}{|l|r|r||r|r||r|r|r||r|r|r|}
\hline
\HeaderDup & \HeaderNumVanilla & \HeaderNumITOnly & \HeaderTimeVanilla & \HeaderTimePerTestVanilla & \HeaderTimeITEnabled & \HeaderTimePerTestITEnabled & \HeaderOverheadITEnabled & \HeaderTimeITDisabled & \HeaderTimePerTestITDisabled & \HeaderOverheadITDisabled \\
\hline
\UseMacro{dup-1}
 & \UseMacro{exp-java-num-vanilla-SUM-dup1}
 & \UseMacro{exp-java-num-inline-SUM-dup1}
 & \UseMacro{exp-java-time-vanilla-SUM-dup1}
 & \UseMacro{exp-java-timept-vanilla-MACROAVG-dup1}
 & \UseMacro{exp-java-time-unit-and-inline-SUM-dup1}
 & \UseMacro{exp-java-timept-unit-and-inline-MACROAVG-dup1}
 & \UseMacro{exp-java-overhead-unit-and-inline-MACROAVG-dup1}
 & \UseMacro{exp-java-time-unit-SUM-dup1}
 & \UseMacro{exp-java-timept-unit-MACROAVG-dup1}
 & \UseMacro{exp-java-overhead-unit-MACROAVG-dup1}
 \\
\UseMacro{dup-10}
 & \UseMacro{exp-java-num-vanilla-SUM-dup10}
 & \UseMacro{exp-java-num-inline-SUM-dup10}
 & \UseMacro{exp-java-time-vanilla-SUM-dup10}
 & \UseMacro{exp-java-timept-vanilla-MACROAVG-dup10}
 & \UseMacro{exp-java-time-unit-and-inline-SUM-dup10}
 & \UseMacro{exp-java-timept-unit-and-inline-MACROAVG-dup10}
 & \UseMacro{exp-java-overhead-unit-and-inline-MACROAVG-dup10}
 & \UseMacro{exp-java-time-unit-SUM-dup10}
 & \UseMacro{exp-java-timept-unit-MACROAVG-dup10}
 & \UseMacro{exp-java-overhead-unit-MACROAVG-dup10}
 \\
\UseMacro{dup-100}
 & \UseMacro{exp-java-num-vanilla-SUM-dup100}
 & \UseMacro{exp-java-num-inline-SUM-dup100}
 & \UseMacro{exp-java-time-vanilla-SUM-dup100}
 & \UseMacro{exp-java-timept-vanilla-MACROAVG-dup100}
 & \UseMacro{exp-java-time-unit-and-inline-SUM-dup100}
 & \UseMacro{exp-java-timept-unit-and-inline-MACROAVG-dup100}
 & \UseMacro{exp-java-overhead-unit-and-inline-MACROAVG-dup100}
 & \UseMacro{exp-java-time-unit-SUM-dup100}
 & \UseMacro{exp-java-timept-unit-MACROAVG-dup100}
 & \UseMacro{exp-java-overhead-unit-MACROAVG-dup100}
 \\
\UseMacro{dup-1000}
 & \UseMacro{exp-java-num-vanilla-SUM-dup1000}
 & \UseMacro{exp-java-num-inline-SUM-dup1000}
 & \UseMacro{exp-java-time-vanilla-SUM-dup1000}
 & \UseMacro{exp-java-timept-vanilla-MACROAVG-dup1000}
 & \UseMacro{exp-java-time-unit-and-inline-SUM-dup1000}
 & \UseMacro{exp-java-timept-unit-and-inline-MACROAVG-dup1000}
 & \UseMacro{exp-java-overhead-unit-and-inline-MACROAVG-dup1000}
 & \UseMacro{exp-java-time-unit-SUM-dup1000}
 & \UseMacro{exp-java-timept-unit-MACROAVG-dup1000}
 & \UseMacro{exp-java-overhead-unit-MACROAVG-dup1000}
 \\
\hline
\end{tabular}

  \end{minipage}
  \end{small}
\end{table*}

\MyPara{\RQCostAlone: cost of running only \inlinetests}
Table~\ref{tab:exp-it-standalone} shows the results of running Python
and Java \inlinetests in the standalone mode.  Without duplicating
the \inlinetests in each example, the average time for running each \inlinetest
is \UseMacro{exp-python-timept-MACROAVG-dup1-its}s for Python and
\UseMacro{exp-java-timept-MACROAVG-dup1-its}s for Java.  As we
duplicate the \inlinetests in each example, the average time for running each
\inlinetest reduces to
\UseMacro{exp-python-timept-MACROAVG-dup1000-its}s for Python
and \UseMacro{exp-java-timept-MACROAVG-dup1000-its}s for Java.
There could be two reasons. First, the cost of reading a file and
extracting \inlinetests is amortized. Second, repeatedly executing
the same \inlinetest is faster.

Figure~\ref{fig:exp-it-standalone} shows how total and per-test
execution time scale as the number of \inlinetests grows.  There,
the total time for running \inlinetests stays almost constant when
duplicating the \inlinetests 10 or 100 times (corresponding to around 10 and
100 \inlinetests per file), and starts to grow dramatically when
duplicating 1000 times.  The Java version of \InlineTest
shows better scalability than the Python-version, as it is slower
initially but faster when duplicating 1000 times, probably due to just-in-time compilation.

\MyPara{\RQCostWithUnitEnabled: overhead of running unit tests with \inlinetests \emph{enabled}}
Table~\ref{tab:exp-ut-and-it} shows the results of running Python and
Java \inlinetests after integrating with the open-source projects and
their unit tests.  There, the \HeaderOverheadITEnabled{} columns show
the overhead when \inlinetests are enabled and executed during the
execution of existing unit tests.  Overall, without duplicating
\inlinetests
(tables~\ref{tab:exp-ut-and-it-python}~and~\ref{tab:exp-ut-and-it-java}),
the overhead of running \inlinetests is negligible compared to unit
tests, and is
\UseMacro{exp-python-overhead-unit-and-inline-MACROAVG-dup1}x for
Python and \UseMacro{exp-java-overhead-unit-and-inline-MACROAVG-dup1}x
for Java.  This observation holds when duplicating \inlinetests
(tables~\ref{tab:exp-ut-and-it-dups-python}~and~\ref{tab:exp-ut-and-it-dups-java});
for example, when duplicating \inlinetests 1000 times, which brings
the number of \inlinetests similar to the number of unit tests, the
overhead is
\UseMacro{exp-python-overhead-unit-and-inline-MACROAVG-dup1000}x for
Python and
\UseMacro{exp-java-overhead-unit-and-inline-MACROAVG-dup1000}x for
Java.  Negligible overhead may be due to \inlinetests running
much faster than unit tests.

\MyPara{\RQCostWithUnitDisabled: overhead of running unit tests with \inlinetests \emph{disabled}}
The \HeaderOverheadITDisabled{} columns in
Table~\ref{tab:exp-ut-and-it} show the overhead when \inlinetests are
disabled during the execution of existing unit tests.  The
\inlinetests are not executed, but having them in the code base may
require unit tests to execute additional no-op statements.
Nevertheless, we found such overhead to be negligible, even when
duplicating the \inlinetests for 10--1000 times; the negative
close-to-zero overhead numbers (e.g.,
\UseMacro{exp-python-overhead-unit-MACROAVG-dup1}x for Python when not
duplicating \inlinetests) are likely due to nondeterminism
during execution.

\section{User Study}
\label{sec:userstudy}

The goals of our study are to evaluate the ease with which developers learn and
use \InlineTest{}, and to obtain their perception about \inlinetesting or how
\InlineTest can be improved.

\subsection{Study Design}

We ask participants to complete three activities: (1)~a short
tutorial to learn about \inlinetesting and \InlineTest{} (expected duration: 20
minutes), (2)~four testing tasks in which they write \inlinetests for four specified
target statements (expected duration: 10 minutes per
task), and (3)~a questionnaire with six questions (unspecified duration).
We suggest a one-hour time limit, but results show
that most participants finish faster. We write scripts to
process the responses, and manually check the correctness of participants' \inlinetests.

We only use \InlineTest{} for Python in our user study to keep
participants focused on \inlinetesting and not on switching between
programming languages. We
plan to do a user study of \InlineTest{} for Java (and other
programming languages) in the future. A sample user study (without
responses) is in our GitHub
repository.
We briefly describe the activities that participants undertake.

\MyItPara{(1)~Tutorial} We provide an overview of \InlineTest's
API (Section~\ref{sec:tech:api}). Then, we ask each participant to run
a provided script to setup the environment. Finally, we illustrate \InlineTest
using three examples. The first example is a toy ``hello world''
example; the other two are examples from our
corpus.  Each example contains a code snippet, specifies a target
statement or two together with one or two \inlinetests per target
statement.  We also describe \InlineTest{}'s API and instructions for
running the \inlinetests.

\MyItPara{(2)~Using \inlinetests} We ask participants to write and run
\inlinetests for four examples from our corpus.  For each example, we present
the participant with the code snippet (without our \inlinetests) and
specify a target statement. Then, we ask participants to write one or
more \inlinetests for the target statement.  We also ask participants
to ensure that their \inlinetests pass.  Finally, we ask participants to
separately report the time taken to understand the target statement
and the time taken to write all \inlinetests.

\MyItPara{(3)~Survey} We ask participants to fill a questionnaire, to
record their experiences with \InlineTest{} and their feedback.
Specifically, we ask participants to (a)~rate the difficulty of
learning \InlineTest{}'s API and of writing \inlinetests, (b)~report
their number of years of general and Python programming experience (to
understand if expertise impacts their experiences), (c)~say whether
they think writing \inlinetests is beneficial for each of the four
tasks compared with unit tests (they can optionally justify their
``yes'' or ``no'' responses), (d)~comment on how to improve \InlineTest{}.

\MyPara{Participants} Our valid user study participants are
\UseMacro{NumUserStudyGrad} graduate students and
\UseMacro{NumUserStudyUnderGrad} undergraduate students from our
institutions and \UseMacro{NumUserStudyIndustry} professional software engineer.
We start with \UseMacro{NumUserStudyInitialParticipants}
participants. Two participants partake in a pilot study, but we
discard their responses after using those responses to refine the user study. We
then send the study to the other participants in batches of
\UseMacro{NumUserStudyFirstGroup} and
\UseMacro{NumUserStudySecondGroup}. No participant is a co-author of
this paper, and we confirm that none of them contributes to the
open-source projects being tested. We got
\UseMacro{NumUserStudyValidResponses} valid responses; participants
report an average \UseMacro{AVG_years_of_programming} years (median:
\UseMacro{MEDIAN_years_of_programming} years) of programming
experience. On a scale of 1 to 5, with 1 being novice and 5 being
expert, participants self-rate their Python expertise as
\UseMacro{AVG_python_expertise} on average (median:
\UseMacro{MEDIAN_python_expertise}).

\MyPara{\Inlinetests vs. unit tests} We did not ask user study
participants to write unit tests or to directly compare them with
\inlinetests for the testing tasks. Rather, we only ask for anecdotal comparisons of
\inlinetests and unit tests in the questionnaire. We chose this study
design for three reasons. First, setting up the unit testing environment
per project is hard (even for us) and
differs across projects. So, asking participants to set up
environments before writing unit tests could be a source of bias. Second,
providing a Docker image (or similar) could induce
bias---installing and running Docker containers could be hard
for participants who are unfamiliar
with Docker. Lastly, we do not assume familiarity with
\pytest, which participants would need to write unit tests in
Python. To work around these three problems, we provide participants with a script that sets up a
minimal Python runtime environment for \inlinetests. It takes only about one minute to run
the script.

\subsection{User Study Results}

\MyPara{Quantitative analysis} Our user study results are shown in
Table~\ref{table:user-study-res}, grouped by the four tasks. For each task, we show the average time (in
minutes) spent by each participant on understanding the target
statement, writing all \inlinetests, and writing each \inlinetest. We
also show the number of \inlinetests that participants write, the
number of participants for whom all \inlinetests pass, and the number
of participants who answer ``yes'' to ``writing \inlinetests is
beneficial compared with just writing unit tests''.
On a scale of 1 to 5 (1 being very difficult and 5 being very easy), participants
rank the difficulty of learning \InlineTest{} as
\UseMacro{AVG_difficulty_of_learning} (median:
\UseMacro{MEDIAN_difficulty_of_learning}) and rank the difficulty of writing
inline tests as \UseMacro{AVG_difficulty_of_using} (median:
\UseMacro{MEDIAN_difficulty_of_using}). On average, participants write
\UseMacro{AVG_num_tests} \inlinetests (median: \UseMacro{MEDIAN_num_tests}) per task,
and spend \UseMacro{AVG_time_per_test} (median:
\UseMacro{MEDIAN_time_per_test}) minutes to understand a target statement and
\UseMacro{AVG_writing_time} (median: \UseMacro{MEDIAN_writing_time}) minutes to
write an \inlinetest.

\MyPara{Qualitative analysis} All participants found \inlinetests to
be beneficial for some of the tasks. In fact, for all four tasks, most
participants think that writing \inlinetests is beneficial, and all
participants agree that \inlinetests are beneficial for Task 4. The
one participant who said that \inlinetesting is not beneficial for
Task 1 preferred to extract the target statement into a function and
then write unit tests. So, while they did not use \inlinetesting for
this task, they still found it important to test the target
statement. For Task 2, the one participant who did not find
\inlinetesting beneficial said that they think that the target
statement is too trivial to test. Lastly, the four participants who
did not find \inlinetesting useful for Task 3 provide two kinds of
reasons: (1)~the variable in the target statement is being returned
from the function, so a unit test would suffice (two participants);
and (2)~the target statement performs sorting, which is easy to
understand and does not warrant \inlinetesting (two participants). The
variance in perceptions on Tasks 1, Task 2, and Task 3, plus the
different reasons given by participants who think that a target statement
does not warrant an \inlinetest shows that developers will likely use
\inlinetests in different ways.

Participants provide feedback on how to further improve
\InlineTest, including by (a)~minimizing the long stack traces that
are shown when \inlinetests fail (\emph{``The stack trace you get when
  a test fails is quite long, but this is an easy fix''});
(b)~allowing \inlinetests to use symbolic variables (\emph{``Having
  tests with symbolic values, meaning that you don't provide values
  for inputs''}); (c)~providing other methods in the API that allow
writing other kinds of oracles beyond equality checks (\emph{``Other
  kinds of checks besides equality''}); (d)~supporting parameterized
\inlinetests, which we have now implemented (\emph{``I would like
  shortcut for checking for multiple inputs''}).

Participants also share feedback on using \InlineTest{}. A participant
liked having \inlinetests in addition to unit tests: \emph{``it is
  quite useful to have an inline testing option available.  Unit
  testing and inline testing don't have to be exclusionary, there are
  some situations where one might be preferable but having both as an
  option is nice''}. Another participant commented that there is a
learning curve: \emph{``I experienced a learning curve to using the
  framework. I was able to understand the structure of how to make
  ... tests much better after doing the first task''}. It will be
important in the future to investigate ways to lower the learning
curve. A participant was curious to know what the overhead is when
\inlinetests are disabled: \emph{``Does inline testing add overhead
  during production runs (i.e. no testing is needed)?''}. We answer
this question in Section~\ref{sec:eval:results}. Also, a participant
thinks \inlinetests may be better than \CodeIn{assert} statements
(\emph{``Inline tests can be good replacement for assertions''}).
Lastly, a participant made the connection to \printfdebugging:
\emph{``I would legitimately want to use a framework like this
  next time I felt the need to do printf debugging''}.

\begin{table}[t]
  \begin{center}
    \vspace{-10px}
\caption{\TitleUserStudyResults}
\vspace{-10px}
\scalebox{0.8}{
\begin{tabular}{|l|l|l|l|l|l|l|l|l|l|l|}
\hline
\HeaderTask & \multicolumn{2}{l|}{\HeaderUnderstandingTime} &  \multicolumn{2}{l|}{\HeaderWritingTime} &  \multicolumn{2}{l|}{\HeaderNumTests} &  \multicolumn{2}{l|}{\HeaderWritingTimePerTest} & \HeaderNumPassedTests & \HeaderBenefits \\
\cline{2-9}
 & \HeaderAvg & \HeaderMed & \HeaderAvg & \HeaderMed & \HeaderAvg & \HeaderMed & \HeaderAvg & \HeaderMed & & \\
\hline
1
 & \UseMacro{AVG_task_1_understanding_time}
 & \UseMacro{MEDIAN_task_1_understanding_time}
 & \UseMacro{AVG_task_1_writing_time}
 & \UseMacro{MEDIAN_task_1_writing_time}
 & \UseMacro{AVG_task_1_num_tests}
 & \UseMacro{MEDIAN_task_1_num_tests}
 & \UseMacro{AVG_task_1_time_per_test}
 & \UseMacro{MEDIAN_task_1_time_per_test}
 & \UseMacro{task_1_correct_times}/\UseMacro{NumUsers}
 & \UseMacro{task_1_benefit}/\UseMacro{NumUsers}
\\
\hline
2
 & \UseMacro{AVG_task_2_understanding_time}
 & \UseMacro{MEDIAN_task_2_understanding_time}
 & \UseMacro{AVG_task_2_writing_time}
 & \UseMacro{MEDIAN_task_2_writing_time}
 & \UseMacro{AVG_task_2_num_tests}
 & \UseMacro{MEDIAN_task_2_num_tests}
 & \UseMacro{AVG_task_2_time_per_test}
 & \UseMacro{MEDIAN_task_2_time_per_test}
 & \UseMacro{task_2_correct_times}/\UseMacro{NumUsers}
 & \UseMacro{task_2_benefit}/\UseMacro{NumUsers}
\\
\hline
3
 & \UseMacro{AVG_task_3_understanding_time}
 & \UseMacro{MEDIAN_task_3_understanding_time}
 & \UseMacro{AVG_task_3_writing_time}
 & \UseMacro{MEDIAN_task_3_writing_time}
 & \UseMacro{AVG_task_3_num_tests}
 & \UseMacro{MEDIAN_task_3_num_tests}
 & \UseMacro{AVG_task_3_time_per_test}
 & \UseMacro{MEDIAN_task_3_time_per_test}
 & \UseMacro{task_3_correct_times}/\UseMacro{NumUsers}
 & \UseMacro{task_3_benefit}/\UseMacro{NumUsers}
\\
\hline
4
 & \UseMacro{AVG_task_4_understanding_time}
 & \UseMacro{MEDIAN_task_4_understanding_time}
 & \UseMacro{AVG_task_4_writing_time}
 & \UseMacro{MEDIAN_task_4_writing_time}
 & \UseMacro{AVG_task_4_num_tests}
 & \UseMacro{MEDIAN_task_4_num_tests}
 & \UseMacro{AVG_task_4_time_per_test}
 & \UseMacro{MEDIAN_task_4_time_per_test}
 & \UseMacro{task_4_correct_times}/\UseMacro{NumUsers}
 & \UseMacro{task_4_benefit}/\UseMacro{NumUsers}
\\
\hline
\HeaderAvg
 & \UseMacro{AVG_understanding_time}
 & \UseMacro{MEDIAN_understanding_time}
 & \UseMacro{AVG_writing_time}
 & \UseMacro{MEDIAN_writing_time}
 & \UseMacro{AVG_num_tests}
 & \UseMacro{MEDIAN_num_tests}
 & \UseMacro{AVG_time_per_test}
 & \UseMacro{MEDIAN_time_per_test}
 & \HeaderNA
 & \HeaderNA
\\
\hline
\end{tabular}
}
\label{table:user-study-res}
\vspace{-10pt}
\end{center}
\end{table}

\section{Limitations}
\label{sec:limit}

We design the \InlineTest API based on \UseMacro{NumTotalFiles} examples that we
select from open-source projects. Also, the \inlinetest inputs and
expected outputs that we use in those tests were neither chosen by the
open-source project developers nor confirmed by them. So, it is not
yet clear if those developers will find our \inlinetests acceptable.

Our own programming experience tells us that more kinds of oracles
will likely need to be supported in \InlineTest. For example, we do not
yet support expected exceptions or allow checking near equality
between floating point values. The current limited set of oracles in \InlineTest results from
using \UseMacro{NumTotalFiles} examples to guide our design. In the
future, by collecting more examples and requirements, \InlineTest{}
can possibly be extended to support more kinds of oracles.

In terms of implementation, Section~\ref{sec:desiderata} shows the
list of language agnostic \desiderata that \InlineTest does not yet
support (\xmark) and those that it only partially supports
(\pmark). This paper motivates, defines, and evaluates \inlinetests as
a way to prove the concept. The engineering effort to fully support
all the \desiderata is a matter of time and resources that we will
invest into seeing that \inlinetests become more mature.

An \inlinetest is inserted as code directly following the code under
test.  In the unlikely case when the code
under test is in a large method or file, inserting \inlinetests may
cause code-too-large errors due to limitations of compilation
tool chains (for example, a Java method can only have a maximum of
65535 bytes of bytecode~\cite{JavaMethodLimit}).

Our current Java \InlineTest implementation is designed to support
language features of Java 8, and it may not work for newer language
features in more recent Java versions. In the opposite direction, our
current Python \InlineTest implementation is designed to support
language features of Python 3.6 and above, so it may not work for
older Python versions.

If a target statement invokes a method with arguments that need to be
assigned in an \inlinetest, then the current \InlineTest
implementation cannot be used to check that target statement (Hence,
the \xmark{} on Requirement~\ref{requirement:arguments} in
Section~\ref{sec:desiderata}). We already observed a consequence of
this limitation in our attempt to write \inlinetests for statements
that use Java's stream API. Most stream operations invoke the kind of
method-with-arguments that we do not yet support. Also, stream
operations typically invoke several methods, so testing them with
\inlinetests can seem like writing unit tests. Finding smart ways to
support the testing of stream operations will be a priority---the
complexity and popularity of stream operations make them attractive
candidates for \inlinetesting.

\Inlinetesting may not generalize well to programming languages that
do not use the imperative style of Java and Python. In particular,
more thoughts need to be given in the future on whether and how
\inlinetesting can be realized effectively for functional languages
like Haskell, logic programming languages like Prolog, or
domain-specific languages like SQL.

We have not investigated how well \inlinetesting can fit into
different software and test design processes. So, it is not yet clear
what impact, if any, \inlinetests will have in the presence of
different testing methodologies. For example, since \inlinetests check
\emph{existing} target statements, its role may be limited in
organizations that follow test-driven development
(TDD)~\cite{astels2003test, beck2003test, santos2021family}. (In TDD, tests are written
prior to writing code.)  As another
example, what role should \inlinetests play during regression testing
and how often should they be re-run during software evolution?
Similarly, it may be that \inlinetests are more useful in systems
where testability~\cite{freedman1991testability} was not a first-class
concern during programming.  That is, \inlinetests may be more helpful
in legacy systems or systems with large monolithic components than in
newer systems that are designed to be unit-testable from the ground
up. We leave the investigation of how to fit \inlinetests into
different software- and test-design processes as future work.

\section{Related Work}
\label{sec:related}

\MyPara{Testing and debugging} Karampatsis and
Sutton~\cite{karampatsis2020often}, and Kamienski et
al.~\cite{kamienski2021pysstubs} curated datasets of single-statement
bugs (SStuBs) in Java and Python, respectively. Also, Latendresse et
al.~\cite{latendresse2021effective} find that continuous integration
(CI) rarely detects SStuBs.  These works on SStuBs further motivate
the need for direct support for checking individual statements, which
\inlinetests provide.

Michael et al.~\cite{michael2019regexes} found that regexes are hard to
read, find, validate, and document.
Eghbali and Pradel~\cite{eghbali2020no} also found that string-related
bugs are common in JavaScript programs.

Section~\ref{sec:example} discussed how \inlinetests can mitigate these problems and
how \InlineTest helped find regex-related and
string-manipulation bugs.

Doctest~\cite{doctest} in Python allows writing tests in function
docstrings.  \Inlinetests are similar to doctests in that both can
help with code comprehension.  But, doctest only supports
function-level testing, while \inlinetests only support statement-level
testing.

Regression test selection (RTS)~\cite{engstrom2010systematic, gligoric2014empirical,
gligoric2015practical, EkstaziToolDemo, ShiEtAlReflectionAwareRTS2019OOPSLA, legunsenextensive, ZhuETAL2019ICSE} speeds up regression testing by only re-running tests that are affected by
code changes.  Section~\ref{sec:eval} showed that
each \inlinetest runs very fast compared to unit
tests, but RTS for \inlinetests may become important as 
\inlinetests usage increases.

In-vivo testing~\cite{murphy2009quality} executes tests in the
deployment environment, to find defects that are hidden by the clean
test environment. In-vivo tests are method-level tests, while
\inlinetests statement-level tests, and \InlineTest currently targets the
test environment.

Fault localization~\cite{wong2016survey, pearson2017evaluating,
  weiser1979program, agrawal1990dynamic, agrawal1995fault,
  liblit2005scalable} helps finds faulty statements that cause a test
failure.  Inaccurate fault localization can occur for unit tests that
cover many statements~\cite{steimann2013threats, lei2018test}.  We expect fault localization for
\inlinetests to be more accurate since they check the immediately
preceding statement that is not an \inlinetest.

\MyPara{Assertions and design by contract}
The \texttt{assert}
construct in many programming languages, e.g.,~\cite{JavaAssert,
  PythonAssert, hoare2003assertions, taylor1980assertions}, allows
checking that a condition holds on the current program state. An
\inlinetests, like an \texttt{assert} ~\cite{voas1994putting}, can be
written after any statement in the production (not test) code.
\Inlinetests allow developer-provided input and are only to be used
for test-time checking, but \texttt{assert}s do not allow developers
to provide arbitrary inputs and can be used for production-time
checking~\cite{rosenblum1995practical}.

There is a lot of work on design-by-contract
(DBC)~\cite{meyer1992applying, PyContracts, pschanely2017,
  Parquery2018, leavens1999jml, bartetzko2001jass,
  milicevic2011unifying, nie2020unifying, rosenblum1995practical} for
specifying preconditions, postconditions, and invariants. DBC tools
include PyContracts~\cite{PyContracts},
Crosshair~\cite{pschanely2017}, Icontract~\cite{Parquery2018} for
Python, and JML~\cite{leavens1999jml}, Jass~\cite{bartetzko2001jass},
Squander~\cite{milicevic2011unifying},
Deuterium~\cite{nie2020unifying} for Java. DBC helps check and
comprehend hard-to-understand programs---goals that \inlinetests also
target.  DBC typically requires developers to use a different
programming language/paradigm developers, so there may be a higher learning
curve. In contrast, \inlinetests are written in the same
language/paradigm as the code.  Also, DBC enables method-level checks
(except for loop invariants~\cite{hoare1969axiomatic,
  floyd1993assigning, furia2014loop}), but \inlinetests check
statements.

\MyPara{Domain specific languages}
We provide \InlineTest as an API in both Python and Java.  However,
the design of our API
was inspired by prior work on domain specific languages
for writing executable comments~\cite{nie2019framework} and
contracts~\cite{nie2020unifying}.

\section{Conclusion}
\label{sec:conclusion}

If developers could write tests for individual program statements,
then they would be able to meet testing needs for which they currently have
little to no support. Such needs are at a lower granularity level
than what today's testing frameworks support, or for which currently supported
levels of test granularity are ill-suited.  We introduced a new
kind of tests, called \inlinetests to help test
individual statements. We implemented the first \inlinetesting
framework, \InlineTest{}, to meet language-agnostic \desiderata that
we define. Our assessment of \InlineTest via a user study and via
performance measurements showed that \inlinetesting is
promising---participants find it easy to learn and use \inlinetesting
and the additional cost of running \inlinetests is tiny. We outline several
directions in which \InlineTest can be extended to make it more mature
and to meet developer needs across programming
languages.

\begin{acks}
We thank Nader Al Awar, Darko Marinov, August Shi, Aditya Thimmaiah, Zhiqiang Zang,
Jiyang Zhang and the anonymous reviewers for their feedback on this
work. We also thank all the user study participants. This work was
partially supported by
a Google Faculty Research Award
and the US National Science Foundation under Grant
Nos.~1652517, 2019277, 2045596, 2107291, 2217696.
\end{acks}

\bibliography{bib}

\end{document}